\documentclass[11pt,a4paper]{article}
\usepackage{jheppub}
\usepackage{multirow,xspace}
\usepackage{amsmath,amsfonts,amssymb}
\usepackage{graphicx}
\usepackage{ulem,fancyvrb}
\usepackage{hyperref}

\usepackage{cleveref}
\crefname{table}{Table}{Tables}
\crefname{figure}{Fig.}{Figs.}
\crefname{equation}{Eq.}{Eqs.}

\usepackage[usenames,dvipsnames,svgnames,table]{xcolor}
\usepackage{xfrac}
\usepackage{cancel}

\let\oldsqrt\sqrt
\def\sqrt{\mathpalette\DHLhksqrt}
\def\DHLhksqrt#1#2{%
\setbox0=\hbox{$#1\oldsqrt{#2\,}$}\dimen0=\ht0
\advance\dimen0-0.2\ht0
\setbox2=\hbox{\vrule height\ht0 depth -\dimen0}%
{\box0\lower0.4pt\box2}}

\DeclareMathOperator{\sgn}{sgn}

\DeclareMathOperator{\SU}{SU}
\DeclareMathOperator{\gU}{U}

\newcommand{\mo}{\ensuremath{m_0}\xspace}
\newcommand{\mhf}{\ensuremath{m_{\sfrac{1}{2}}}\xspace}
\newcommand{\az}{\ensuremath{A_0}\xspace}
\newcommand{\tb}{\ensuremath{\tan\beta}\xspace}
\newcommand{\sgnmu}{\ensuremath{\sgn\left(\mu\right)}\xspace}

\newcommand{\cm}{\ensuremath{\sqrt{s}}\xspace}

\newcommand{\TeV}{\ensuremath{\,\text{Te\kern -0.10em V}}\xspace}
\newcommand{\GeV}{\ensuremath{\,\text{Ge\kern -0.10em V}}\xspace}
\newcommand{\MeV}{\ensuremath{\,\text{Me\kern -0.10em V}}\xspace}
\newcommand{\fb}{\ensuremath{\,\text{fb}}\xspace}
\newcommand{\ifb}{\ensuremath{\,\text{fb}^{-1}}\xspace}

\newcommand{\gev}{\ensuremath{\text{Ge\kern -0.08em V}}\xspace}
\newcommand{\tev}{\ensuremath{\text{Te\kern -0.08em V}}\xspace}

\newcommand{\hz}{\ensuremath{h^0}\xspace}
\newcommand{\Hz}{\ensuremath{H^0}\xspace}
\newcommand{\Az}{\ensuremath{A^0}\xspace}
\newcommand{\Hpm}{\ensuremath{H^\pm}\xspace}

\newcommand{\g}{\ensuremath{\widetilde{g}}\xspace}

\newcommand{\na}{\ensuremath{{\chi}_1^0}\xspace}
\newcommand{\nb}{\ensuremath{{\chi}_2^0}\xspace}
\newcommand{\nc}{\ensuremath{{\chi}_3^0}\xspace}

\newcommand{\cha}{\ensuremath{{\chi}_1^\pm}\xspace}

\newcommand{\chabar}{\ensuremath{{\chi}_1^\mp}\xspace}
\newcommand{\chaminus}{\ensuremath{{\chi}_1^-}\xspace}
\newcommand{\chaplus}{\ensuremath{{\chi}_1^+}\xspace}

\newcommand{\sta}{\ensuremath{\widetilde{\tau}_1}\xspace}

\newcommand{\stopa}{\ensuremath{\widetilde{t}_1}\xspace}

\newcommand{\sbota}{\ensuremath{\widetilde{b}_1}\xspace}

\newcommand{\q}{\ensuremath{\widetilde{q}}\xspace}

\newcommand{\stopx}{\ensuremath{\widetilde{t}}\xspace}

\newcommand{\bsmumu}{\ensuremath{B^0_s\to \mu^+\mu^-}\xspace}

\newcommand{\bsg}{\ensuremath{B\to X_s\gamma}\xspace}
\newcommand{\br}[1]{\ensuremath{\mathcal{B}r\left(#1\right)}\xspace}

\newcommand{\order}[1]{\ensuremath{\mathcal{O}\left(#1\right)}\xspace}

\DeclareRobustCommand{\[}{\begin{equation}}
\DeclareRobustCommand{\]}{\end{equation}}

\newcommand{\et}{{\ensuremath{E_{\rm T}^\text{miss}}\xspace}}
\newcommand{\pt}{{\ensuremath{p_{\rm T}}\xspace}}
\newcommand{\ptvec}{{\ensuremath{\vec{p}_{\rm T}}\xspace}}

\newcommand{\ST}{S_{\textrm{\footnotesize T}}}
\newcommand{\etmiss}{E_{\textrm{\footnotesize T}}^{\textrm{\footnotesize miss}}}
\newcommand{\htj}{H_{\textrm{\footnotesize T}}^{\textrm{\footnotesize 4j}}}
\newcommand{\meffj}{m_{\textrm{\footnotesize eff}}^{\textrm{\footnotesize 4j}}}
\newcommand{\meffincl}{m_{\textrm{\footnotesize eff}}^{\textrm{\footnotesize incl}}}
\newcommand{\mSFOS}{m_{\textrm{\footnotesize SFOS}}}

\newcommand{\NEUAffil}{Department of Physics, Northeastern University, Boston, MA 02115, USA}
\newcommand{\OKAffil}{Homer L. Dodge Department of Physics and Astronomy, \\
University of Oklahoma, Norman, OK 73019, USA}
\newcommand{\MTAAffil}{MTA-DE Particle Physics Research Group, 
University of Debrecen,\\ H-4010 Debrecen P.O. Box 105, Hungary}
\newcommand{\MonashAffil}{ARC Centre of Excellence for Particle Physics at the Terascale, \\ 
School of Physics, Monash University, Melbourne VIC 3800, Australia}

\title{
Sparticle Mass Hierarchies, Simplified Models from SUGRA Unification, and
Benchmarks for LHC Run-II SUSY Searches
}

\author[a]{David Francescone}
\author[b,c]{Sujeet Akula}
\author[d]{Baris Altunkaynak}
\author[a]{Pran Nath}
\affiliation[a]{\NEUAffil}
\affiliation[b]{\MTAAffil}
\affiliation[c]{\MonashAffil}
\affiliation[d]{\OKAffil}
\emailAdd{d.francescone@neu.edu}
\emailAdd{sujeet.akula@coepp.org.au}
\emailAdd{baris.altunkaynak@ou.edu}
\emailAdd{p.nath@neu.edu}

\abstract{
Sparticle mass hierarchies contain significant information regarding the origin
and nature of supersymmetry breaking. The hierarchical patterns are severely
constrained by electroweak symmetry breaking as well as by the astrophysical and
particle physics data. They are further constrained by the Higgs boson mass
measurement. The sparticle mass hierarchies can be used to generate simplified
models consistent with the high scale models. In this work we consider
supergravity models with universal boundary conditions for soft parameters at
the unification scale as well as supergravity models with nonuniversalities and
delineate the list of sparticle mass hierarchies for the five lightest
sparticles.  Simplified models can be obtained by a truncation of these,
retaining a smaller set of  lightest particles. 
The mass hierarchies and their truncated versions enlarge significantly the list of simplified
models currently being used in the literature.
Benchmarks for a variety of supergravity unified models
appropriate for SUSY searches at future colliders
are also presented. The signature
analysis of two benchmark models has been carried out and a discussion of the 
searches needed for their discovery at LHC RUN-II is given.
{An analysis of the spin-independent neutralino-proton cross section exhibiting
the Higgs boson mass dependence and the hierarchical patterns is also carried
out. It is seen that a knowledge of the spin-independent neutralino-proton cross
section and the neutralino mass will narrow down the list of the allowed
sparticle  mass hierarchies. Thus dark matter experiments along with analyses
for the LHC Run-II will provide  strong clues to the nature of symmetry breaking at the
unification scale.}
}

\arxivnumber{1410.4999}

\begin{document}
\maketitle

\section{Introduction \label{sec1}}
The discovery~\cite{Chatrchyan:2012ufa,Aad:2012tfa} of the Higgs
boson~\cite{Englert:1964et,Higgs:1964pj,Guralnik:1964eu} and the measurement of
its mass at $\sim\! 126\GeV$ have strong implications for discovery of
supersymmetry.  In  {the minimal supersymmetric standard model (MSSM)} one identifies the observed Higgs boson as the
lightest $CP$-even state $h^0$ (see e.g., \cite{Akula:2011aa,Baer:2011ab,
Arbey:2011ab,Draper:2011aa,Carena:2011aa,Akula:2012kk,
Arbey:2012dq,Strege:2012bt}).
It is noteworthy that the observed Higgs boson mass lies below but close to the
upper limit on the Higgs boson mass predicted in supergravity grand unified
models~\cite{Chamseddine:1982jx,Nath:1983aw,Hall:1983iz,Arnowitt:1992aq}
with radiative breaking of the electroweak symmetry (for a review
see\cite{Ibanez:2007pf}), and this upper limit is well known to be around
$130\GeV$~\cite{Akula:2011aa,Baer:2011ab,Arbey:2011ab,Draper:2011aa,Carena:2011aa,Akula:2012kk,Arbey:2012dq,Strege:2012bt,Buchmueller:2011ab,Baer:2012mv}.
Further, in supersymmetric models and specifically those within supergravity
grand unification, one finds that a Higgs mass of $\sim\! 126\GeV$ implies the
scale of supersymmetry to be large with the squark masses typically lying in
the few \TeV region~\cite{Akula:2011aa,Baer:2011ab, Arbey:2011ab,Draper:2011aa,Carena:2011aa,Akula:2012kk, Arbey:2012dq,Strege:2012bt,Buchmueller:2011ab,Baer:2012mv,Akula:2013ioa}.
The largeness of the SUSY scale explains the non-observation of sparticles in
searches in Run-I of the LHC.  However, the LHC energy is being ramped up to
$\cm=13\TeV$ and one expects some of the light sparticles to show up in
Run-II\footnote{Although the current plan is for the LHC to operate at
$\cm=13\TeV$ for Run-II, we will carry out the analysis at $\cm=14\TeV$, using
the Snowmass~\cite{SMSnowmass} Standard Model {backgrounds.}}.
 
The nature of the observed sparticles, and more generally the hierarchical mass
patterns,  hold a key to the nature of symmetry breaking at high scales in
unified models.  {Given that} there are 31 additional particles beyond the
spectrum of the Standard Model, there are {\itshape a priori}
$31!\sim8\times10^{33}$ ways in which these particles can arrange themselves.
This is the landscape of {possible} mass hierarchies of the new particles\footnote{We
would loosely call these the sparticle mass hierarchies even though they contain
the Higgs boson states, $H^0$, $A^0$, $H^{\pm}$, which are $R$ parity
even.}\textsuperscript{,}\footnote{The landscape is even larger in that the mass
gaps among the sparticles  can vary continuously which makes the allowed sparticle
landscape larger than {even} the string landscape which has as many as
$10^{500}$  possible vacua.}. The number of allowed possibilities is significantly reduced in
supergravity grand unification with radiative breaking of the electroweak
symmetry\cite{Feldman:2007zn,Feldman:2007fq,Feldman:2008hs}.  {Additionally},
the accelerator and dark matter constraints further reduce the allowed number of
possibilities. The landscape of supergravity based models was analyzed in  a
number of
works~\cite{Feldman:2007zn,Feldman:2007fq,Feldman:2008hs,Chen:2010kq,Berger:2008cq,Conley:2010du,Altunkaynak:2010tn,Nath:2010zj} 
which, however, were all before the discovery of the Higgs boson and a
measurement of its mass.

Regarding the Higgs boson mass of $126\GeV$, there are a limited number of ways
in which one can lift its tree mass which lies below $M_Z$ to the observed
value. These include $D$-term contributions from extra $\gU(1)$'s, loop
contributions from extra matter~\cite{Feng:2013mea} or {large loop corrections 
from within the MSSM}. The latter possibility implies a relatively high scale of
supersymmetry, which explains in part the reason for its non-observation thus
far. {In this work,} we revisit the sparticle landscape analysis taking into
account the constraint from the Higgs boson mass measurement on the sparticle
landscape. We analyze several different classes of high scale models: these
include mSUGRA (also called CMSSM) and supergravity models with nonuniversal
boundary conditions at the grand unification scale which include
nonuniversalities in the $\SU(2)_L$ and $\SU(3)_C$ gaugino sector, in the Higgs
sector and in the third generation sfermion sector.  The most dominant
hierarchical patterns that emerge are identified.  The hierarchical patterns
provide a simple way to connect the simplified
models~\cite{ArkaniHamed:2007fw,Alwall:2008ag,Alwall:2008va,Alves:2011sq,Alves:2011wf,Papucci:2011wy,Mahbubani:2012qq,Chatrchyan:2013sza,Cohen:2013xda}
with grand unified models. Specifically, we consider five particle mass
hierarchies where the various combinations of the five lightest particles that
originate in supergravity models are investigated. These five particle mass
hierarchies {effectively} constitute {existing and novel} simplified
models.  {The hierarchy of five particles} can be further truncated to give
simplified models with three or four lightest particles {as has been more
common}. It should be noted that these simplified models are obviously part of a
UV complete theory since they are obtained by truncation of the spectrum arising
from a high scale model. 

The outline of the rest of the paper is as follows: In \cref{sec2} we give
details of the analysis and  a cartography of the allowed 5 particle mass
hierarchies including  the LSP. Five different classes of high scale boundary
conditions are analyzed.  In \cref{sec3} we connect simplified models to the
sparticle mass hierarchies.  
 In \cref{sec5} we discuss the generic signatures that one
expects for the mass patterns.  In \cref{sec6} we give benchmarks for future searches for 
supersymmetry at colliders. We also give a signature analysis of two benchmark models 
with a discussion of cuts needed for their discovery at  the LHC Run-II.
  {In \cref{sec4} we carry out an analysis of spin
independent neutralino-proton cross section in terms of the sparticle mass
patterns to see how a measurement of the spin-independent cross
section along with  a knowledge of the neutralino mass can narrow down the
possible hierarchical patterns.}
 Conclusions are given in \cref{sec7}.

\section{Details of the analysis\label{sec2}}
We consider several different supergravity  models of soft breaking. These
include supergravity models with (1) universal boundary conditions as well as
supergravity models with nonuniversal boundary conditions (nuSUGRA) at the grand
unification scale\footnote{The literature on supergravity models with
nonuniversalities is vast. For a sample of works  on nonuniversalities the
reader may refer
to~\cite{Anderson:1996bg,Nath:1997qm,Ellis:2002wv,Anderson:1999uia,Huitu:1999vx,Corsetti:2000yq,Chattopadhyay:2001mj,Chattopadhyay:2001va,Martin:2009ad,Feldman:2009zc,Gogoladze:2012yf,Ajaib:2013zha,Kaufman:2013oaa}
and for a review see~\cite{Nath:2010zj}.}. {Within nuSUGRA, we consider (2)
nonuniversalities  in the $\SU(2)_L$ gaugino mass sector,  (3) nonuniversalities
in the $\SU(3)_C$ gaugino mass  sector, (4)  nonuniversalities in the flavor
sector with the squark masses for the third generation being different from the
masses in the first two generations, and (5) nonuniversalities in the Higgs
sector}. The parameter space  probed for each of these models is discussed
below. \\

{
\noindent
Model [1]: mSUGRA: $\mo \in
[0.1,10] \TeV$, $\mhf \in [0.1,1.5] \TeV$, $\frac{\az}{\mo} \in [-5,5]$,
$\tan\beta \in [2,50]$. \\

\noindent
Model [2]: nuSUGRA, light chargino: \\
\indent $\mo \in [0.1,10] \TeV$, $M_1=M_3=\mhf
\in [0.1,1.5] \TeV$, $M_2 = \alpha \mhf$, $\alpha \in [\frac{1}{2},1]$,
$\frac{\az}{\mo} \in [-5,5]$, $\tan\beta \in [2,50]$. \\

\noindent
Model [3]: nuSUGRA, light gluino: \\
\indent $\mo \in [0.1,10] \TeV$, $M_1=M_2=\mhf \in
[0.1,1.5] \TeV$, $M_3 = \alpha \mhf$, $\alpha \in [\frac{1}{6},1]$,
$\frac{\az}{\mo} \in [-5,5]$, $\tan\beta \in [2,50]$. \\

\noindent
Model [4]: nuSUGRA, nonuniversal Higgs: \\
\indent $\mo \in [0.1,10] \TeV$, $\mhf \in
[0.1,1.5] \TeV$, $\frac{\az}{\mo} \in [-5,5]$, $\tan\beta \in [2,50]$,
$m_{H_i}(\textrm{M}_G)= \mo(1+\delta_i)$, where $\delta_i \in
[-0.9,1]$.\\

\noindent
Model [5]: nuSUGRA, light 3rd generation: \\
\indent $\mo^{(1)}=\mo^{(2)}=\mo \in [0.1,10]
\TeV$, $\mo^{(3)}=\frac{\mo}{1\TeV+\mo}$, $\mhf \in [0.1,1.5] \TeV$,
$\frac{\az}{\mo} \in [-5,5]$, $\tan\beta \in [2,50]$.\\
}

In each of the above cases, $\mu$ was taken to be positive.  The overview of the
parameter scan for the five classes of models listed above is given in
\cref{tab1}.  The scans were
performed using {\scshape\ttfamily SusyKit}~\cite{susykit} which employs
{\scshape\ttfamily SOFTSUSY}~\cite{Allanach:2001kg} for 2-loop RG evolution including
sparticle thresholds and for sparticle mass calculations,
{\scshape\ttfamily FeynHiggs}~\cite{Heinemeyer:1998yj, Hahn:2010te} {for}
computing the Higgs boson masses, {including the recently added resummation
to all orders of leading and subleading logs of type
$\log(m_{\stopx}/m_t)$~\cite{Hahn:2013ria}}, and {\scshape\ttfamily
micrOMEGAs}~\cite{Belanger:2010st} to calculate the relic density and flavor
observables. We only considered points where the LSP is the neutralino, the
thermal relic density of the LSP is not overabundant, i.e., $\Omega_\chi h^2 <
0.12$, and the Higgs boson mass is not too light with ${m_{h^0}} > 120\GeV$
\footnote{
 We have chosen a bit generous  window on the Higgs boson mass for the
 following reason: aside from the possible errors in theory computations which
 are now reduced since we use FeynHiggs, in extended models with extra $\gU(1)$
 corrections to the Higgs mass can arise of the size  of $\order{1\GeV}$ from
 extra D-term contributions. Also  extra vector-like matter if it exists could
 make a contribution of the same size {(see,e.g., \cite{Feng:2013mea}}).  So in
 order that our analysis also apply  to such models, we consider a bit generous
 window but the results for the narrower window on the Higgs mass are easily
 extracted from the analysis presented in the paper.  For this reason we have
 color-coded  \cref{fig3} for the mSUGRA case and \cref{fig4} for the nuSUGRA
 cases with the Higgs mass. 
}. (For some recent works on mSUGRA
see~\cite{Fowlie:2014awa,Kim:2013uxa,Fowlie:2014faa,Roszkowski:2014wqa}).
These constraints immediately reduce the available number of sparticle mass
hierarchies.  Of the large number of model points in the scan of \cref{tab1} two
models were investigated in detail for their discovery potential at RUN II of
LHC.  One of these is an mSUGRA model whose parameters are listed in
\cref{tabbenchmark1} and the other is an nuSUGRA model whose parameters are
listed in \cref{tabbenchmark2}.  The signatures for these models were
investigated using the ATLAS cuts at $\cm=8\TeV$ as given in \cref{tab-atlas}.

{We begin the study of the hierarchies by first identifying the next-to-lightest
sparticle (NLSP). For each NLSP case, we next determine the possible
4-sparticle hierarchies (LSP, NLSP, and two heavier sparticles). Within these,
we additionally identify the next lightest sparticle, giving a hierarchy of the
5 lightest sparticles. These hierarchies are labeled by the following scheme: we 
begin with a symbol for the NLSP, followed by a number for one of the possible
4-sparticle hierarchies for the given NLSP, and lastly append a letter for the
5\textsuperscript{th} lightest sparticle. {We note  that when the mass gap $\Delta
m(H^0, A^0)$ is $\order{1\GeV}$, $H^0$ and $A^0$  are treated as degenerate in mass in the
classification scheme.}}
In \cref{msugra_mass_patterns} the hierarchical mass patterns for the mSUGRA {case} are
exhibited.  In all cases the exhibited mass hierarchies are those that remain
after all the constraints have been imposed. 

The sparticle mass hierarchies in the nuSUGRA {Models} [2]-[5] are exhibited in   
\cref{nusugra_lightchargino_mass_patterns,nusugra_lightgluino_mass_patterns,nusugra_nuhiggs_mass_patterns,nusugra_light3rdgen_mass_patterns}.
The sparticle mass hierarchies for nuSUGRA Model [2] with nonuniversalities in the $\SU(2)_L$ gaugino mass
sector are given in \cref{nusugra_lightchargino_mass_patterns}, while those for the nuSUGRA Model [3] with
nonuniversality in the gluino mass sector are given in \cref{nusugra_lightgluino_mass_patterns}.
In \cref{nusugra_nuhiggs_mass_patterns} we give an analysis of the sparticle mass hierarchies 
for the  nuSUGRA Model [4] with nonuniversality in the Higgs boson mass sectors, and in \cref{nusugra_light3rdgen_mass_patterns}
an analysis is given of the sparticle mass hierarchies for nuSUGRA Model~[5] for the case when nonuniversality is in the
third generation sfermion sector. {In \cref{benchmarks_HEparams} a number
of benchmarks are presented for both universal and nonuniversal  SUGRA cases.} These benchmarks respect all the
collider, flavor and cosmological constraints and are not excluded by the
{by Run-I of the LHC}. 

 \begin{table}[t!]
  \renewcommand{\arraystretch}{1.2}
 \begin{center}
 \begin{tabular}{|l|c|c|c|c|}
 \hline
 
 		& Parameter 
 			& NLSP 	 
 				& 4-sparticle  
 					& 5-sparticle 	\\
 \multicolumn{1}{|c|}{SUGRA Model}
 	& Points 
 		& Patterns 
 			& Patterns 
 				& Patterns	\\
 \hline
 [1]  mSUGRA
 		& 5453
 			& 5 
 				& 15
 					& 	  32\\
 
 [2] Light Chargino
 		& 8000
 			& 4  
 				&  26
 					&		89\\
 
 [3] Light Gluino
 		& 8008
 			& 7  
 				&  44
 					&		85\\
 
 [4]  Nonuniversal Higgs 
 		& 4738
 			&  5
 				&  26
 					&	59	\\
 
 [5] Light 3rd Generation
 		& 2668
 			& 6 
 				&  26
 					&	43	\\
 \hline
\end{tabular}
\caption{\label{msp_count}\label{tab1}
An overview of the parameter scans, detailing the number of parameter points simulated, the number of NLSP patterns, the number of 4-sparticle  and the number of 5-sparticle mass hierarchies {for Models [1]-[5]}.
}
\end{center}
\end{table}
 
\section{Sparticle mass hierarchies and simplified models\label{sec3}}

Recently a new avenue for the exploration of new physics at colliders has been
explored via the {so-called} simplified
models~\cite{ArkaniHamed:2007fw,Alwall:2008ag,Alwall:2008va,Alves:2011sq,Papucci:2011wy,Mahbubani:2012qq,Chatrchyan:2013sza,Cohen:2013xda}.
For example, one might consider a {system of 3 particles:} $A$, $B$, $C$
with masses $m_A$, $m_B$, $m_C$ and the hierarchy \[ m_A > m_B > m_C\ .
\label{sm1} \] One further assumes that the branching ratio of the decay of $A$
into the state $B$ is $100\%$, {and likewise} the branching ratio of $B$ to $C$
is also $100\%$. More {generally,} one could also have $A$ going directly to
{$C$,} but often in simplified models the direct decay of $A$ to $C$ is ignored.
One could also consider simplified models including  four particles
$A$, $B$, $C$, $D$ with the mass hierarchy \[ m_A > m_B > m_C > m_D\ .\label{sim2} \]
Here one has six allowed branchings $A\to B, C, D$, $B\to C,D$ and $C\to D$.
Again the simplified assumption would be to consider just a cascade type decay
$A\to B\to C\to D$ which involves only three branchings which can all be assumed
to be 100\%.  A specific example of a three particle simplified model is given
by  the decay chain $\nb \to \cha\to \na$, and examples of 4 and 5 particle
simplified models are given by the decay chains $\g\to \nb \to \cha\to\na$ and
$\nc\to \g\to \nb \to \cha\to \na$.

In simplified models one makes an {\itshape ad hoc} choice of the lightest
particles and their decay chains.  However, a simplified model with particles
chosen in an {\itshape ad hoc} fashion may not be embeddable in a high scale
model.  Thus, it is worthwhile to investigate the classes of simplified {models}
that can arise from truncation of sparticle mass hierarchies that arise from a
high scale model. In this way one can work with simplified models at the level
of a preliminary investigation keeping in mind that it is a truncation of a UV
complete model which would eventually replace the simplified model by a more
complete one.  In this work we give a fairly exhaustive analysis of the class of
simplified models that arise in truncation of sparticle mass hierarchies in
supergravity  unified models using universal as well as nonuniversal boundary
conditions, i.e., in mSUGRA as well as in nuSUGRA models as discussed in
\cref{sec2} \footnote{Some of the preliminary results were presented at SUSY2014
\cite{susy2014}.}.  The results of the mass hierarchies are given in \cref{tab2}
for the mSUGRA case and in
\cref{nusugra_lightchargino_mass_patterns,nusugra_lightgluino_mass_patterns,nusugra_nuhiggs_mass_patterns,nusugra_light3rdgen_mass_patterns} 
for the nonuniversal SUGRA case.  Thus, \cref{tab2} gives the mass hierarchies
that arise for five particles including the LSP. These can be truncated to give
mass hierarchies for just three particle mass hierarchies.

\begin{figure}[t!]
  \begin{center}
  	\includegraphics[width=15cm]{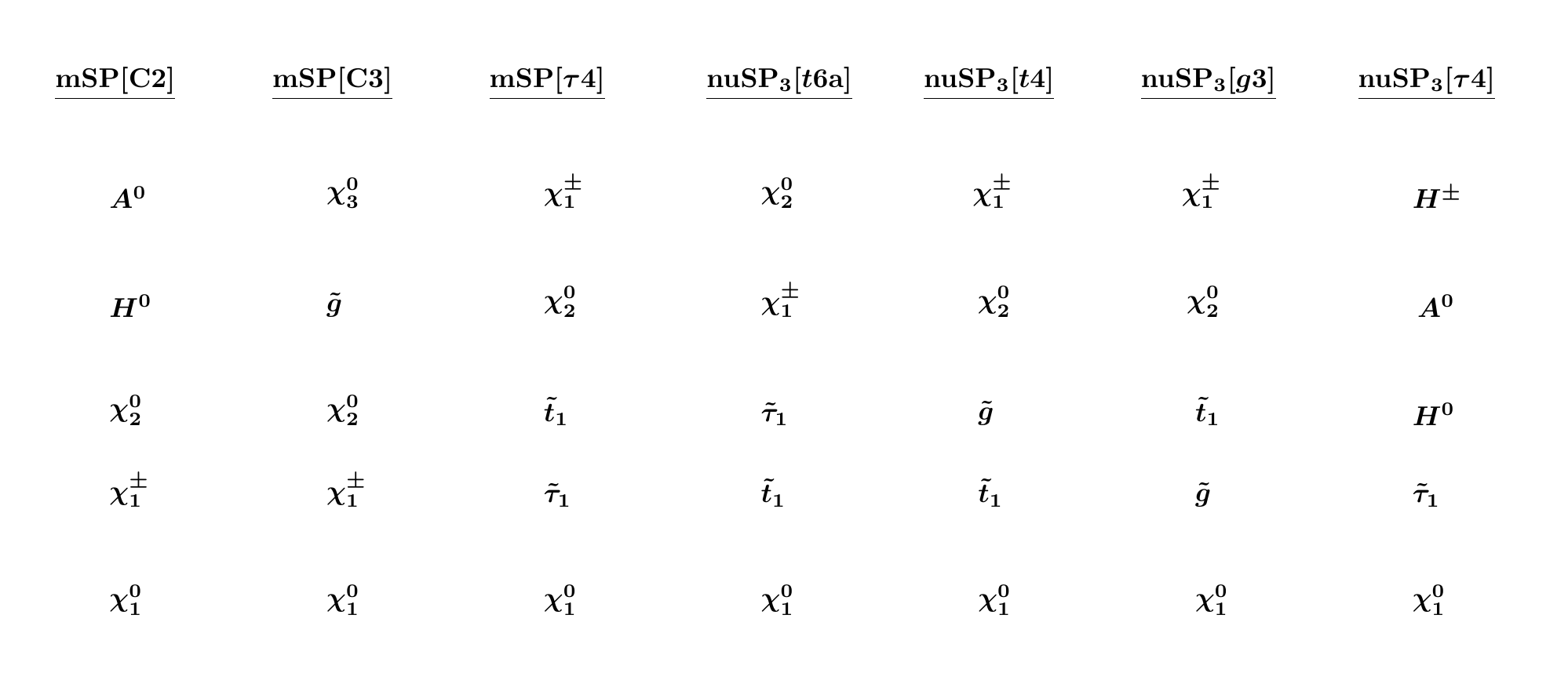}
		\vspace{0cm}
			\caption{\label{sim1}\label{fig1}
 An illustration of 
  simplified models with five lightest particles  arising from SUGRA unified models
 where the masses are in ascending order. Simplified models with a lower number of particles can be
 {obtained} by retaining the appropriate lower number of particles in the mass hierarchy.
 A more complete set of three, four and five particle simplified models can be {obtained} by retaining three, four 
 and five particles in 
 \cref{tab2} and in \cref{tab4}
 -\cref{tab7}.}
  \end{center}
\end{figure}

An illustration of several simplified models arising from mSUGRA and {from}
nonuniversal SUGRA models is given in \cref{fig1} where the masses of the
particles are in ascending order. One can use each of the columns to generate
three, four or five particle simplified models. Most of the current work centers
around keeping just three particles in the analysis. Such an approximation is
valid if the mass gaps between the first three and the remaining higher ones is
appreciable so that their inclusion would not radically change the signature
calculus. However, often this is not the case. In mSUGRA, often \cha
and $\nb$ are essentially degenerate, as are \Hz, \Az, \Hpm; thus, a truncation
will lead to significant errors. Additionally, if there is a strongly
interacting particle lying just above the lowest three, a neglect of this
particle will also lead to erroneous results.

We discuss further the relative advantages and disadvantages of the simplified
models.  In the analysis of the simplified models, the coupling and the masses
of the particles can be taken to be adjustable parameters.  {This procedure}
circumvents carrying out radiative breaking of the electroweak symmetry every
time the parameters are varied. Here one already has the masses and the
couplings and so one can get down to the task of fitting the data.  Further, the
simplified models can explore the phase space of signatures which may otherwise
be forbidden from the point of view of well-motivated models with constraints
such as, for example, the constraint of radiative breaking of the electroweak
symmetry.  However, the disadvantage is that some of the models may not be
embeddable in a UV complete theory and the connection of such models with
fundamental physics one is trying to explore becomes tenuous. On the other hand,
if one uses the simplified models arising from the supergravity unified models,
the connection with a high scale theory is much stronger. 

\begin{figure}[t]
  \begin{center}
   \includegraphics[width=15cm]{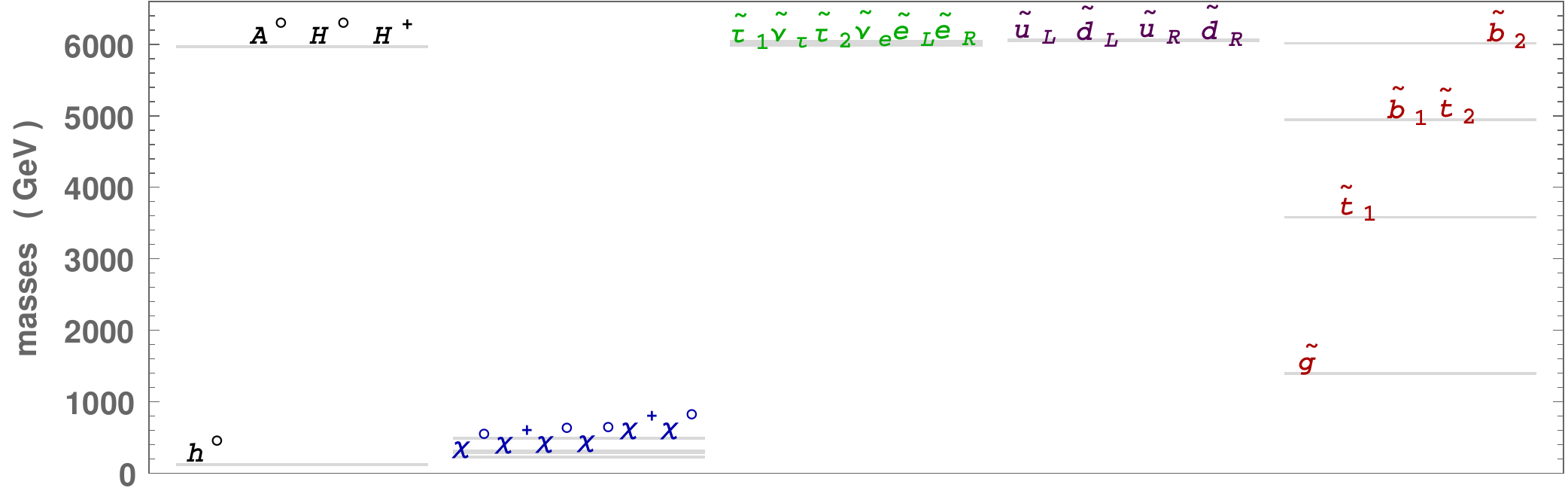}
   \end{center}
   \vspace{-.5cm}
   \caption{An exhibition of clustering of particles arising from a  high scale model 
   where the 
   inputs for the model point are $m_0=6038\GeV$, $m_{1/2}=538\GeV$, $A_0= -2734\GeV$,
   $\tan\beta=10$ and $\mu$ is positive.  The particles are arranged according to their masses as indicated by 
   the y-axis.     }
   \label{clustering}\label{fig2}
\end{figure}

While the simplified models  are much easier to deal with, the approximation of
using just three particles can be very limiting as noted earlier. Thus, for
example, for the case {where several particles are clustered}, the
approximation of just keeping three lowest ones is not justified. An example of
this is given in \cref{clustering} where the electroweak gauginos are packed
closely together.  Also as noted above, if there is a strongly interacting
particle in the vicinity, then an analysis {that ignores it} would lead to
erroneous results. Additionally, of course any randomly chosen three light
particles would not necessarily arise from the spectrum of a UV complete
theory. 

{
The observant reader may have noticed that {in \cref{tab2} as well in \cref{tab4}-\cref{tab7}} the first five particles are mostly electroweak.
The colored particles when present appear with low frequency  as given by the  percentage of  occurrence  shown in the last column 
of the tables. The reason for this phenomenon can be traced in part to the largeness of the Higgs mass and in part to the lower {experimental} 
limit on the gluino mass of around 1 \TeV.  Thus the Higgs boson mass of $\sim\! 125-126\GeV$ requires that the average 
{SUSY} scale $M_s\sim \sqrt {m_{\tilde t_1} m_{\tilde t_2}\!}$ be in the few \TeV range to produce the loop correction necessary to raise the Higgs
boson mass from the tree value of $\leq M_Z$ to the experimentally observed value. In mSUGRA a large $M_s$ can arise from 
either a large $m_0$ or a large gluino mass or by a combination of both. For large $m_0$ all the squarks and sleptons 
except for the possibility of a light stop will be above the \TeV region. Combined with the current experimental lower limit on
the gluino mass of roughly a \TeV, {these constraints}
imply that all of the colored particles except for the possibility of a light stop or gluino
will not populate the lightest set of sparticle patterns.  There is, however, an alternate possibility of generating a large $M_s$
needed for producing the large loop correction to lift the tree level Higgs boson mass. Here one can choose $m_0$ to be
 in the low $\order{100\GeV}$  region, and choose the gluino mass at the GUT scale to be in the several \TeV region {\cite{Akula:2013ioa}}. In this case
 the RG running drives the squark masses to high values while the slepton masses are low since they do not receive large
 corrections from RG evolution. {In each of these cases one finds that the low mass sparticle spectrum will contain mostly the 
 electroweak particles and color particles would be rare, often consisting of just the stop or the gluino consistent with 
 the  {experimental} lower  mass limits.} The result of this analysis obviously has important implications regarding sparticle searches,
 i.e., the focus of the searches be geared to {look} for the electroweak particles and in the color sector for the stop and the gluino
 which are often the lightest colored particles. 
}

\section{Model Signatures\label{sec5} }

The {most effective signatures for the LHC Run-II will be correlated
strongly with the type of mass hierarchical patterns considered}.  Below we
discuss a few illustrative examples and the possible signatures associated with
the chosen patterns.  We consider first  the  sparticle production for the
hierarchical patterns mSP[C1] - mSP[C4]. If we retain only the three lightest
particles then the patterns mSP[C1] - mSP[C4] are indistinguishable. 
To
discriminate among the subcases, for example, between mSP[C1] and mSP[C2]
{we need to consider at least a four particle mass hierarchy. For both of these 
patterns the first three particles have the hierarchy $\na<\cha<\nb$ but differ 
in the placement of the 4th lightest particle. For mSP[C1] it is \nc while for  mSP[C2], it is \Hz. }
A similar situation arises
for the hierarchical mass patterns nuSP$_2[$C1] - nuSP$_5$[C5].  Here also
these five patterns can only be discriminated by considering more than three
particles.  Further, the only difference between mSP[C1] - {mSP[C4]} vs
nuSP$_2$[C1] - {nuSP$_2$[C4]} relates to the mass gaps between the three sparticle
states.
If we keep only the first three particles in the mass hierarchies, then the
particles likely to be produced {at the colliders} for the patterns     mSP[C1] - mSP[C4]  and
nuSP$_2$[C1] - nuSP$_2$[C5] are as follows:

\begin{equation*}
{\rm mSP[C1]-{mSP[C4]}/nuSP_2[C1]-nuSP_2[C5]:}\ \cha\chabar,\ \cha\na,\ \na\nb,\
\cha\nb,\ \nb\nb\,.
\end{equation*}

These sparticle pairs would arise from the parton-level processes $q\bar q\to
\chi_i^\pm \chi_j^\mp, \chi_k^0
\chi_\ell^0$, and $u\bar d \to \chi_i^+\chi_k^0$ etc.  The
chargino can decay so that $\chaminus\to W^- \na$ with $W^-\to \ell^- \bar \nu$
so we will have $\ell^- + E_T^{\rm miss}$ in the decay of the chargino. Thus,
\cha\na will produce a  charged lepton and missing energy.  Next
\chaplus\chaminus  will produce  two charged leptons and $E_T^{\rm miss}$.
Since a chargino can also decay {via a $W^*$} into $q_1\bar q_2 +\na$, we
can have in addition a single charged lepton plus jets and \et. Now \nb
will have decays such as $\ell^+\ell^- + \et$. Thus, \cha\nb will have
trileptonic decays~\cite{Chamseddine:1983eg,Dicus:1983cb,Baer:1986vf,Nath:1987sw}
$\ell_1^{\pm} \ell_2^{+} \ell_2^-+$ \et. The production channels
\na\nb will have decays of the type $\ell^+\ell^-+$\et and 
\nb\nb will have 4-lepton decays $\ell_1^+\ell_1^- \ell_2^+\ell_2^-+$ \et as
well as decays of the type $\ell^+\ell^- +{\rm jets}+$ \et.  For cases where the
\nb mass is large enough so that $m_{\nb} > m_{\na} + m_{\hz}$, we can have
on-shell decays $\nb\to\na \hz$, and \hz predominantly decays via the mode
$\hz\to b\bar b$.  This gives rise to important new signatures such as
$\cha\nb\to\ell^{\pm} b\bar b+$\et and $\nb\nb\to\ell^+\ell^- b\bar b$.
 
Next we consider the pattern {nuSP$_2$[C6] and nuSP$_2$[C7]} which have the same lowest
three particle pattern, i.e., $\chi_1^0, \chi_1^{\pm}, \tilde \tau_1$
with the mass hierarchy $m_{\chi_1^0} < m_{\chi_1^{\pm}} < m_{\tilde
\tau_1}$.  Again to distinguish between {nuSP$_2$[C6] and nuSP$_2$[C7]}  we would need to
consider a four particle mass hierarchy. For these mass patterns   we can produce
a chargino-neutralino, two charginos or two staus in $pp$ collisions:
\begin{equation*}
 {\text{nuSP$_2$[C6], nuSP$_2$[C7]}:}\ \cha\na,\ \chaplus\chaminus,\ \sta\sta^*\ . 
\end{equation*}
 
The signatures arising from $\cha\na$ and $\chaplus\chaminus$ production are as
discussed for mSP[C1]-mSP[C5] and {nuSP$_2$[C1]-nuSP$_2$[C5]}.  Here, we additionally have
$\sta\sta^*$ production. Since $\sta$ has the decay $\sta\to\tau\na$, we will have a signature of the
type $\tau^+\tau^-+ \et$. Additionally, since \sta is heavier than the chargino,
we will have the decay $\sta^-\to \chaminus\nu_\tau$ with $\chaminus\to
\ell^-\bar\nu$. This will lead to  signatures such {as} $\ell_i^+\ell_j^-+\et$
where $\ell_i= e, \mu, \tau$.  Closely related to the signatures arising from
{nuSP$_2$[C6] and nuSP$_2$[C7]}  are the signatures arising from {nuSP$_2$[$\tau 2$] and
nuSP$_2$[$\tau 3$]} where we have the same three particles but the mass hierarchy is
inverted for the top two, i.e., we $m_{\na} <m_{\sta}< m_{\cha}$.  {Thus,} the
final states produced for these patterns will be the same as in {nuSP$_2$[C6] and
nuSP$_2$[C7]} except that here the chargino is heavier than the stau. {Thus,} instead
of stau decay into a chargino we have a chargino decaying into a stau which
gives $\chaminus \to \sta^- \bar\nu_\tau$.

Next we consider the pattern {nuSP$_2$[C9]} where the three lightest particles are
\na,\cha,\stopa with the mass hierarchy $m_{\na} < m_{\cha} < m_{\stopa}$. The
sparticle states produced in $pp$ collisions are as follows
\begin{equation*}
  {\text{nuSP$_2$[C9]}:}\ \cha\na,\ \chaplus\chaminus,\ \stopa\stopa^*\ . 
\end{equation*}
The new {production mode} here is $\stopa\stopa^*$, where \stopa has the
decay channels \[ \stopa \to t\na,\ b\cha,\ c\na\ .\label{stopdecay} \] Further,
the chargino will have the decay $\cha \to  W^{\pm}\na\to\ell\nu\na,\ jj\na$ and
the top quark has the decays \[t\to jjb,\ \ell\nu b\ .\label{topdecay} \] Thus 
$\stopa\stopa^*$ production will {{lead} to a} variety of signals involving
leptons and jets and missing energy such as the signals $\ell^+\ell^-b\bar b
+\et$, $\ell b\bar b+\text{jets}+\et$, etc.
 
Another illustrative example is the mass pattern {{nuSP$_3$[$g$1]}} where the lightest
three particles are \na, \g, and \nb, with the mass hierarchy $m_{\na}< m_{\g}
<m_{\nb}$. The sparticles states produced in $pp$ collisions consist of the
following 
\begin{equation*}
 {{\text{nuSP$_3$[$g$1]}}:}\ \g\g,\ \nb\nb\ .
\end{equation*}
The new {production mode} here is $\g\g$ which will dominate the signatures
since the gluino {interacts strongly} and the production cross section for
$\g\g$ {will be much greater than that of electroweak gaugino production}. The
gluino will have the decay $\g\to t\stopa^*$ where the decays of \stopa are
given in \cref{stopdecay} and the decays of the top in \cref{topdecay}.  
{These will generate a variety of signatures involving leptons, at least 2-$b$ jets, light jets and missing energy.}

An overall issue in the analysis of signatures concerns the NLSP and LSP mass
difference.  This mass difference determines the $\pt$ of the {jets and
leptons} and the $\et$ in the NLSP decay. A small mass difference between the  NLSP
and the LSP will lead to softer jets and leptons and a small $\et$ may not
pass the cuts or be distinguishable from the background.  {However,} there are a variety of
other signatures that can be investigated.  For recent analyses {relating} to
sparticle signature identification at the upgraded LHC see
\cite{Baer:2012vr,Altunkaynak:2013xya}.  An interesting issue
relates to {how one may discriminate  among ``mirror patterns''.}  Thus consider, for example,
the third and the fourth columns of \cref{fig1}.  Retaining only the three
lowest mass {particles,} column three has the hierarchy $\na<\sta<\stopa$ while
column four has the hierarchy $\na<\stopa<\sta$. Thus aside from the LSP the
spectrum for column four is inverted relative to that for  column three.  In a
similar fashion the lightest three particle spectrum arising from column five is
{$\na<\tilde t_1<\g$} while that from column six is $\na<\g<\stopa$.  One may call
such pairs mirror patterns. It should be interesting to investigate the
characteristic signals that can discriminate between the {two mirror patterns}.   
 
{As mentioned already,} a study of signatures based on three lowest lying
particles would not lift the degeneracy among those patterns which have
the same three lowest mass particles and we would need to include higher lying
particles to discriminate among the patterns. Further, as discussed in
\cref{sec3} keeping just the three lowest lying particles is inadequate when
there is a clustering of particles as illustrated in \cref{clustering}.  In this
case, all particles within the cluster must be taken account of. {Another}
example where truncation {to} three or four particles is {inadequate is} when
there is a strongly interacting particle lying close above. In this case again
one may be lead to erroneous results by the truncation procedure {of
constructing a simplified model}.

\section{Benchmarks for future SUSY searches at colliders\label{sec6} }

Benchmarks are useful {as illustrative examples of signature analyses that
can lead to new discovery channels for superpartner particles.} Here we give a
few benchmarks which satisfy all the current collider, flavor and
cosmological constraints. Specifically, we impose the following set of
constraints in choosing the benchmarks: $m_{\hz}\in[123,127]\GeV$,
{$\Omega_{\na}h^2 < 0.12$}, {$\br\bsmumu < 6.2 \times 10^{-9}$,
$\br\bsg<4.27\times 10^{-4}$}. {Additionally, we only selected benchmarks
that have sparticle mass hierarchies that are frequently observed in the
constrained parameter space.} 
{In \cref{benchmarks_HEparams} we give a set of benchmarks for both universal and nonuniversal SUGRA cases. Here we
also identify the corresponding SUGRA pattern to which they belong.}  The
benchmarks that are displayed  satisfy the LHC Run-I exclusion constraint in
the \mo--\mhf plane. These benchmarks should  be useful for future SUSY searches
at colliders. An interesting feature of \cref{benchmarks_HEparams} is that most of the benchmarks have relatively small $\mu$
compared to $m_0$ which points to the fact that they lie on the hyperbolic branch of radiative breaking of the electroweak symmetry
\cite{Chan:1997bi,Chattopadhyay:2003qh,Baer:2003wx,Akula:2011jx,Liu:2013ula} and are thus  natural according to the 
criteria discussed in \cite{Chan:1997bi}.  

\subsection{Signature analysis for an mSUGRA Benchmark}

First we consider {a benchmark within mSUGRA.} Our model's GUT scale
parameters and observables are given in \cref{tabbenchmark1}. {The particle mass hierarchy for this model point is $m_{\chi^{\pm}_{1}} < m_{\chi^{0}_{2}} < m_{\chi^{0}_{3}}< m_{\chi^{0}_{4}}$.} This benchmark
has a light LSP of mass $m_{\na}=199\GeV$, a chargino NLSP of mass
$m_{\cha}=261\GeV$ and a second neutralino of mass $m_{\nb}=271\GeV$. The first
and the second generation squarks are heavy, $m_{\q}\gtrsim6\TeV$, the stop and
the sbottom are $m_{\stopa}=3.6\TeV$, $m_{\sbota}=4\TeV$, and a gluino at
$m_{\g}=1257\GeV$. With the light neutralinos/charginos and a relatively light
gluino, the total SUSY cross section is 740\fb at 14\TeV. Since the rest of the
spectrum is heavy, \cha and \nb decay through an off shell $W$ and $Z$
generating relatively soft jets and leptons.

To test the visibility of this model at 14\TeV, we study the trilepton final
state and follow the ATLAS search~\cite{Aad:2014nua}. This analysis is based
on a simplified model with only \na, \nb, and \cha. It is optimized for 8\TeV
and should be re-tuned for 14\TeV, but we use similar cuts as our starting
point. The ATLAS analysis defines the signal region SR0$\tau$a which is the most
sensitive to \cha and \nb decays through $W$ and $Z$ bosons. This signal region
requires 3 leptons ($e$ or $\mu$) including a same flavor opposite sign (SFOS)
pair. The SFOS pair giving the invariant mass closest to the $Z$ mass is then
identified and the remaining lepton's momentum is used to calculate the
transverse mass defined as \[  \label{mt} 
  m_{\rm T}^2 (\ptvec^{\,\ell}, \ptvec^\text{ miss}) = 
    2\pt^{\ell} \et -2\ptvec^{\,\ell} \cdot \ptvec^\text{ miss}\ .
\] The main irreducible background for this channel is diboson ($WZ$ and $ZZ$),
$t\bar{t}V$ and $tZ$ productions. The reducible backgrounds include single and
pair production of top quarks. We follow the ATLAS analysis and veto events with
$b$-tagged jets to suppress the top quark production.

In a previous work on signature analysis by two of the authors (BA and PN) the
standard model backgrounds at LHC energy of $\cm=7\TeV$  were generated
in-house~\cite{Altunkaynak:2010we}. Fortunately, for the current analysis this
computer intensive process was circumvented by employing the Snowmass Standard
Model background~\cite{SMSnowmass} normalized to NLO and generated with five
flavor MLM matching and in bins of $\ST$ which is the scalar sum of $\pt$ of all
generator level particles. For signal event generation, we use
{\scshape\ttfamily Pythia}~\cite{Sjostrand:2006za} for hard scattering and
showering/hadronization, and {\scshape\ttfamily
Delphes}~\cite{deFavereau:2013fsa} for detector simulation with the same card
used in the Snowmass background that was tuned according to the detector
performances in the last run. We normalize our signal cross section to match NLO
cross section obtained by {\scshape\ttfamily Prospino}~\cite{Beenakker:1996ed}.

\begin{figure}[!t]
\begin{center}
\includegraphics[width=7.5cm]{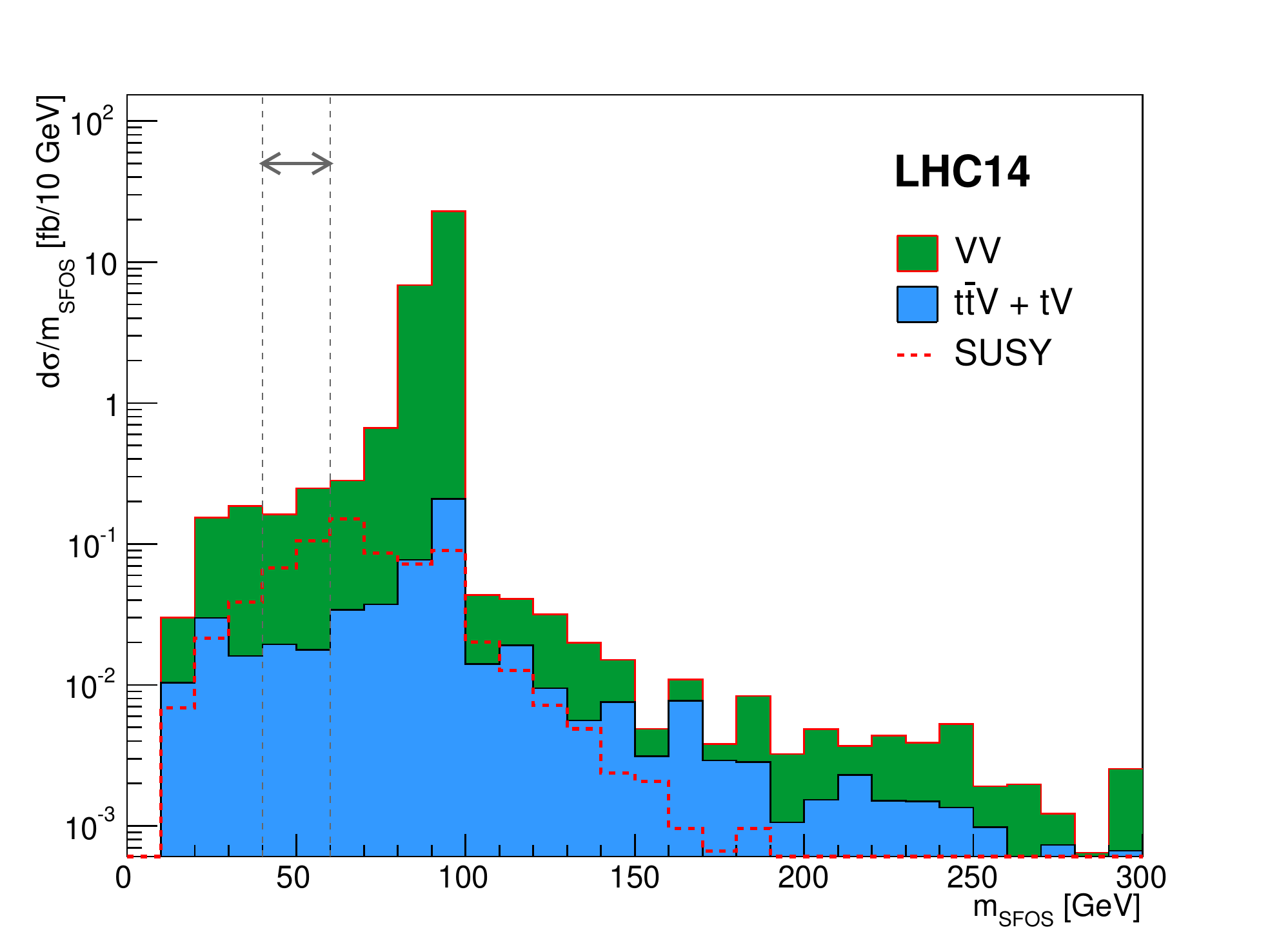}
\includegraphics[width=7.5cm]{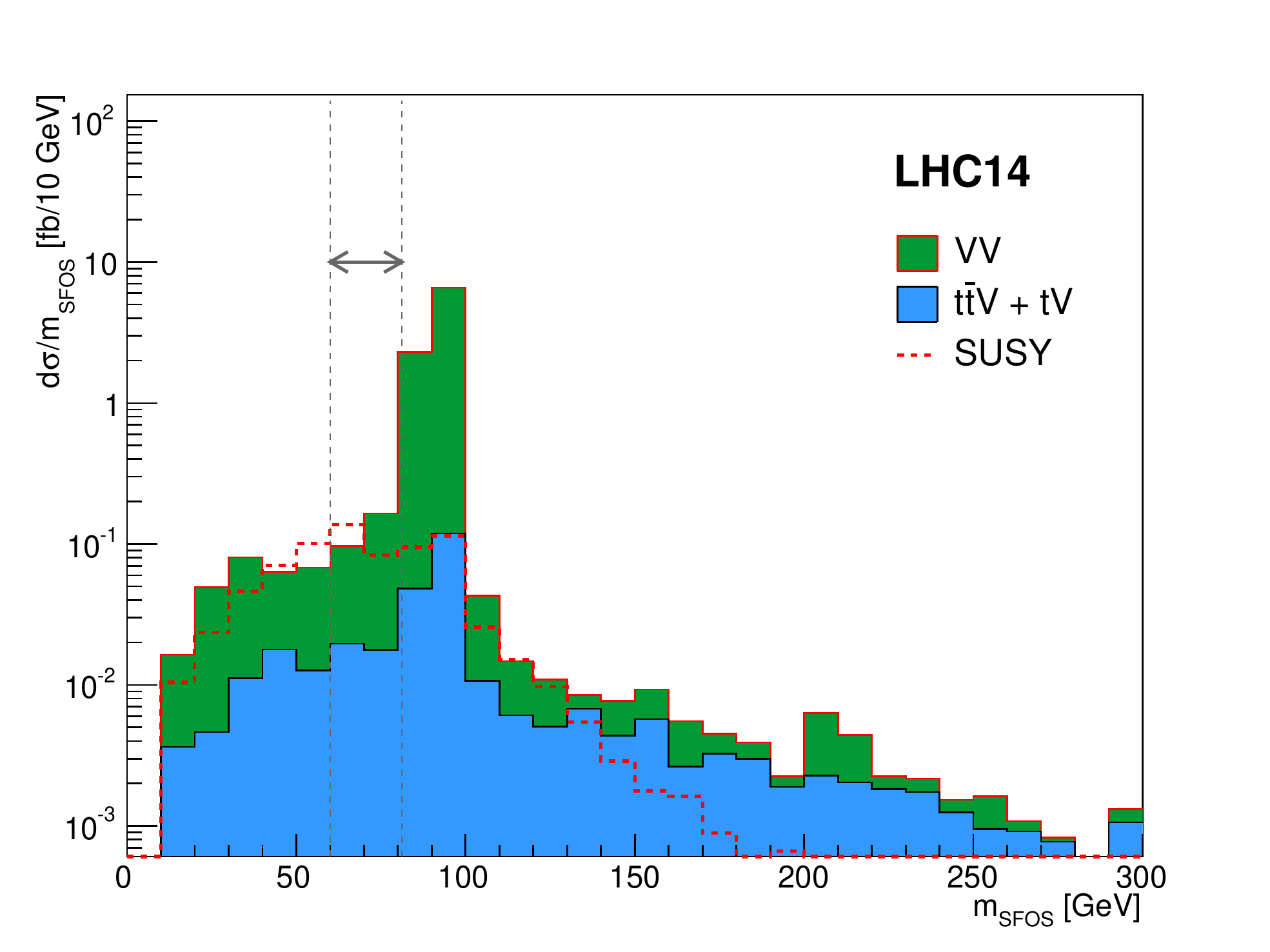}
\caption{\label{mSUGRA1}{Distribution of $\mSFOS$ in signal region SR0$\tau$a for bins 6 (on the left with cuts $m_T < 80\GeV$, $\etmiss > 75\GeV$) and 12 (on the right with cuts $m_T > 110\GeV$, $\etmiss > 75\GeV$) {prior to the requirements on these variables.}  Main SM background consisting of $VV$ (green) and $t\bar{t}V+tV$ (green) are shown. The signal is shown in dashed red. {The remaining invariant mass cuts used to define each signal region in the ATLAS analysis are also displayed. Our calculations show that $5\sigma$ significance is obtained with an integrated {luminosity of $L \gtrsim 340 \ifb$ (left) and  $L \gtrsim 135 \ifb$ (right).} }}}
\end{center}
\end{figure}

{In the ATLAS analysis, the signal region SR0$\tau$a is composed of 20 disjoint bins with varying ranges of $m_{\textrm{SFOS}}$, $m_T$ and $E_T^{\textrm{miss}}$. We follow a similar approach and simply study two of those bins that offer the best discrimination of signal from the SM background for our benchmark point. These are bin-6 with cuts $m_T < 80\GeV$, $\etmiss > 75\GeV$ and bin-12 with cuts $m_T > 110\GeV$, $\etmiss > 75\GeV$. The distributions of $m_{\textrm{SFOS}}$ in those bins are displayed in \cref{mSUGRA1} prior to the cut on that variable.  A further cut to {constrain} $m_{\textrm{SFOS}}$ into a mass window following the ATLAS analysis is also displayed. Our calculations show that the SUSY signal produced by our mSUGRA benchmark point will be discoverable at the LHC Run-II at $5\sigma$ significance defined by $S/\sqrt{B} = 5$ with an integrated luminosity of $L \gtrsim 340 \, \fb^{-1} (135 \, \fb^{-1})$ by using the cuts of bin 6(12).}

\begin{table}[!t]
\renewcommand{\arraystretch}{1.8}
\begin{center}
\begin{tabular}{|c|c||c|c|}
\hline
\multicolumn{2}{|c||}{Model Parameters} & \multicolumn{2}{c|}{Observables} \\
\hline
\mo     & 6183\GeV    & $m_{\hz}$ & 126.1\GeV \\
\mhf    & 470\GeV     & $\left(m_{\g}, m_{\tilde{t}_1}\right)$ & $(1257, 3601)\GeV$ \\
\az     & $-4469\GeV$ & $\left(m_{\na},m_{\nb},m_{\cha}\right)$ & $(199, 271, 261)\GeV$ \\
\tb     & 52.1        & $\Omega_{\na}h^2$ & {$< 0.12$} \\
\sgnmu  & $+1$        & $\sigma\left(pp\to\nb\cha\right)$ & 134.7\fb \\
\hline
\end{tabular}
\caption{\label{tabbenchmark1}Model parameters and observables of our mSUGRA benchmark model.}
\end{center}
\end{table}
   
\subsection{Signature analysis for an nuSUGRA Benchmark}

Next, we consider a nonuniversal SUGRA benchmark with {nonuniversality} in the
gaugino sector as given in \cref{tabbenchmark2}. {The particle mass hierarchy for this model point is $m_{\chi^{\pm}_{1}} < m_{\chi^{0}_{2}} < m_{\tilde g}< m_{\tilde t_1}$.} Our benchmark has a neutralino
LSP of mass {$m_{\na}=441\GeV$,} \nb and \cha are within 20\GeV of the LSP, with a
gluino of mass 1446\GeV and a stop of mass 1481\GeV. These masses are beyond the
current limits obtained by ATLAS and CMS experiments with simplified models.
With a relatively light LSP and NLSP, this model can have considerable
electroweak production of neutralinos and charginos. For example we obtain
$\sigma(pp\to\nb\cha) = 69.7\fb$ and $\sigma(pp\to\chaplus\chaminus) = 33.2\fb$
at NLO. But the small mass gap between LSP and NLSP/NNLSP make the final decay
products quite soft resulting in some or most of them being missed by the
triggers. One possibility of obtaining {sufficiently energetic} final
states to pass the triggers is via a high-$\pt$ ISR jet that will boost the event
and provide large momenta to the final decay products. But with all these
challenges, it  will be quite difficult to observe this model at the LHC solely
from electroweak production of neutralinos and charginos.

This model, however, has a gluino of mass $1446\GeV$ which gives rise to a
gluino pair production cross section of $\sigma(pp\to \g\g)=23.0\fb$ at NLO.
Although this is smaller than the chargino-neutralino production cross section,
the large mass gap provides energetic final states and large missing momentum
which can easily get triggered. Since our squarks are all heavy except the stop,
the gluino in this benchmark decays mostly into $t\bar{t}\na$, producing 4
b-jets in the final state.  To check the viability of detecting this benchmark 
at the LHC Run-II, we considered the ATLAS analysis~\cite{Aad:2014lra} for
gluino pair production. Like the trilepton analysis we used for our mSUGRA
benchmark model, this ATLAS analysis is also optimized for 8\TeV, but we use the
exact same cuts as the ATLAS analysis~\cite{Aad:2014lra} as our starting point.
The ATLAS analysis introduces 9 signal regions demanding at least 4/6/7 jets
with at least 3 of them being tagged as a b-jet and 0 or 1 lepton. The main
reducible background for this process is $t\bar{t}$ production where a c-jet or
a hadronically decaying $\tau$ lepton is mis-tagged as a b-jet. The irreducible
backgrounds from $t\bar{t} + b /b\bar{b}$ and $t\bar{t} + Z/h(\to b\bar{b})$ are
dominant.

\begin{table}[!t]
\renewcommand{\arraystretch}{1.8}
\begin{center}
\begin{tabular}{|c|c||c|c|}
\hline
\multicolumn{2}{|c||}{Model Parameters} & \multicolumn{2}{c|}{Observables} \\
\hline
\mo & 4\TeV & $m_{\hz}$ & 124.7\GeV \\
$(M_1, M_2, M_3)$ & $(980, 520, 550)\GeV$ & $(m_{\g}, m_{\stopa})$ & $(1446, 1481)\GeV$ \\
\az & $-7\TeV$ & $(m_{\na},m_{\nb},m_{\cha})$ & $(441, 461, 462)\GeV$ \\
\tb & 30 & $\Omega_{\na}h^2$ & 0.12 \\
\sgnmu & $+1$ & $\sigma(pp\to\g\g)$ & $23.0\fb$ \\
\hline
\end{tabular}
\caption{\label{tabbenchmark2}Model parameters and observables of our nuSUGRA benchmark.}
\end{center}
\end{table}
\begin{table}[t!]
\renewcommand{\arraystretch}{1.8}
\begin{center}
\begin{tabular}{|l||c|c|c|c|c|}
\hline
Signal Region  &  N jets (\pt) &  \et &  $m_{\rm eff}$   & $\et/\sqrt{H_T^{6j}}$\\
\hline
 SR- $0\ell-4j-$A & $\geq 4 (50)$   &  $> 250$  & $> 1300$ & -  \\
 \hline
 SR- $0\ell-4j-$B & $\geq 4 (50)$   &  $> 350$  & $> 1100$ & -  \\
 \hline
 SR- $0\ell-4j-$C & $\geq 4 (30)$   &  $> 400$  & $> 1000$ & $>16$  \\
\hline  \hline
Signal Region  &  N jets (\pt) &  \et &  $m_{\rm eff}^{\rm inc}$   & $\et/\sqrt{H_T^{6j}}$\\
\hline
 SR- $0\ell-7j-$A & $\geq 7 (30)$   &  $> 200$  & $> 1000$ & -  \\
 \hline
 SR- $0\ell-7j-$B & $\geq 7 (30)$   &  $> 350$  & $> 1000$ & -  \\
 \hline
 SR- $0\ell-7j-$C & $\geq 7(30)$   &  $> 250$  & $> 1500$ & -  \\
 \hline \hline
Signal Region  &  N jets (\pt) &  \et &  $m_{\rm T}$   & $m_{\rm eff}^{\rm inc}$  \\
\hline
 SR- $1\ell-6j-$A & $\geq 6 (30)$   &  $> 175$  & $> 140$ &  $>700$ \\
 \hline
 SR- $1\ell-6j-$B & $\geq 6 (30)$   &  $> 225$  & $> 140$ & $>800$  \\
 \hline
 SR- $1\ell-6j-$C & $\geq 6(30)$   &  $> 275$  & $> 160$ & $>900$  \\
 \hline
\end{tabular}
\caption{\label{tab-atlas}
A summary of signal regions (SR) with cuts used in the signature analysis of the
model benchmark.  The top two blocks in the table have the additional
{constraints of  $\pt(j_1)> 90\GeV$, $\et>150\GeV$, $\geq 4$ jets with
$\pt>30\GeV$, $\Delta \phi_{\rm min}^{4j}>0.5$, $\et/m_{\rm eff}^{4j} >0.2$,
$\geq 3$ b-jets with $\pt>30\GeV$.  The bottom block in the table has the
constraints $\pt(j_1)> 90\GeV$, $\et>150$, $\geq 4$ jets with $\pt > 30\GeV$,
$\geq 3$ b-jets with $\pt>30\GeV$. These constraints are adopted from the ATLAS
analysis at $\cm=8\TeV$~\cite{Aad:2014lra} and applied to our analysis at
$\cm=14\TeV$.}}
\end{center}
\end{table}

The following {kinematic} variables correlated with the overall mass scale
are introduced: $\htj$ which is the scalar sum of the transverse momenta of the
four leading jets, $\meffj$ which is the scalar sum of the \et and the
transverse momenta of the four leading jets, and $\meffincl$ which is the scalar
sum of $\etmiss$ and the  transverse momenta of all jets with $\pt > 30\GeV$. A
cut on the minimum azimuthal separation between any of the four leading jets and
the missing transverse momenta is also used to remove the multi-jet events. And
finally transverse mass (defined in \cref{mt}) computed from the leading lepton
and the missing transverse momenta is used to cut down the $t\bar{t}$ events
where one of the $W$ bosons decays leptonically.  In the analysis we use the
signal regions SR-0$\ell$-4j-[A,B,C], SR-0$\ell$-7j-[A,B,C] and
SR-1$\ell$-6j-[A,B,C] as defined by ATLAS in their analysis of SUSY signals at
$\cm=8\TeV$~\cite{Aad:2014lra} and summarized in \cref{tab-atlas}.

\begin{figure}[!ht]
  \begin{center}
    \includegraphics[width=7.5cm]{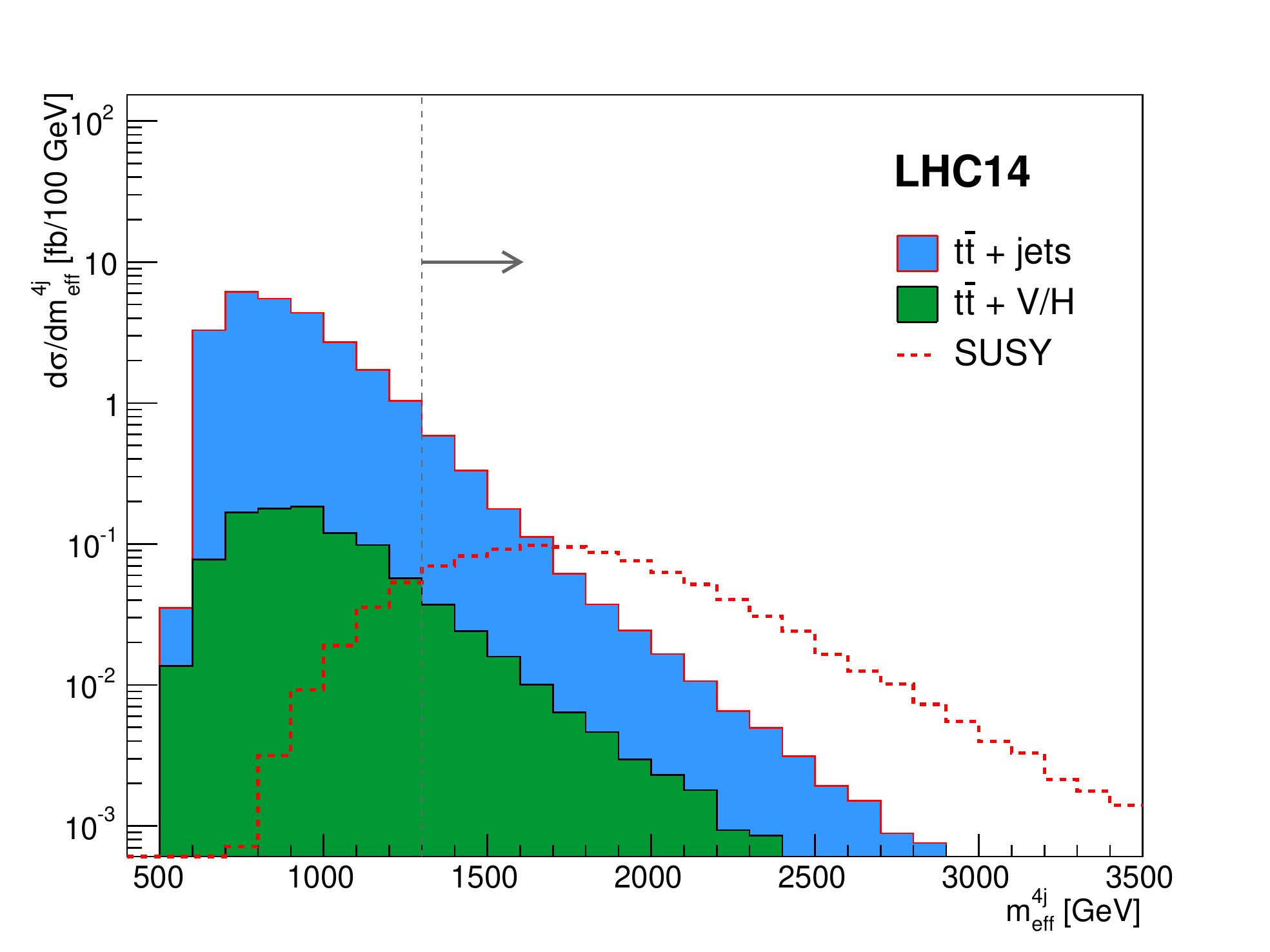}
    \includegraphics[width=7.5cm]{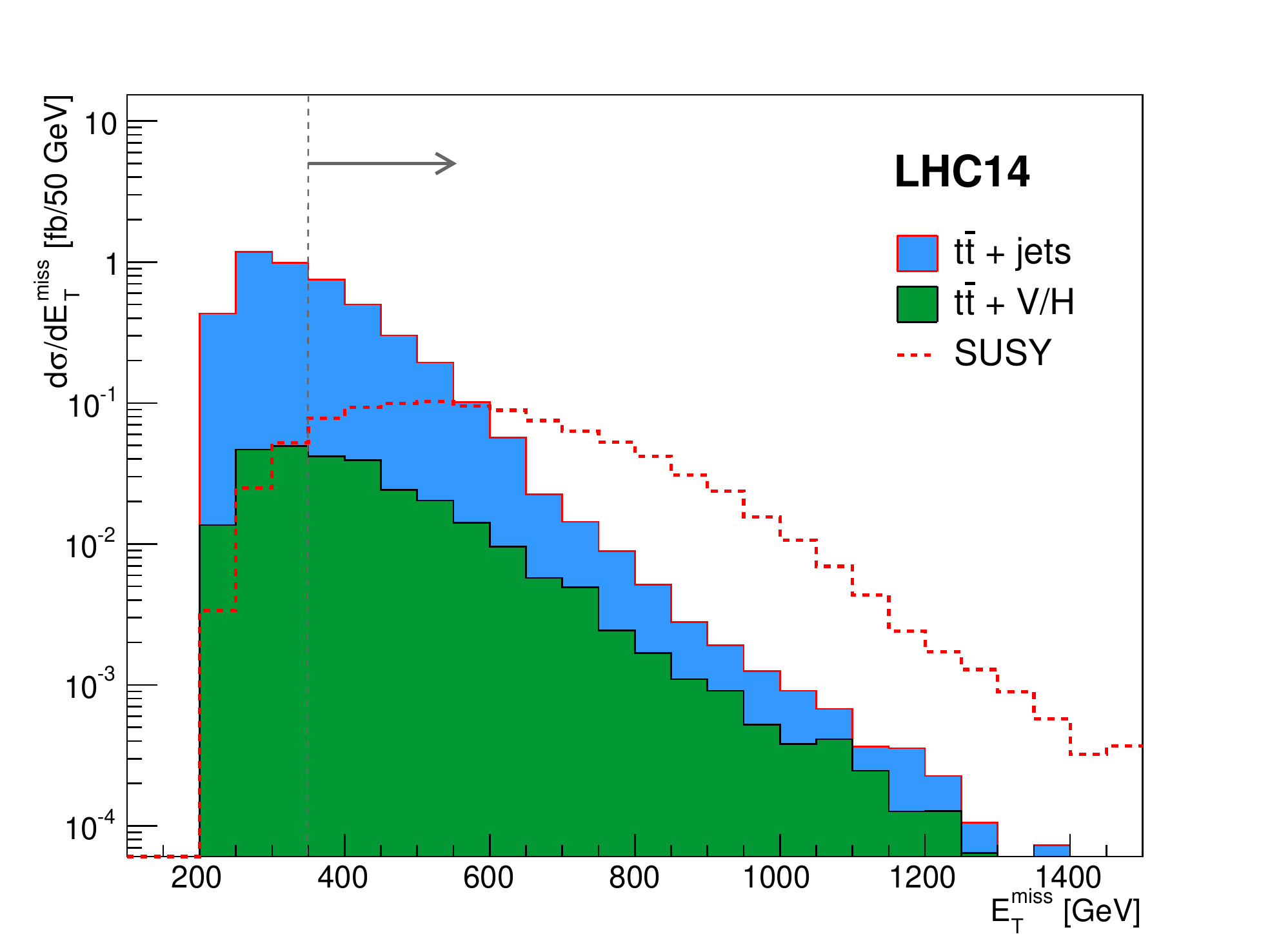}
    \includegraphics[width=8.7cm]{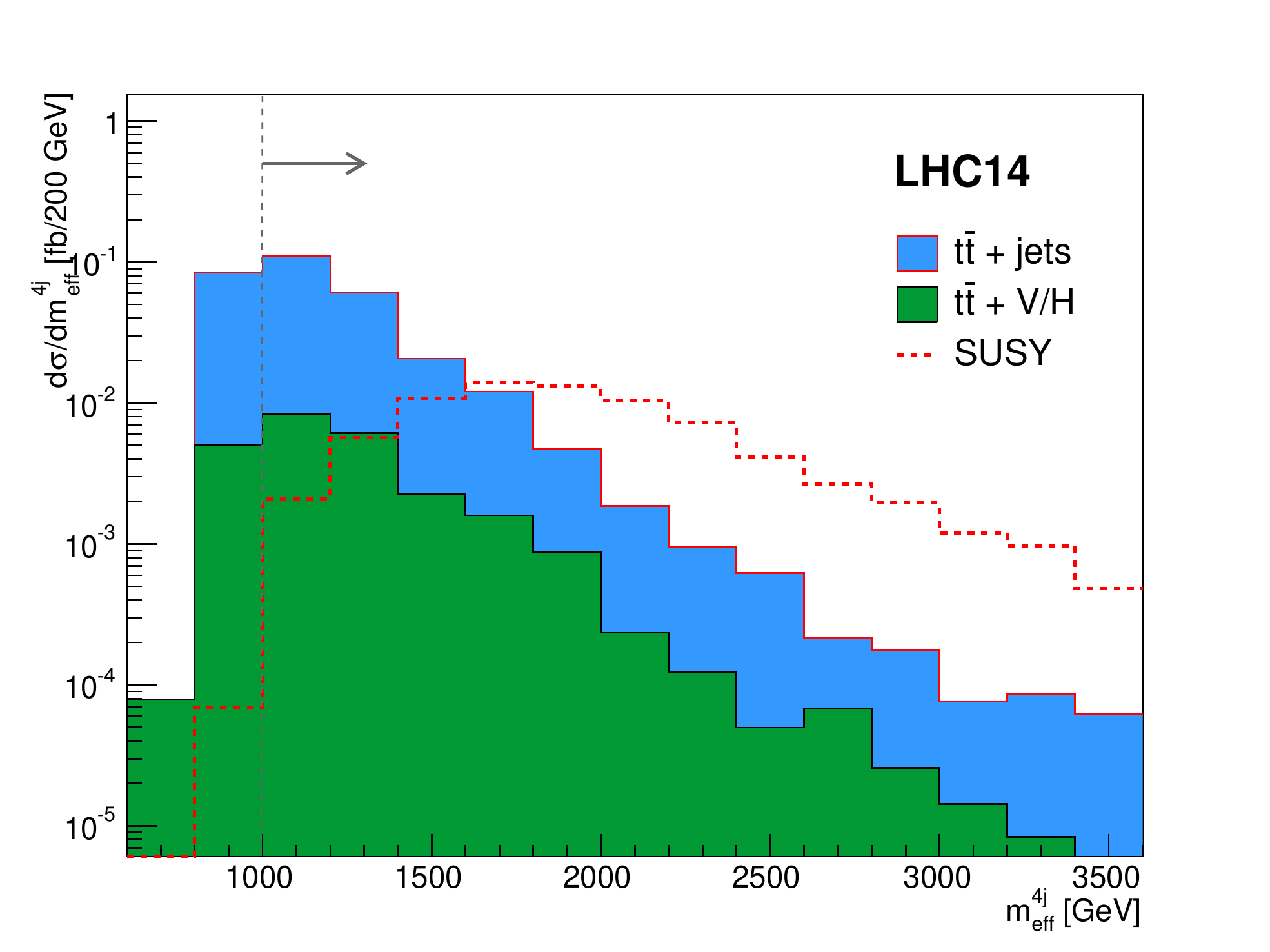}
    \caption{\label{nuSUGRA1}
      Distribution of $\meffj$ and \et in the signal regions
      SR-0$\ell$-4j-[A,B,C] {prior to the requirements on these variables. Arrows indicate the remaining cuts used to define the signal} {regions. Main} SM background consisting of $t\bar{t}$ + jets
      (blue) and $t\bar{t} + V/H$ (green) are shown. The signal is shown in
      dashed red. {Our calculations show that $5\sigma$ significance is obtained with an integrated {luminosity of $L \gtrsim 45 \ifb$ (left),  $L \gtrsim 60 \ifb$ (right) and $L \gtrsim 920 \ifb$ (bottom).}}
    }
  \end{center}
\end{figure}

\begin{figure}[!ht]
  \begin{center}
    \includegraphics[width=7.5cm]{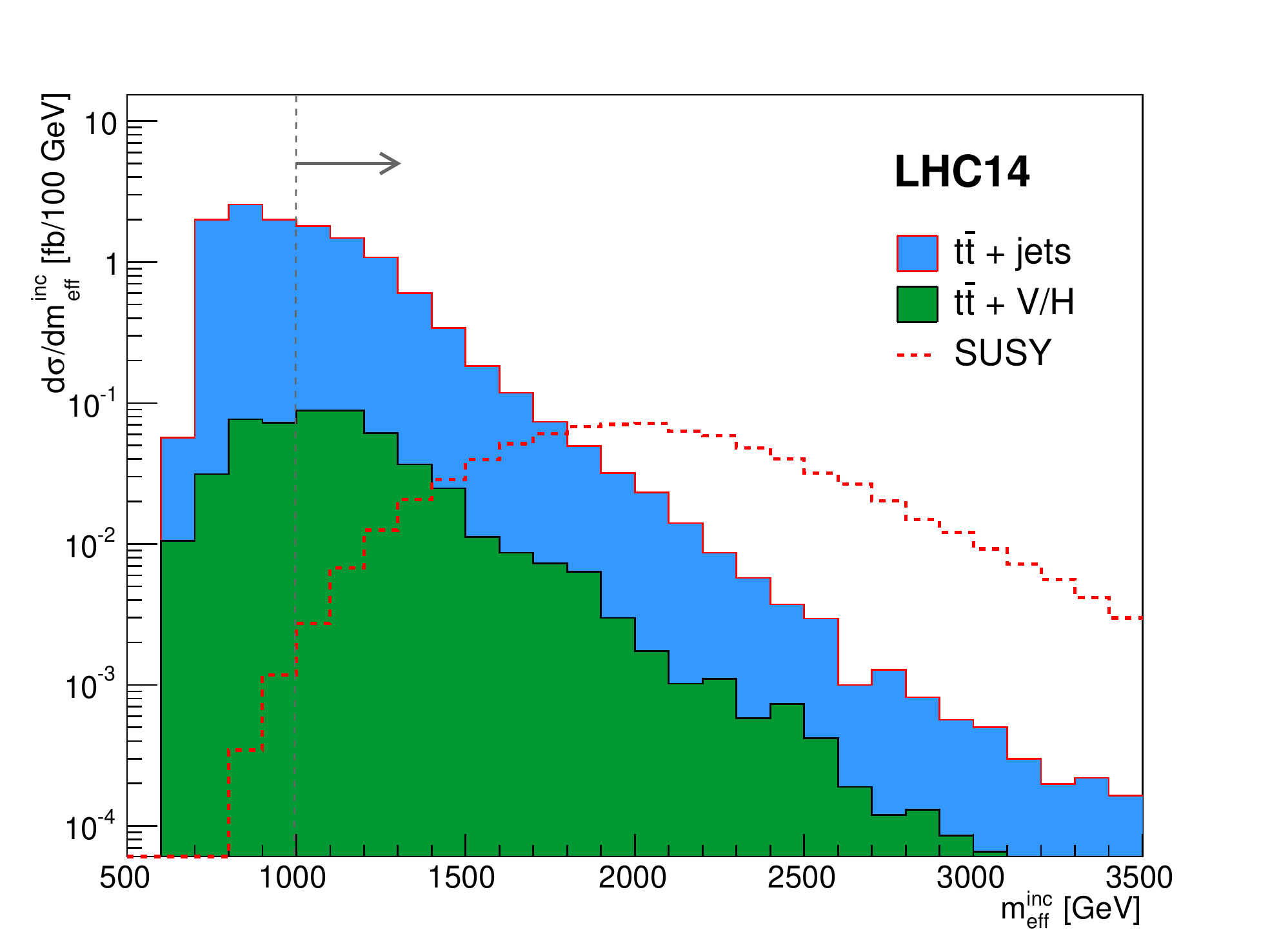}
    \includegraphics[width=7.5cm]{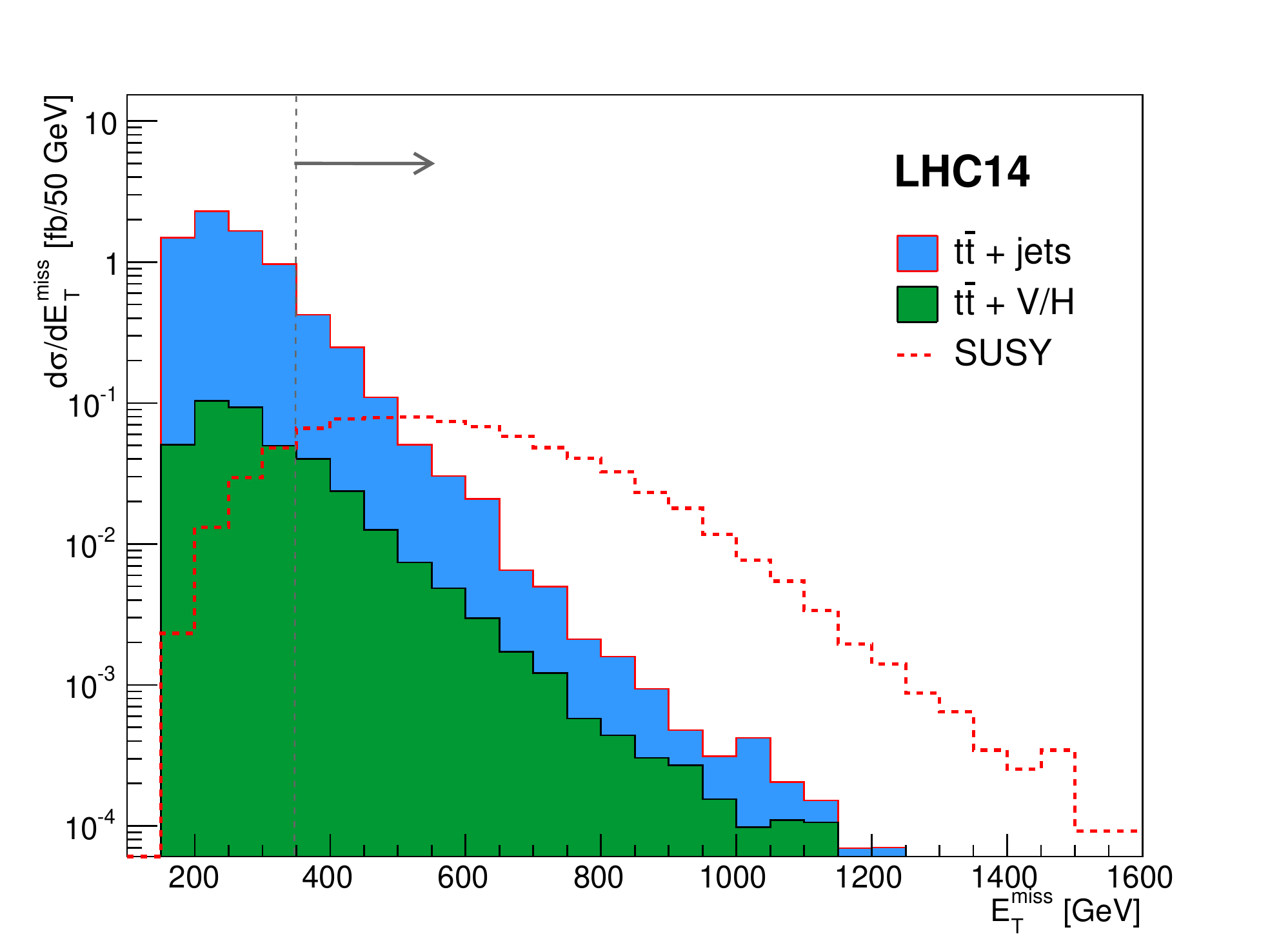}
    \includegraphics[width=8.7cm]{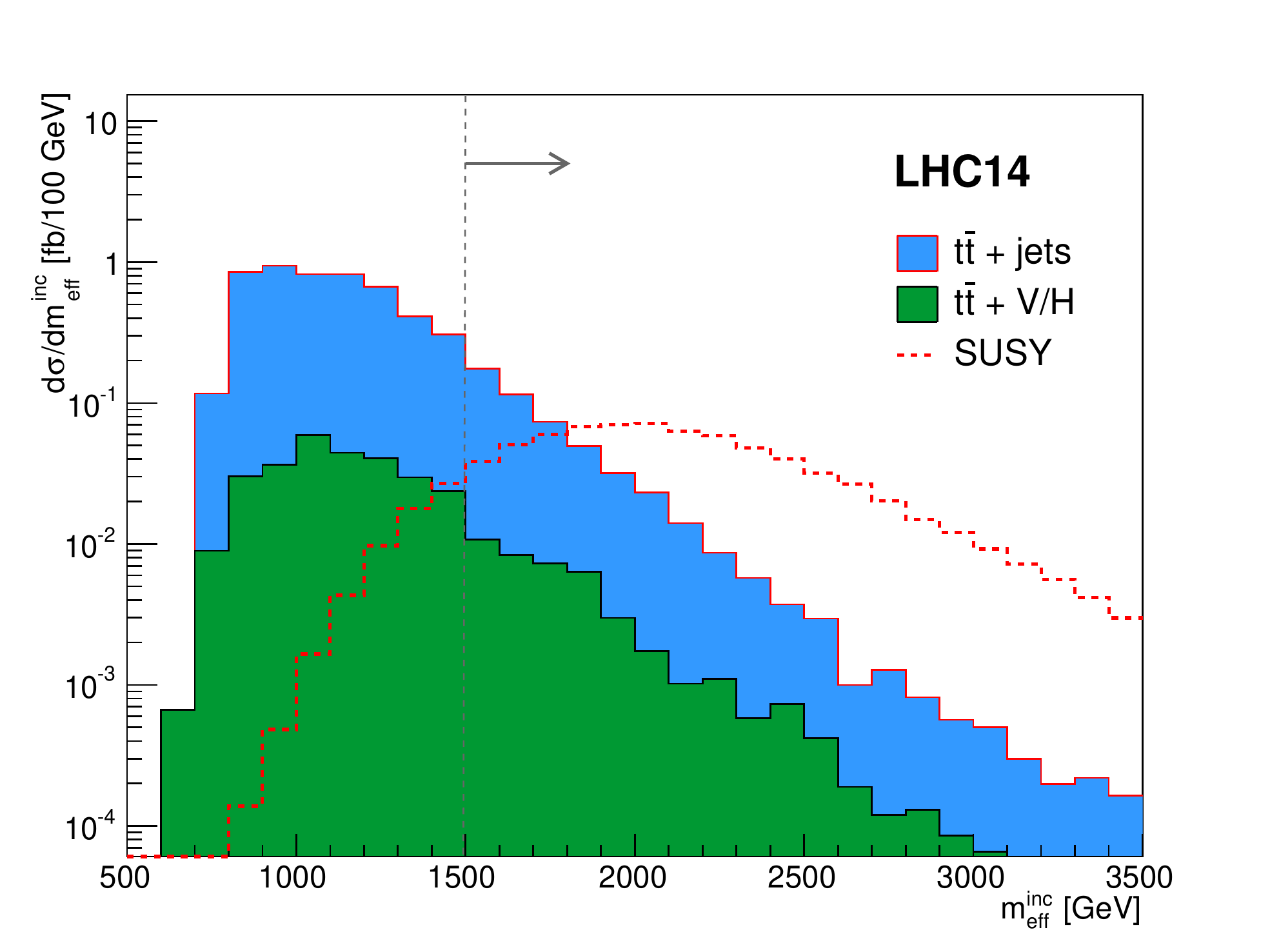}
    \caption{\label{nuSUGRA2}
      Distribution of $\meffincl$ and $\etmiss$ in the signal regions
      SR-0$\ell$-7j-[A,B,C] {prior to the requirements on these variables. Arrows indicate the remaining cuts used to define the signal regions.} Main SM background consisting of $t\bar{t}$ + jets
      (blue) and $t\bar{t} + V/H$ (green) are shown. The signal is shown in
      dashed red. {Our calculations show that $5\sigma$ significance is obtained with an integrated {luminosity of $L \gtrsim 235 \ifb$ (left),  $L \gtrsim 45 \ifb$ (right) and $L \gtrsim 25 \ifb$ (bottom).}}
    }
  \end{center}
\end{figure}

\begin{figure}[!ht]
\begin{center}
\includegraphics[width=7.5cm]{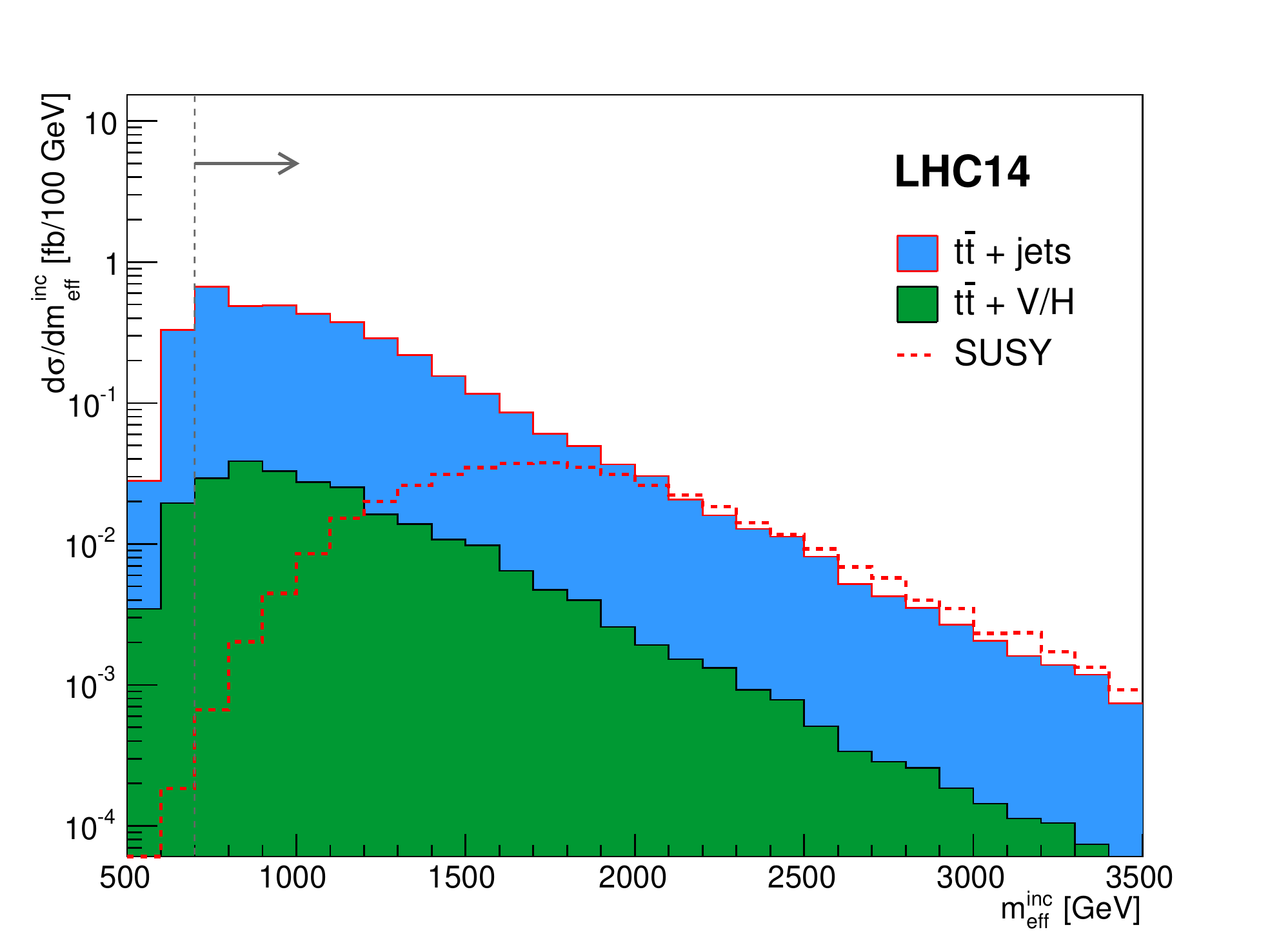}
\includegraphics[width=7.5cm]{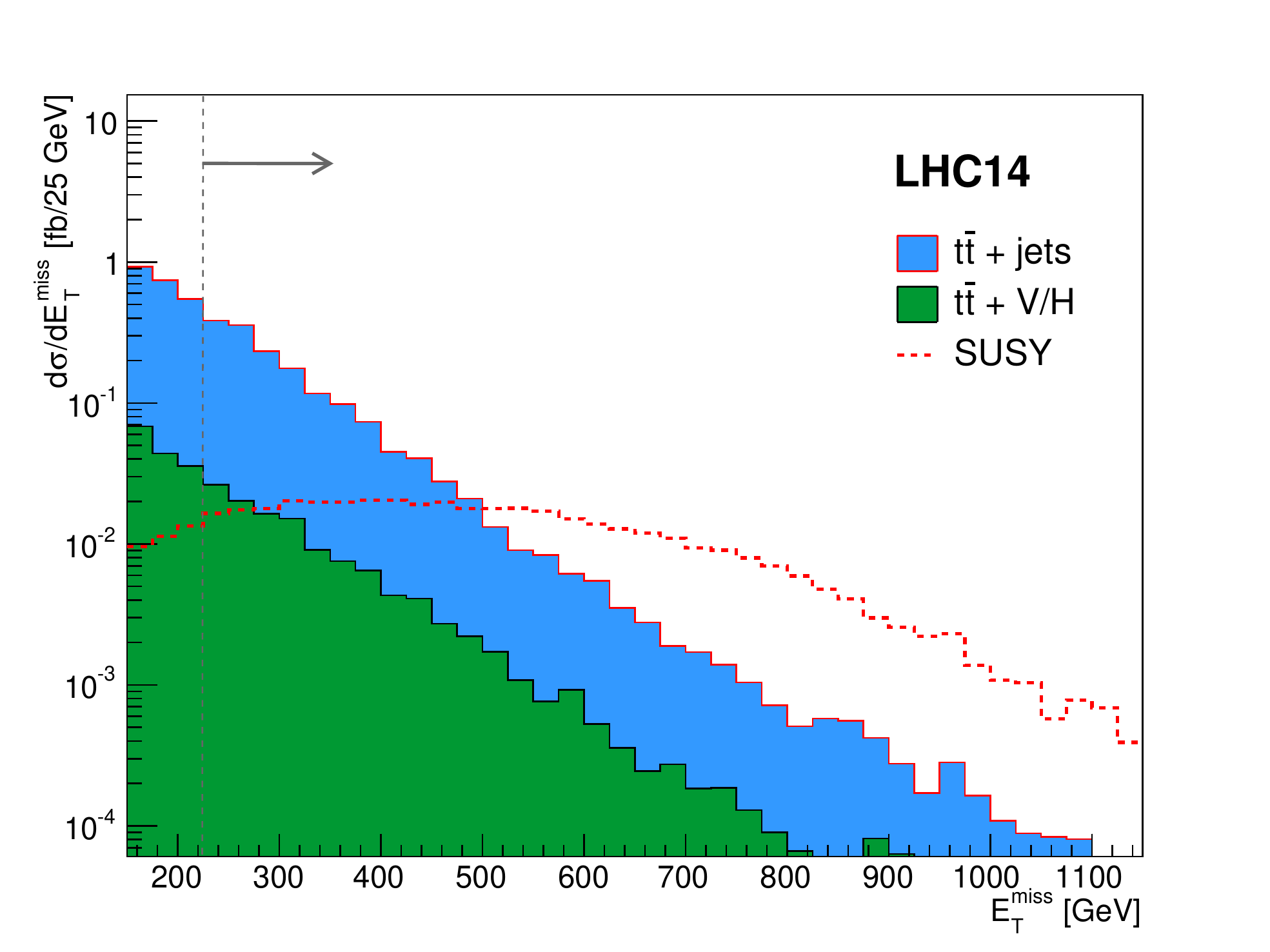}
\includegraphics[width=8.7cm]{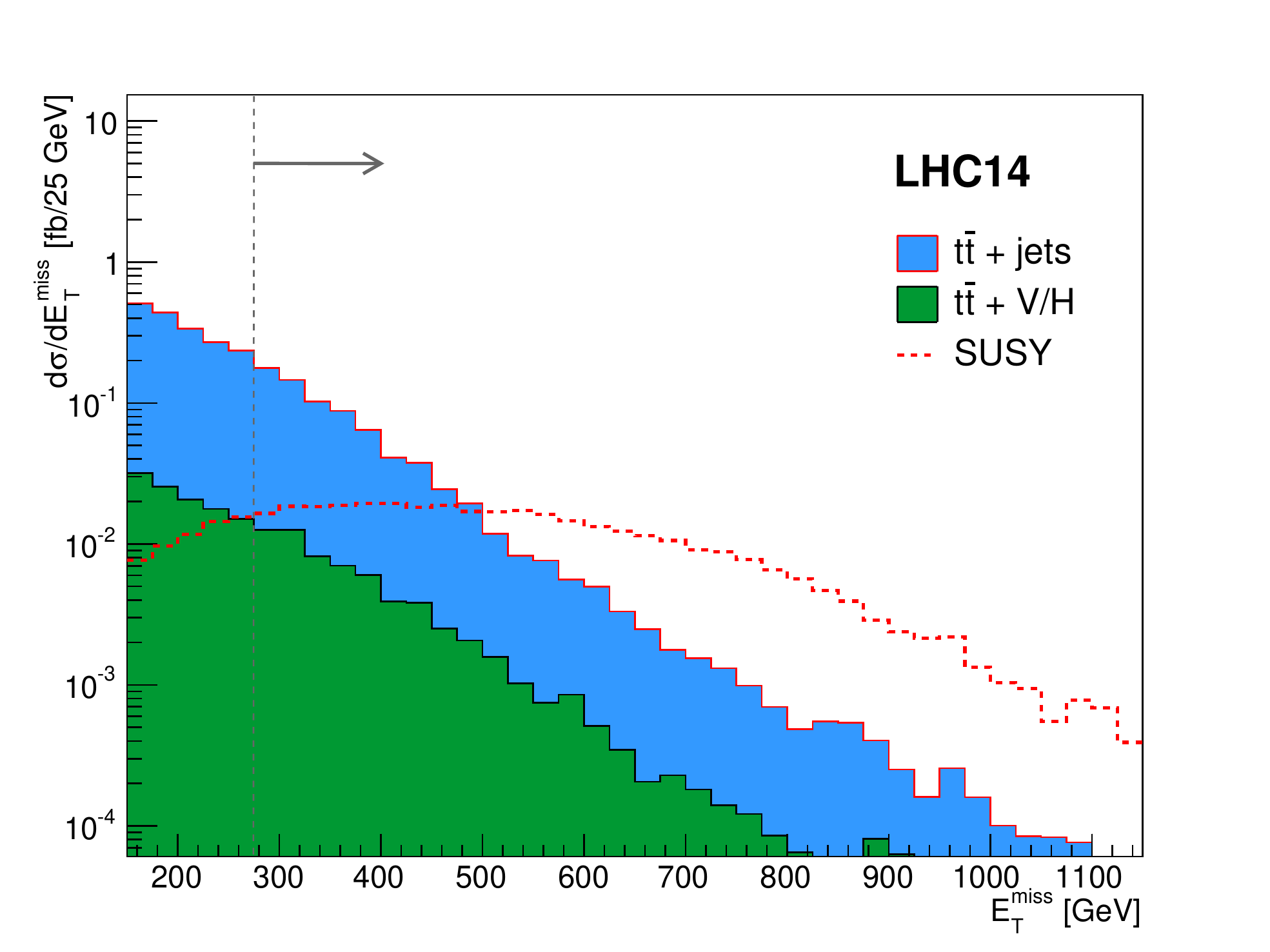}
\caption{\label{nuSUGRA3}Distribution of $\meffincl$ and $\etmiss$ in the signal regions SR-1$\ell$-6j-[A,B,C] {prior to the requirements on these variables. Arrows indicate the remaining cuts used to define the signal regions.} Main SM background consisting of $t\bar{t}$ + jets (blue) and $t\bar{t} + V/H$ (green) are shown. The signal is shown in dashed red. {Our calculations show that $5\sigma$ significance is obtained with an integrated {luminosity of $L \gtrsim 510 \ifb$ (left),  $L \gtrsim 265 \ifb$ (right) and $L \gtrsim 160 \ifb$ (bottom).}}}
\end{center}
\end{figure}

{Distributions of $\meffj$, $\meffincl$ and $\etmiss$ for all the signal regions  SR-0$\ell$-4j-[A,B,C],
SR-0$\ell$-7j-[A,B,C] and SR-1$\ell$-6j-[A,B,C] are shown in
\cref{nuSUGRA1,nuSUGRA2,nuSUGRA3} prior to the requirements on these variables. We follow the ATLAS analysis and apply all
the cuts including the ones shown with arrows in the
distributions and calculate the minimum integrated luminosities required for a $5\sigma$ discovery. The analysis presented in
\cref{nuSUGRA1,nuSUGRA2,nuSUGRA3} shows that for our benchmark model, almost all
the signal regions are effective for discovery {of sparticles at the LHC
Run-II.} Specifically we find that the signal regions SR-0$\ell$-4j-[A,B,C] will require an integrated luminosity of $L \gtrsim 45/60/920$ fb$^{-1}$, the signal regions SR-0$\ell$-7j-[A,B,C] will require an integrated luminosity of $L \gtrsim 235/45/25$ fb$^{-1}$, and the signal regions SR-1$\ell$-6j-[A,B,C] will require an integrated luminosity of $L \gtrsim 510/265/160$ fb$^{-1}$.}

\section{Dark Matter  \label{sec4}}

\begin{figure}[t!]
  \begin{center}
    \includegraphics[width=15cm]{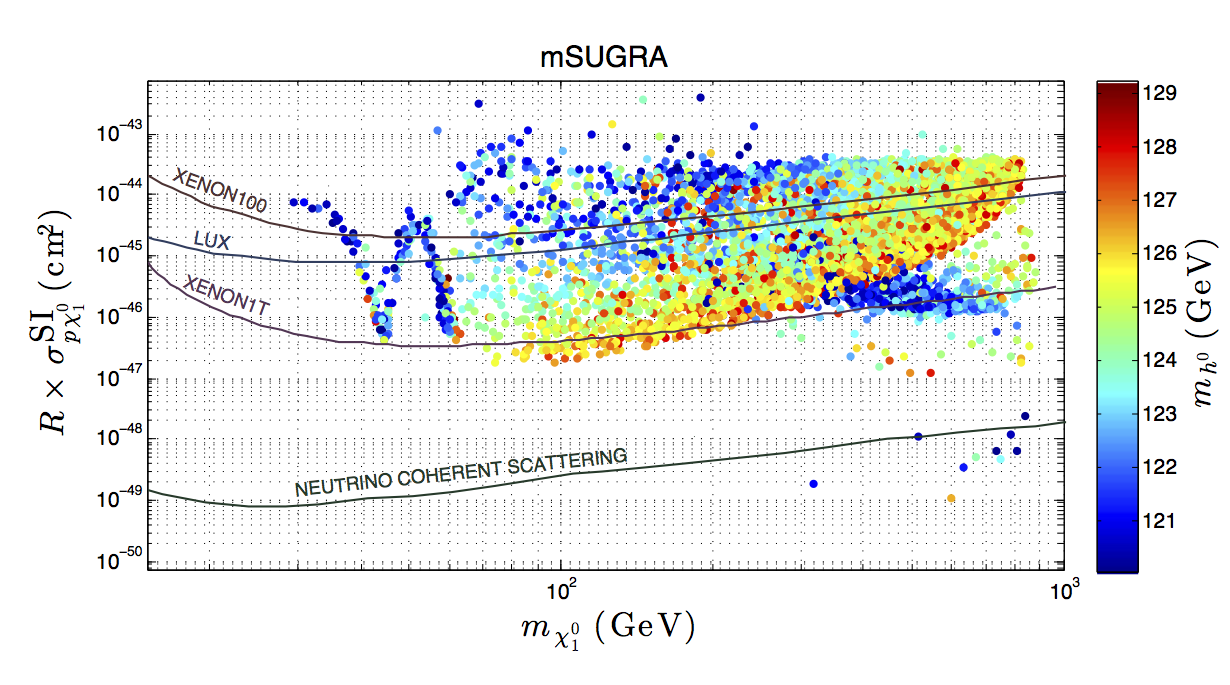}\\
  	\includegraphics[width=15cm]{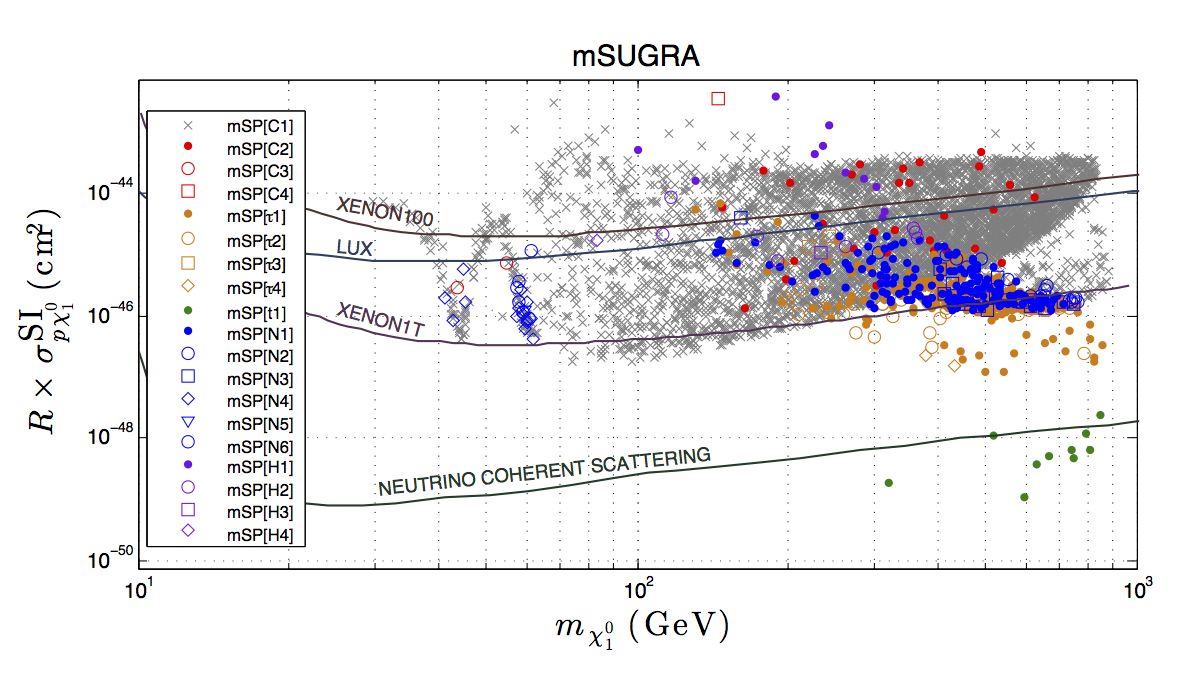}
	\caption{\label{fig3}
	Top panel: A display of $R\times \sigma^\text{SI}_{p,\na}$ vs the neutralino mass for mSUGRA (Model [1])
	where the colors exhibit the Higgs boson mass.  Bottom panel:  
	 Exhibited are various hierarchical patterns contributing to the
	allowed parameter space where  the constraints
	$\Omega h^2 < 0.12$, $m_{h^0} > 120 \GeV$ hold.
	  The analysis indicates that a knowledge of the spin-independent neutralino-proton 
	cross section along with the neutralino mass will allow one to identify the possible underlying sparticle mass hierarchy.}
  \end{center}
\end{figure}

Each of the Models [1]-[5] in \cref{sec2} gives rise to spin-independent
neutralino-proton cross sections which can have a significant range {and,} in some
cases, span several orders of magnitude depending on the content of the
neutralino being a bino, Higgsino, wino, or an admixture. The Higgs boson mass
measurement is also a strong constraint on the spin-independent
neutralino-proton cross section. In the top panel of {\cref{fig3}}
we  display $R\times \sigma^\text{SI}_{p,\na}$ vs the neutralino mass for mSUGRA
Model [1] where $\sigma^\text{SI}_{p,\na}$ is the {spin-independent}
neutralino-proton cross section and $R= \rho_{\na}/\rho_c$ where
$\rho_{\na}$ is the neutralino relic density and $\rho_c$ is the critical relic
density needed to close the universe.  The allowed parameter space in the
$R\times \sigma^\text{SI}_{p,\na}-m_{\na}$ plane is colored according to the Higgs
boson mass.  In the bottom panel of \cref{fig3} we give an analysis of $R\times
\sigma^\text{SI}_{p,\na}$ vs the neutralino mass {for mSUGRA} where we exhibit
the sparticle mass patterns.

A very similar analysis for nuSUGRA Models  is given in \cref{fig4,fig5,fig6,fig7}.
Thus in \cref{fig4} an analysis is given of $R\times \sigma^\text{SI}_{p,\na}$ vs the
neutralino mass for {nuSUGRA} Models [2]-[5] where the colors specify the
Higgs boson mass. In \cref{fig5} we give a composite of all cases considered,
i.e., Models [1]-[5] with the additional constraint that $m_{\hz} \in [123,127]
\GeV$, $\Omega_{\na} h^2 < 0.12$, $\br\bsmumu < 6.2 \times 10^{-9}$, $\br\bsg <
4.27 \times 10^{-4}$.  In the figures here and also in later figures, we have
exhibited the line where the signals from coherent scattering of solar,
atmospheric and diffuse supernova neutrinos will begin to appear. The test of
the neutrino cross sections is interesting in itself.  However, for the purpose
of WIMP detection the neutrino background must be subtracted or it would require
a directional analysis in direct detection experiments to separate the neutrino
backgrounds from the dark matter signals~\cite{Cushman:2013zza}.

In \cref{fig6} we give an analysis of the mass patterns for the nuSUGRA Models
[2]-[3] and in \cref{fig7} we give an analysis of mass patterns for the nuSUGRA
models [4]-[5].  A very interesting phenomenon relates to the fact that dark
matter searches can be used in part as a diagnostic for the type of underlying
sparticle pattern and thus of the sparticle mass hierarchy.  From the bottom
panel of \cref{fig3} and from \cref{fig6} and \cref{fig7}, we note that the
models which dominate the parameter space {mostly have chargino as NLSP.}
However, there are corners where some patterns are more frequent than others.
Thus, for example, in the bottom panel of  \cref{fig3} all the models that lie
close to the neutrino coherent scattering line are the ones where the stop is
the NLSP and the ones above those are mostly  where the lightest stau is the
NLSP. Further, the region between neutralino mass of 60-105\GeV which lies below
the LUX limit consists exclusively of chargino NLSP patterns.  In the analysis
of nuSUGRA models as shown in \cref{fig6,fig7} one  finds that there are certain
regions where only very specific patterns appear while in others only a
combination of two or three patterns appear. In these cases a knowledge of the
spin-independent cross section along with a knowledge of  the neutralino mass
hopefully from collider data will allow us to narrow down the possibilities for
the allowed hierarchical  patterns.  This would help delineate the nature of
high scale boundary conditions for the underlying supergravity grand unification
model. 

 The analysis of  {\cref{fig3,fig4,fig5,fig6,fig7}} shows that certain sparticle patterns {often give too small 
a cross section which are below} the reach of XENON1T and other
similar size dark matter experiments while some of the other patterns have cross sections 
which lie {even} below the neutrino floor. Specifically from the lower panel of \cref{fig3} we see that 
the stop patterns, i.e., mSP$[t1]$ have cross sections which cross the neutrino coherent scattering line
and thus the cross sections from these would be extremely challenging to observe. The reason for
the extreme smallness of the cross sections in that the neutralino in these case is essentially 
almost 100\% bino-like, and do not have light first and second generation
squark states available which suppresses the strength of interaction in the
scattering off nuclei.  
Similar observations apply to the analysis of \cref{fig4}-\cref{fig7}. In
\cref{sugra_nmix}, we display the bino and Higgsino fractions of such model
points along with {the corresponding} spin-independent neutralino-proton cross sections. {To explain the smallness of the spin-independent cross section further for some of the simplified  models,
 we consider the explicit form of the scalar cross section for the neutralino-nucleus scattering which is given by
\begin{equation}
\sigma^{\rm SI}_{\chi_1^0 N}  =  \left(4 \mu^2_{r}/\pi\right) \left(Z f_p +(A-Z)f_n\right)^2~.
\end{equation}
where $Z$ is the  total number of protons, $A$ the total number of protons and neutrons in the nucleus,
and $\mu_{r}$ is  the neutralino reduced mass.
For the case when the squarks are very heavy, the s-channel pole diagrams in neutralino-quark scattering
give a relatively small contribution which is dominated by the t-channel Higgs-boson exchanges.
In this circumstance $f_{p/n}$ are given by 
\begin{equation}
f_{p/n}=\sum_{q=u,d,s} f^{(p/n)}_{T_q} {C}_q \frac{m_{p/n}}{m_q} + \frac{2}{27} f^{(p/n)}_{TG} \sum_{q=c,b,t} {C}_q \frac{m_{p/n}}{m_q}, 
\end{equation}
where  the form factors $f^{(p/n)}_{T_q}$  and  $f^{(p/n)}_{TG} $  are given in 
\cite{Chattopadhyay:1998wb,Ellis:2000ds,Belanger:2008sj} and  the couplings $C_i$ are given by 
\begin{multline}
{C}_q   =    - \frac{g_2 m_{q}}{4 m_{W} \delta_3} \left[  \left( 
g_2 n_{12} - g_Y n_{11} \right) \delta_{1} \delta_4\delta_5 \left( - \frac{1}{m^{2}_{H}} + 
\frac{1}{m^{2}_{h}} \right) \right. 
\\
 +  \left. \left(g_2 n_{12} - g_Y n_{11} \right)  \delta_{2}\left(  \frac{\delta_4^{2}}{m^{2}_{H}} +\frac{\delta_5^{2}}{m^{2}_{h}}\right) \right]~.
 \label{spinind-3}
\end{multline} 
In the above $n_{11}$, etc are defined so that $\chi_1^0= n_{11} \tilde B + n_{12} \tilde W_3  + n_{13} \tilde H_1 + n_{14} \tilde H_2$, where $\tilde B, \tilde W_3, \tilde H_1, \tilde H_2$ are the bino, wino, Higgsino 1 and Higgsino 2 fields. Further, 
 $\delta_i$ are defined so that  for up quarks 
$\delta_i = (n_{13},n_{14}, \sin\beta, \sin\alpha, \cos\alpha)$ and  for down quarks 
$\delta_i=
(n_{14},-n_{13}, \cos\beta,\cos\alpha,-\sin\alpha)$,  where $i$ runs from $1$ to $5$ and $\alpha$ is the neutral Higgs mixing 
parameter.  From Eq.(\ref{spinind-3}) we see that the cross section depends directly on $\delta_1, \delta_2$ and 
consequently on $n_{13}, n_{14}$ and hence on the Higgsino fraction. From the last column of \cref{sugra_nmix}
we see that the Higgsino content is indeed very small for the patterns listed in \cref{sugra_nmix} which explains
the smallness of spin-independent cross sections listed in the 10th column of \cref{sugra_nmix}
and also explains the smallness of the cross sections that appear in \cref{fig3}-\cref{fig7}.}

\section{Conclusion\label{sec7}}

In this work we have analyzed the {landscape of sparticle mass hierarchies
that are consistent with the Higgs boson mass measurement.} Sparticle mass
hierarchies are {crucial for} understanding the nature of symmetry breaking
at the high scale. {As the soft breaking parameters at the grand unification
scale will in general be nonuniversal, the} different patterns of soft
parameters as well as their specific values  lead to a variety of different
hierarchical patterns for sparticle masses.  For 31 particle masses beyond the
standard model one finds that {a landscape of $10^{33}$ mass hierarchies
arises}.  A large number of these are eliminated under the constraints of
electroweak symmetry breaking, and collider and dark matter constraints.
However, the residual number of patterns is still large.  This number is
drastically reduced when one considers the hierarchical patterns for the case
when  the number of particles taken into account is five{, including the
neutralino LSP.} 

The analysis of the hierarchical patterns is done for {several} different high
scale models. These include [1] the mSUGRA model with universal boundary
conditions, [2] {the nonuniversal} SUGRA model with nonuniversality in the $\SU(2)_L$
gaugino mass sector, [3] {the nonuniversal} SUGRA model with nonuniversality in the
$\SU(3)_C$ gaugino mass sector, [4] {the nonuniversal} SUGRA model with
nonuniversality in the Higgs boson mass sector, and [5]  {the nonuniversal} SUGRA
model with nonuniversality in the third generation sfermion sector. 
Further, we have produced lists of  simplified models with 3, 4 or 5 sparticle mass
hierarchies along {with} their relative occurrences that arise from supergravity models
with universal and {nonuniversal} boundary conditions. These hierarchical
patterns will be helpful in establishing the nature of the high scale models
which give rise to the hierarchical patterns. 
 In addition
we have provided benchmarks for SUGRA Models, {which span the
parameter space of supergravity models with universal as well as nonuniversal
boundary conditions. These benchmarks should prove to be useful for SUSY
searches at the LHC Run-II}.
We have also carried out an explicit signature analysis for two benchmark
models, one for a supergravity model with universal boundary conditions and the
other for a supergravity model with {nonuniversal} boundary conditions.  

\begin{figure}[!]
  \begin{center}
  	\includegraphics[scale=0.45]{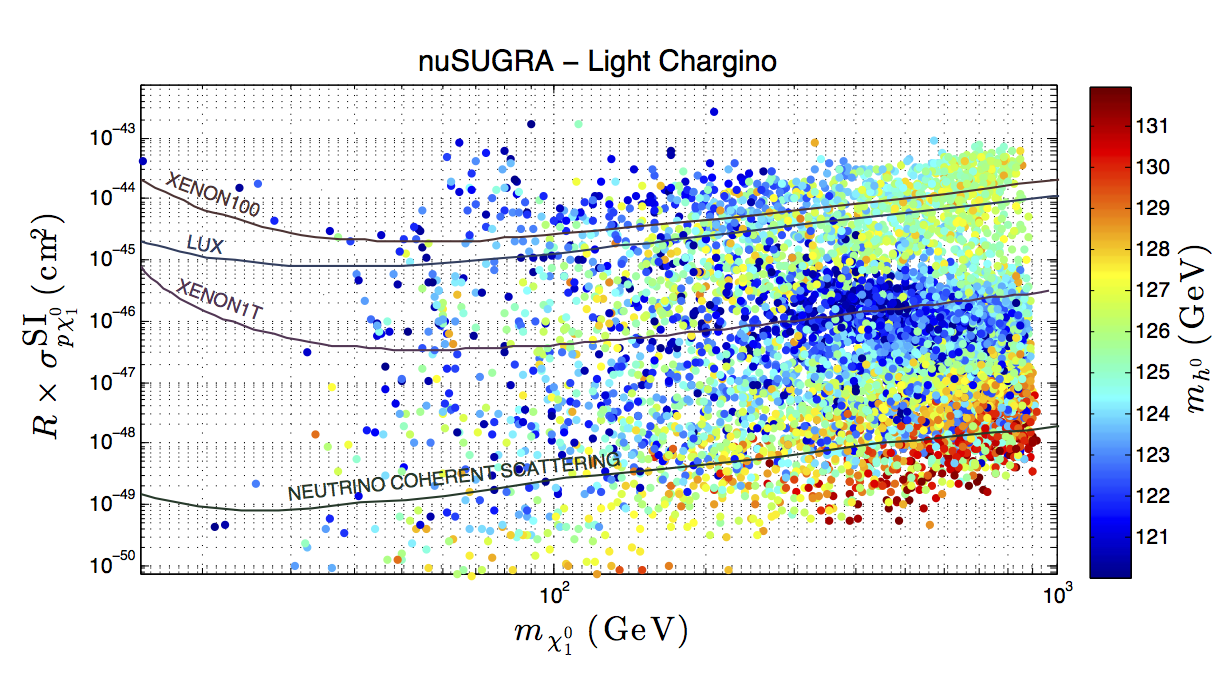}
  	\includegraphics[scale=0.45]{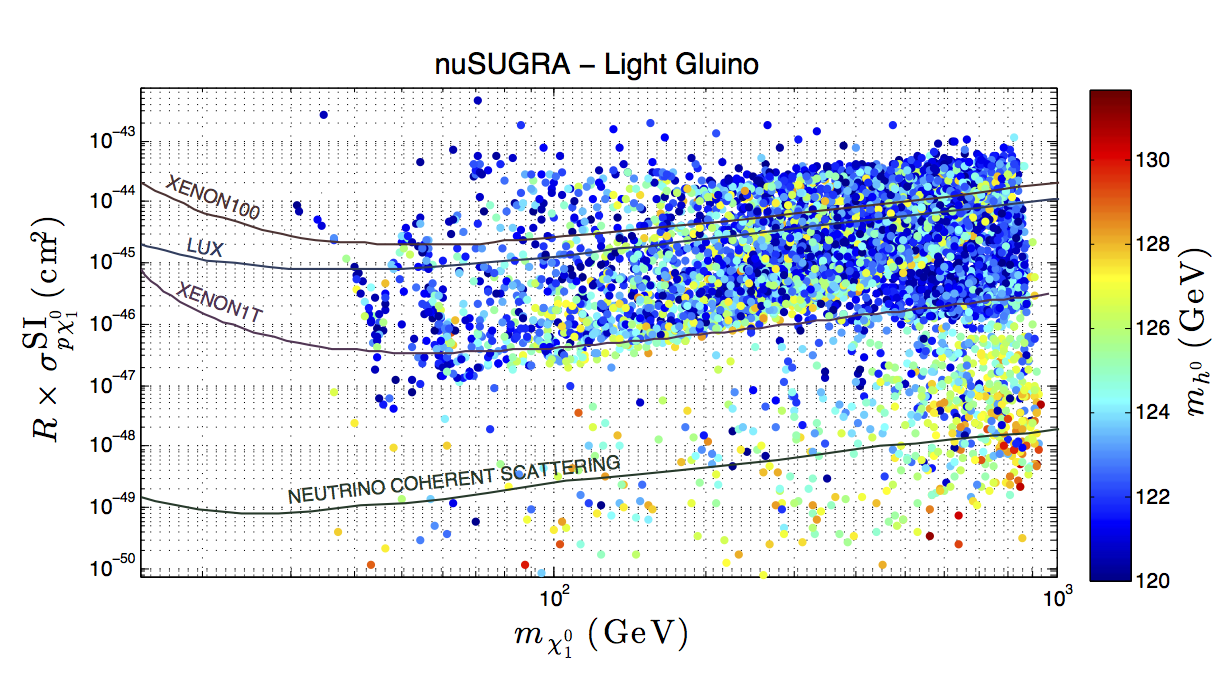}
	  \includegraphics[scale=0.45]{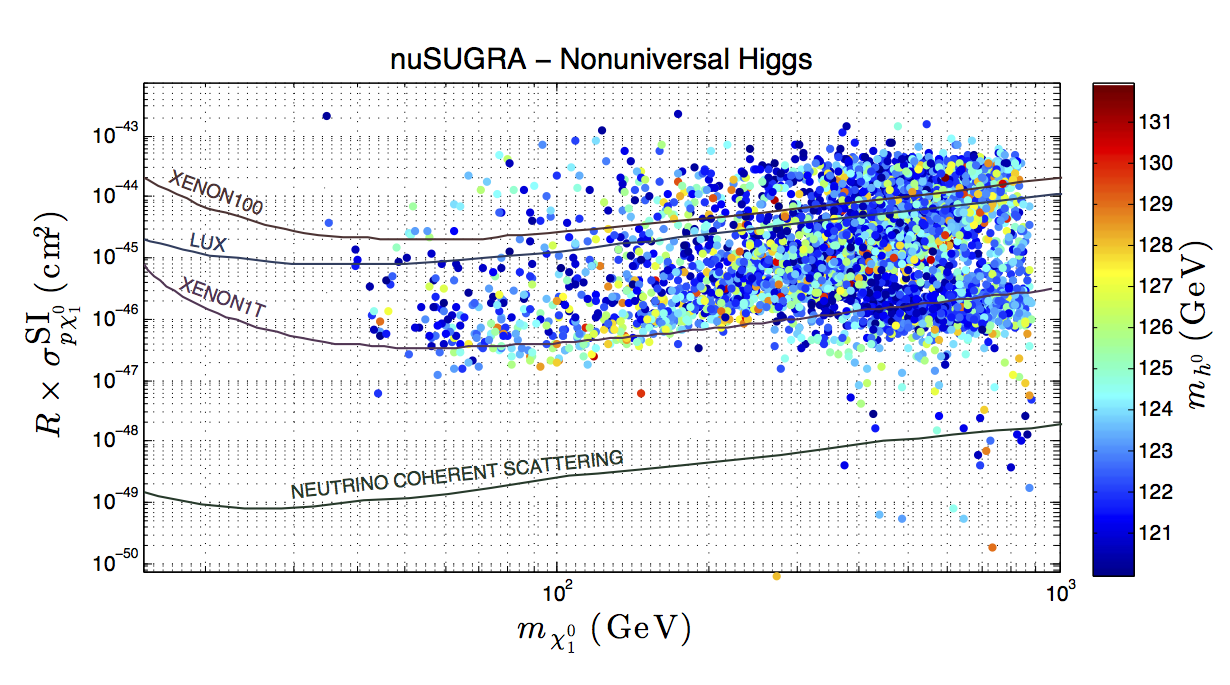}
  	\includegraphics[scale=0.45]{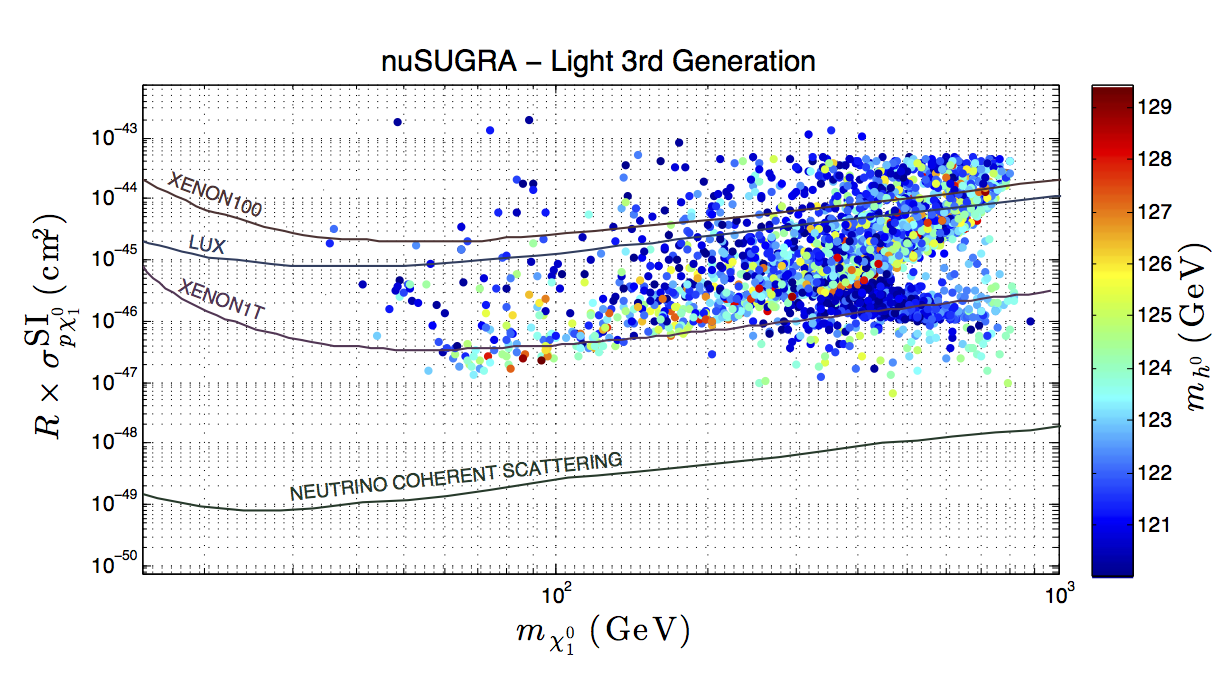}
	\caption{\label{fig4}
	Top to bottom:
  A display of $R\times \sigma^\text{SI}_{p,\na}$ vs the neutralino mass  for
  nuSUGRA Models [2]-Model [5].  The color specifies the mass of the Higgs boson.
 The cross section for neutrino-coherent scattering (the neutrino floor) is also plotted~\cite{Cushman:2013zza}.}
  \end{center}
\end{figure}

\begin{figure}[!]
  \begin{center}
	\includegraphics[width=15cm]{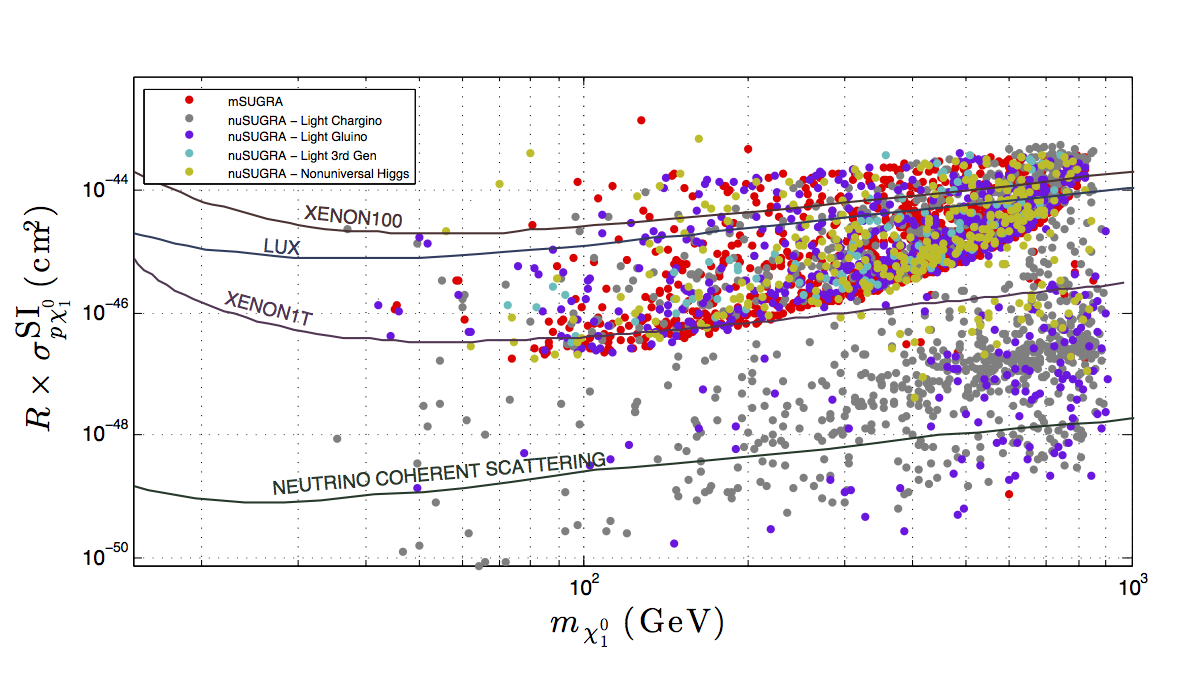}
		\caption{\label{fig5}
  A display of $R\times \sigma^\text{SI}_{p,\na}$ vs the neutralino mass 
  which is  a composite for  Models [1]-[5] with the additional constraints, ${m_{h^0}} \in [123,127] \GeV, \Omega h^2 < 0.12, \bsmumu < 6.2 \times 10^{-9}, \bsg < 4.27 \times 10^{-4}$.}
  \end{center}
\end{figure}

\begin{figure}[!]
  \begin{center}
	\includegraphics[width=15cm]{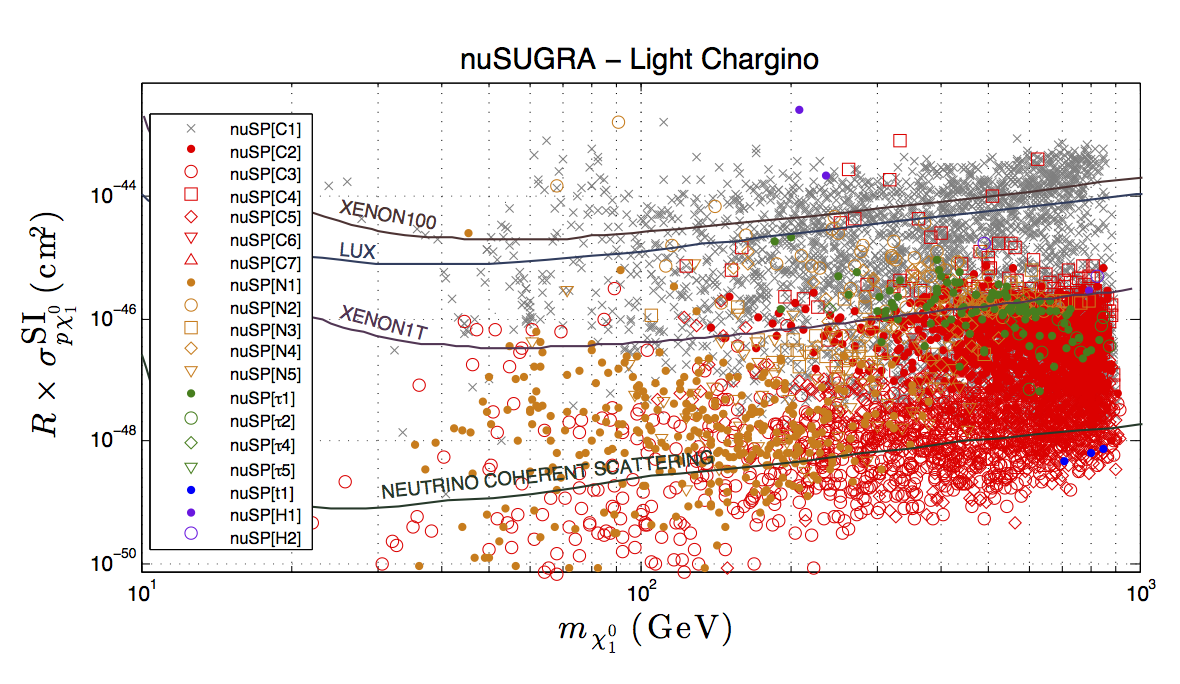}
  \includegraphics[width=15cm]{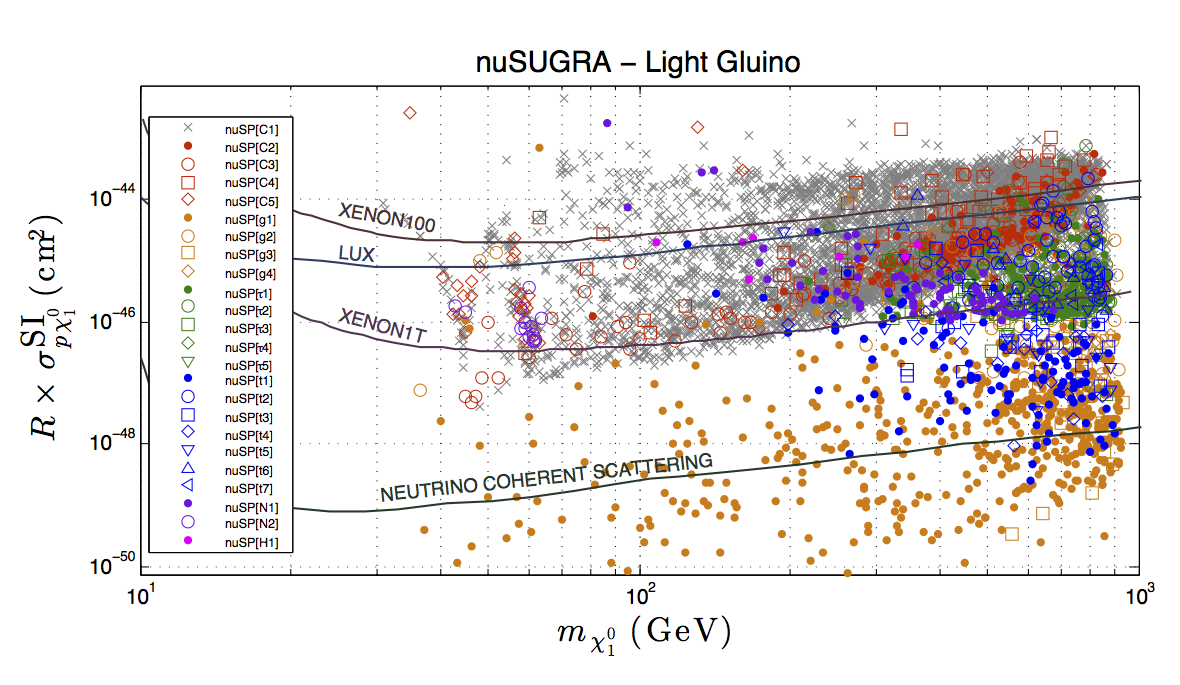}	
	\caption{\label{fig6}
      Dark matter plots of the nuSUGRA models, including the {light} chargino
      case {(Model [2]) and} the light gluino case (Model [3]).  The color and
      shape specifies the sparticle mass pattern.
    }
  \end{center}
\end{figure}

\begin{figure}[!]
  \begin{center}
	\includegraphics[width=15cm]{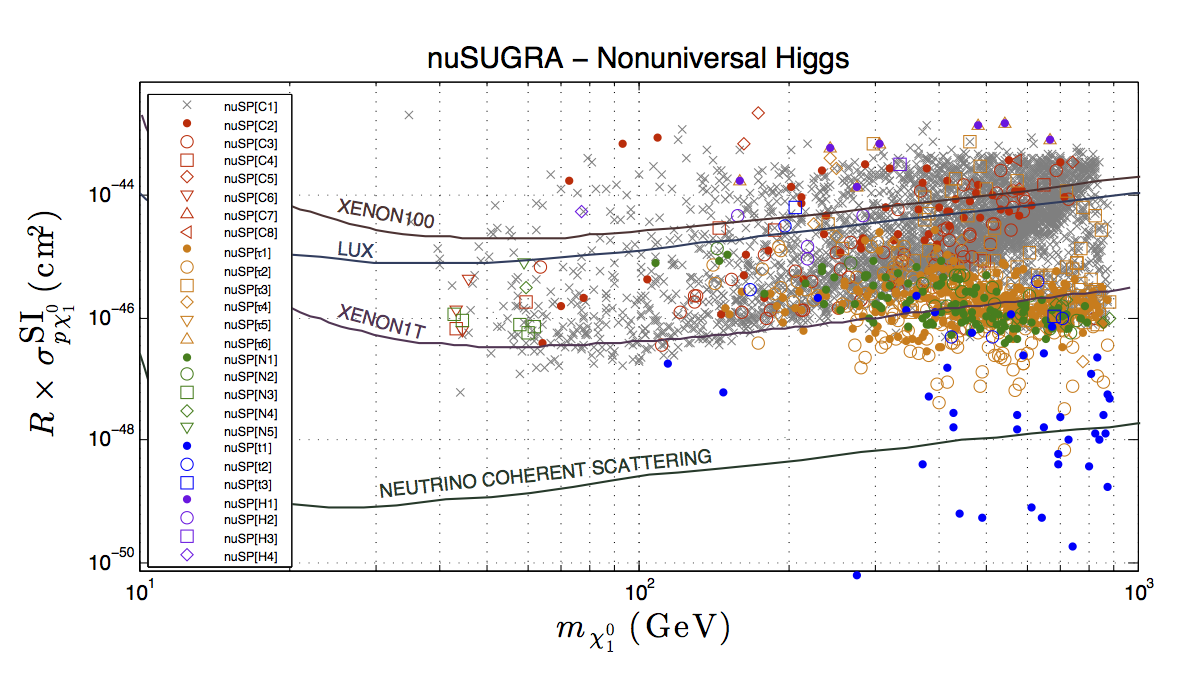}
  \includegraphics[width=15cm]{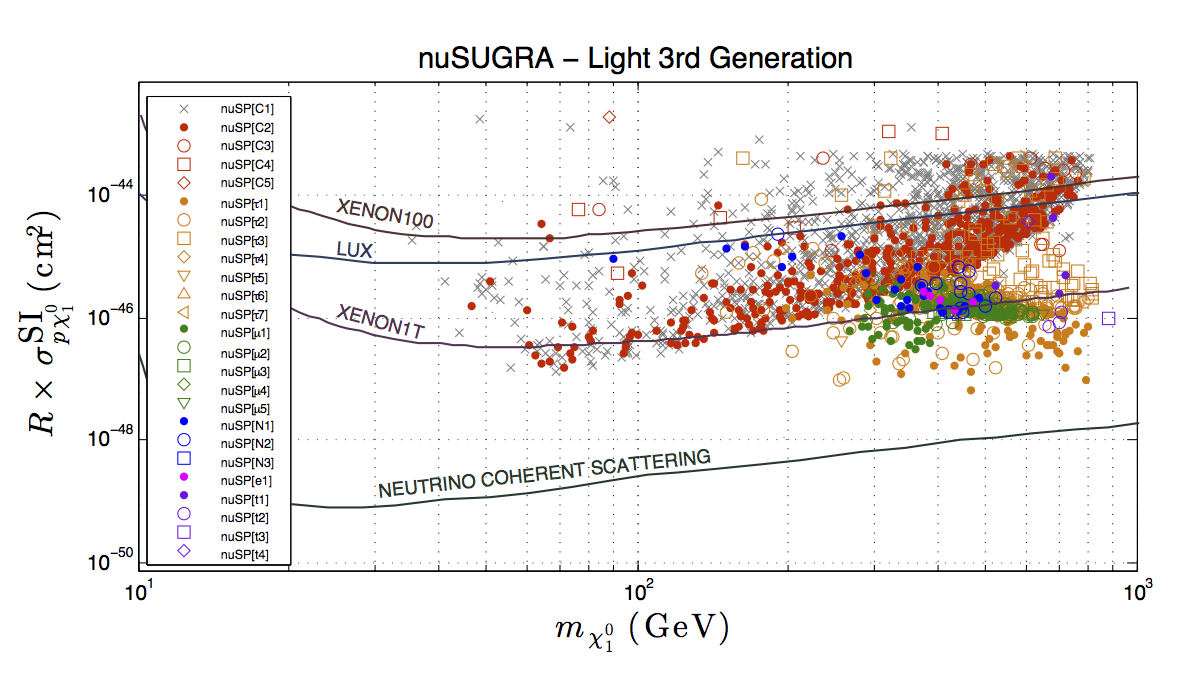}
	\caption{\label{dm_nusugra_colorMassPatterns}\label{fig7}
      Dark matter plots of the nuSUGRA models, including the nonuniversal {Higgs} case (Model [4]) and 
      the light third generation case (Model [5]).       
       The color and the shape specifies the sparticle mass pattern.
    }
  \end{center}
\end{figure}

{
 In addition
to the analysis of  sparticle mass hierarchies we have also analyzed the
spin-independent neutralino-proton cross section for the five classes of supergravity Models discussed
in  \cref{sec2}. It is
found that the  latest limits from the LUX dark matter experiment probe a
significant part of the parameter space of models, and the  XENON1T and
SuperCDMS will be able to exhaust significantly more of the parameter space {of
many} of the SUGRA models. However, for the case of the SUGRA models with
nonuniversalities especially in the $\SU(2)_L$  and $\SU(3)_C$ gaugino sectors,
the {neutralino-proton} cross {sections extend} downwards even past the
{so-called} neutrino floor as shown in \cref{fig4,fig5,fig6,fig7}. 
Another important aspect of the analysis relates to the diagnostic of the
{spin-independent cross section vs the neutralino mass}. An analysis of the patterns in
the {spin-independent cross section vs the neutralino mass-plane} one finds that certain regions of
the plane are populated dominantly by one or two patterns.  Thus a simultaneous
measurement of the spin-independent cross section and a knowledge of the
neutralino mass, such as from collider experiments, could {isolate} the likely
sparticle mass hierarchies and thus provide strong clues to the nature of
symmetry breaking in high scale models. Thus dark matter analyses along with
analyses of the LHC Run-II can allow one to pin down in a concrete way the
nature of the high scale models leading to the sparticle mass patterns. }

\acknowledgments
Discussions with Zuowei Liu and Meenakshi Narain are acknowledged. The research of D.F. and P.N. was
supported in part by NSF grants PHY-1314774 and PHY-0969739,  and by XSEDE grant
TG-PHY110015.  The research of S.A. was supported by European Union and the
European Social Fund through LHCPhenoNet network PITN-GA-2010-264564. The
research of B.A. was supported in part by the U.S. Department of Energy under
Grant No. DE-FG02-13ER41979.  This research used resources of the National
Energy Research Scientific Computing Center, which is supported by the Office of
Science of the U.S. Department of Energy under Contract No.  DE-AC02-05CH11231,
and {of the} OU Supercomputing Center for Education \& Research (OSCER) at the University
of Oklahoma (OU).

\vspace{1.0cm}

\vspace{-1cm}

\begin{table}[h!]
\begin{center}
\begin{tabular}{|l|l|c|}
\hline
Pattern Label
		&  \multicolumn{1}{c|}{Mass Hierarchy}
			& \%	 \\
\hline
\hline
mSP[C1a]
		& $\chi^{\pm}_{1} < \chi^{0}_{2} < \chi^{0}_{3}< \chi^{0}_{4}$
			&  83.8 \\
mSP[C1b]
		& $\chi^{\pm}_{1} < \chi^{0}_{2} < \chi^{0}_{3}< H^0$
			&  2.49 \\
mSP[C1c]
		& $\chi^{\pm}_{1} < \chi^{0}_{2} < \chi^{0}_{3}< \chi^{\pm}_{2} $
			&  1.62 \\
\hline
mSP[C2]
		& $\chi^{\pm}_{1} < \chi^{0}_{2} < H^0< A^0$
			&  0.65 \\
\hline
mSP[C3]
		& $\chi^{\pm}_{1} < \chi^{0}_{2} < g< \chi^{0}_{3} $
			&  0.04 \\
\hline
mSP[C4]
		& $\chi^{\pm}_{1} < \chi^{0}_{2} < A^0< H^0 $
			&  0.02 \\
\hline
\hline
mSP[$\tau$1a]
		& $\tau_{1} < \chi^{0}_{2} <\chi^{\pm}_{1}< H^0 $
			&  3.89 \\
mSP[$\tau$1b]
		& $\tau_{1} < \chi^{0}_{2} <\chi^{\pm}_{1}< \mu_r $
			&  0.89 \\
mSP[$\tau$1c]
		& $\tau_{1} < \chi^{0}_{2} <\chi^{\pm}_{1}< \nu_\tau $
			&  0.15 \\
mSP[$\tau$1d]
		& $\tau_{1} < \chi^{0}_{2} <\chi^{\pm}_{1}< t_1 $
			&  0.09 \\
\hline
mSP[$\tau$2a]
		& $\tau_{1} < \mu_r < e_r < \chi^{0}_{2}$
			&  0.69 \\
mSP[$\tau$2b]
		& $\tau_{1} < \mu_r < e_r < \nu_\tau$
			&  0.52 \\
\hline
mSP[$\tau$3a]
		& $\tau_{1} < H^0 < A^0 < \chi^{0}_{2}$
			&  0.04 \\
mSP[$\tau$3b]
		& $\tau_{1} < H^0 < A^0 < H^{\pm}$
			&  0.02 \\
\hline
mSP[$\tau$4]
		& $\tau_{1} < t_1 < \chi^{0}_{2} < \chi^{\pm}_{1}$
			&  0.04 \\
\hline
\hline
mSP[t1a]
		& $t_1 < \chi^{0}_{2} < \chi^{\pm}_{1} < g$
			&  0.11 \\
mSP[t1b]
		& $t_1 < \chi^{0}_{2} < \chi^{\pm}_{1} < \tau_1$
			&  0.06 \\
mSP[t1c]
		& $t_1 < \chi^{0}_{2} < \chi^{\pm}_{1} < b_1$
			&  0.02 \\
\hline
\hline
mSP[N1a]
		& $ \chi^{0}_{2} < \chi^{\pm}_{1} < H^0 < A^0$
			&  3.31 \\
mSP[N1b]
		& $ \chi^{0}_{2} < \chi^{\pm}_{1} < H^0 < \chi^{0}_{3} $
			&  0.02 \\
mSP[N1c]
		& $ \chi^{0}_{2} < \chi^{\pm}_{1} < H^0 < \tau_1$
			&  0.02 \\
\hline
mSP[N2a]
		& $ \chi^{0}_{2} < \chi^{\pm}_{1} < \chi^{0}_{3} < H^0$
			&  0.24 \\
mSP[N2b]
		& $ \chi^{0}_{2} < \chi^{\pm}_{1} < \chi^{0}_{3} < \chi^{0}_{4}$
			&  0.20 \\
\hline
mSP[N3]
		& $ \chi^{0}_{2} < \chi^{\pm}_{1} < \tau_1 < H^0$
			&  0.39 \\
\hline
mSP[N4]
		& $ \chi^{0}_{2} < \chi^{\pm}_{1} < g < \chi^{0}_{3}$
			&  0.26 \\
\hline
mSP[N5]
		& $ \chi^{0}_{2} < \chi^{\pm}_{1} < t_1 < g$
			&  0.02 \\
\hline
mSP[N6]
		& $ \chi^{0}_{2} < H^0 < \chi^{\pm}_{1} < A^0$
			&  0.02 \\
\hline
\hline
mSP[H1a]
		& $ H^0 < A^0 < H^{\pm} < \chi^{\pm}_{1} $
			&  0.15 \\
mSP[H1b]
		& $ H^0 < A^0 < H^{\pm} < \chi^{0}_{2} $
			&  0.06 \\
\hline
mSP[H2]
		& $ H^0 < A^0 < \chi^{0}_{2} < \chi^{\pm}_{1}$
			&  0.15 \\
\hline
mSP[H3]
		& $ H^0 < \chi^{0}_{2} < A^0 < \chi^{\pm}_{1}$
			&  0.02 \\
\hline
mSP[H4]
		& $ H^0 < \chi^{0}_{2} < \chi^{\pm}_{1} < A^0$
			&  0.02 \\
\hline
\end{tabular}
\caption{\label{msugra_mass_patterns}\label{tab2}
Sparticle mass hierarchies for the  mSUGRA parameter space (Model [1]), where $\na$ is the LSP. The high scale parameters 
lie in the range  $\mo \in [0.1,10] \TeV$, $\mhf \in [0.1,1.5] \TeV$, $\frac{\az}{\mo} \in [-5,5]$, $\tan\beta \in [2,50]$, $\mu > 0$, with the constraints $\Omega h^2 < 0.12$, {$m_{h^0} > 120 \GeV$}.
{Here and in \cref{tab4,tab5,tab6,tab7} the last column gives the percentage with which the patterns appear
in the scans given in \cref{tab1}.}
}
\end{center}
\end{table}

\begin{table}[h!]
\begin{center}
{\footnotesize
\begin{tabular}{|l|l|c|}

\hline
Pattern Label
		&  \multicolumn{1}{c|}{Mass Hierarchy}
			& \%	 \\
\hline
\hline
nuSP$_2$[C1a]
		& $ \chi^{\pm}_1 < \chi^{0}_2 < \chi^{0}_3 < \chi^{0}_4 $
			&  35.71 \\
nuSP$_2$[C1b]
		& $ \chi^{\pm}_1 < \chi^{0}_2 < \chi^{0}_3 < \chi^{\pm}_2 $
			&  14.57 \\
nuSP$_2$[C1c]
		& $ \chi^{\pm}_1 < \chi^{0}_2 < \chi^{0}_3 < H^0 $
			&  0.188 \\
nuSP$_2$[C1d]
		& $ \chi^{\pm}_1 < \chi^{0}_2 < \chi^{0}_3 < g $
			& 0.010  \\
nuSP$_2$[C1e]
		& $ \chi^{\pm}_1 < \chi^{0}_2 < \chi^{0}_3 < \tau_1 $
			& 0.009  \\
nuSP$_2$[C1f]
		& $ \chi^{\pm}_1 < \chi^{0}_2 < \chi^{0}_3 < t_1 $
			&  0.003 \\
nuSP$_2$[C1g]
		& $ \chi^{\pm}_1 < \chi^{0}_2 < \chi^{0}_3 < A^0 $
			& 0.001  \\
\hline
nuSP$_2$[C2a]
		& $ \chi^{\pm}_1 < \chi^{0}_2 < \tau_1 < \nu_\tau $
			& 12.25  \\ 
nuSP$_2$[C2b]
		& $ \chi^{\pm}_1 < \chi^{0}_2 < \tau_1 < H^0 $
			& 1.571  \\ 
nuSP$_2$[C2c]
		& $ \chi^{\pm}_1 < \chi^{0}_2 < \tau_1 < t_1$
			& 0.907  \\ 
nuSP$_2$[C2d]
		& $ \chi^{\pm}_1 < \chi^{0}_2 < \tau_1 < \chi^{0}_3 $
			& 0.778  \\ 
nuSP$_2$[C2e]
		& $ \chi^{\pm}_1 < \chi^{0}_2 < \tau_1 < \mu_r $
			& 0.289  \\ 
nuSP$_2$[C2f]
		& $ \chi^{\pm}_1 < \chi^{0}_2 < \tau_1 < g $
			&  0.002 \\ 
\hline
nuSP$_2$[C3a]
		& $ \chi^{\pm}_1 < \chi^{0}_2 < g < \chi^{0}_3 $
			&  6.237 \\ 			
nuSP$_2$[C3b]
		& $ \chi^{\pm}_1 < \chi^{0}_2 < g < t_1 $
			&  4.820 \\ 
nuSP$_2$[C3c]
		& $ \chi^{\pm}_1 < \chi^{0}_2 < g < H^0 $
			&  0.083 \\ 
nuSP$_2$[C3d]
		& $ \chi^{\pm}_1 < \chi^{0}_2 < g < H^{\pm} $
			& 0.016  \\ 
\hline
nuSP$_2$[C4a]
		& $ \chi^{\pm}_1 < \chi^{0}_2 < H^0 < A^0$
			&  2.767 \\ 
nuSP$_2$[C4b]
		& $ \chi^{\pm}_1 < \chi^{0}_2 < H^0 < H^{\pm}	 $
			&  0.074 \\ 
nuSP$_2$[C4c]
		& $ \chi^{\pm}_1 < \chi^{0}_2 < H^\pm < H^0 $
			&  0.025 \\ 
nuSP$_2$[C4d]
		& $ \chi^{\pm}_1 < \chi^{0}_2 < A^0 < H^0$
			&  0.003 \\ 
nuSP$_2$[C4e]
		& $ \chi^{\pm}_1 < \chi^{0}_2 < A^0 < \tau_1$
			&  0.001 \\ 
\hline
nuSP$_2$[C5a]
		& $ \chi^{\pm}_1 < \chi^{0}_2 < t_1 < g$
			&  6.015 \\ 
nuSP$_2$[C5b]
		& $ \chi^{\pm}_1 < \chi^{0}_2 < t_1 < \tau_1$
			&  2.746 \\ 
nuSP$_2$[C5c]
		& $ \chi^{\pm}_1 < \chi^{0}_2 < t_1 < \chi^{0}_3$
			&  1.490 \\ 
nuSP$_2$[C5d]
		& $ \chi^{\pm}_1 < \chi^{0}_2 < t_1 < b_1 $
			& 0.222  \\ 
nuSP$_2$[C5e]
		& $ \chi^{\pm}_1 < \chi^{0}_2 < t_1 < H^0$
			&  0.039 \\ 
nuSP$_2$[C5f]
		& $ \chi^{\pm}_1 < \chi^{0}_2 < t_1 < H^\pm$
			&  0.008 \\ 
\hline
nuSP$_2$[C6a]
		& $ \chi^{\pm}_1 < \tau_1 < \chi^{0}_2 < \nu_\tau$
			&  0.274 \\ 
nuSP$_2$[C6b]
		& $ \chi^{\pm}_1 < \tau_1 < \chi^{0}_2 < H^0$
			&  0.008 \\ 
nuSP$_2$[C6c]
		& $ \chi^{\pm}_1 < \tau_1 < \chi^{0}_2 < \mu_r$
			&  0.004 \\ 
nuSP$_2$[C6d]
		& $ \chi^{\pm}_1 < \tau_1 < \chi^{0}_2 < t_1$
			&  0.001 \\ 
\hline
nuSP$_2$[C7a]
		& $ \chi^{\pm}_1 < \tau_1 < \nu_\tau < \nu_\mu$
			&  0.023 \\ 
nuSP$_2$[C7b]
		& $ \chi^{\pm}_1 < \tau_1 < \nu_\tau < \chi^{0}_2 $
			&  0.004 \\ 
\hline
nuSP$_2$[C8a]
		& $ \chi^{\pm}_1 < H^0 < A^0 < H^\pm   $
			&   0.010 \\ 
nuSP$_2$[C8b]
		& $ \chi^{\pm}_1 < H^0 < A^0 < \chi^0_2   $
			&  0.006 \\ 
nuSP$_2$[C8c]
		& $ \chi^{\pm}_1 < A^0 < H^0 < \chi^0_2   $
			& 0.001  \\ 
\hline
nuSP$_2$[C9a]
		& $ \chi^{\pm}_1 < t_1 < \chi^0_2 < g  $
			&  0.003 \\ 
nuSP$_2$[C9b]
		& $ \chi^{\pm}_1 <  t_1 < \chi^0_2 < \tau_1  $
			&  0.003 \\ 
\hline
nuSP$_2$[C10]
		& $ \chi^{\pm}_1 < H^0 < \chi^0_2 < A^0  $
			&  0.001 \\ 
\hline
\hline
nuSP$_2$[t1a]
		& $ t_1 < \chi^0_2 < \chi^{\pm}_1 < g $
			&  0.030 \\ 
nuSP$_2$[t1b]
		& $ t_1 < \chi^0_2 < \chi^{\pm}_1 < \tau_1 $
			&  0.009 \\ 
nuSP$_2$[t1c]
		& $ t_1 < \chi^0_2 < \chi^{\pm}_1 < b_1 $
			&  0.004  \\ 
\hline
nuSP$_2$[t2a]
		& $ t_1 < \chi^0_2 < \chi^{\pm}_1 < g $
			&  0.001 \\ 
nuSP$_2$[t2b]
		& $ t_1 < \chi^0_2 < \chi^{\pm}_1 < \tau_1 $
			&  0.001 \\ 
\hline
\end{tabular}
\begin{tabular}{|l|l|c|}
\hline
Pattern Label
		&  \multicolumn{1}{c|}{Mass Hierarchy}
			& \%	 \\
\hline
\hline
nuSP$_2$[H1a]
		& $ H^0 < A^0 < H^\pm < \chi^\pm_1 $
			&  0.016		 \\
nuSP$_2$[H1b]
		& $ H^0 < A^0 < H^\pm < \chi^0_2 $
			&  0.003		 \\
nuSP$_2$[H1c]
		& $ A^0 < H^0 < H^\pm < \chi^0_2 $
			&  0.001		 \\
nuSP$_2$[H1d]
		& $ A^0 < H^0 < H^\pm < \chi^\pm_1 $
			&  0.001		 \\
\hline
nuSP$_2$[H2a]
		& $ H^0 < A^0 < \chi^\pm_1 < \chi^0_2 $
			&  0.006		 \\
nuSP$_2$[H2b]
		& $ H^0 < A^0 < \chi^\pm_1 < H^\pm $
			&  0.002		 \\
nuSP$_2$[H2c]
		& $ A^0 < H^0 < \chi^\pm_1 < \chi^0_2 $
			&  0.001		 \\
\hline
nuSP$_2$[H3]
		& $ H^0 < A^0 < \chi^0_2 < \chi^\pm_1 $
			&  	0.002	 \\
\hline
nuSP$_2$[H4]
		& $ H^0 < \chi^\pm_1< A^0  < \chi^0_2$
			&  	0.001	 \\
\hline
\hline
nuSP$_2$[N1a]
		& $ \chi^0_2 < \chi^\pm_1 < g < \chi^0_3$
			&  	2.246	 \\
nuSP$_2$[N1b]
		& $ \chi^0_2 < \chi^\pm_1 < g < t_1 $
			&  1.774		 \\
nuSP$_2$[N1c]
		& $ \chi^0_2 < \chi^\pm_1 < g < H^0 $
			&  0.065		 \\
\hline
nuSP$_2$[N2a]
		& $ \chi^0_2 < \chi^\pm_1 < H^0 < A^0 $
			&  	0.976	 \\
nuSP$_2$[N2b]
		& $ \chi^0_2 < \chi^\pm_1 <  H^0 < \tau_1 $
			&  	0.001	 \\
nuSP$_2$[N2c]
		& $ \chi^0_2 < \chi^\pm_1 < A^0 <  H^0 $
			&  	0.001	 \\
\hline
nuSP$_2$[N3a]
		& $ \chi^0_2 < \chi^\pm_1 < \chi^0_3 < \chi^0_4$
			&  	0.448	 \\
nuSP$_2$[N3b]
		& $ \chi^0_2 < \chi^\pm_1 < \chi^0_3 < \chi^\pm_2$
			&  	0.049	 \\
nuSP$_2$[N3c]
		& $ \chi^0_2 < \chi^\pm_1 < \chi^0_3 < H^0$
			&  0.017		 \\
nuSP$_2$[N3d]
		& $ \chi^0_2 < \chi^\pm_1 < \chi^0_3 < g$
			&  	0.003	 \\
nuSP$_2$[N3e]
		& $ \chi^0_2 < \chi^\pm_1 < \chi^0_3 < t_1$
			&  	0.001	 \\
\hline
nuSP$_2$[N4a]
		& $ \chi^0_2 < \chi^\pm_1 < \tau_1 < H^0$
			&  	0.442	 \\
nuSP$_2$[N4b]
		& $ \chi^0_2 < \chi^\pm_1 < \tau_1 < \nu_\tau$
			&  	0.427	 \\
nuSP$_2$[N4c]
		& $ \chi^0_2 < \chi^\pm_1 < \tau_1 < t_1 $
			&  	0.090	 \\
nuSP$_2$[N4d]
		& $ \chi^0_2 < \chi^\pm_1 < \tau_1 < \chi^0_3 $
			&  	0.010	 \\
nuSP$_2$[N4e]
		& $ \chi^0_2 < \chi^\pm_1 < \tau_1 < \mu_r $
			&  	0.006	 \\
\hline
nuSP$_2$[N5a]
		& $ \chi^0_2 < \chi^\pm_1 < t_1 < g $
			&  	0.613	 \\
nuSP$_2$[N5b]
		& $ \chi^0_2 < \chi^\pm_1 < t_1 < \tau_1 $
			&  	0.248	 \\
nuSP$_2$[N5c]
		& $ \chi^0_2 < \chi^\pm_1 < t_1 < \chi^0_3$
			&  	0.067	 \\
nuSP$_2$[N5d]
		& $ \chi^0_2 < \chi^\pm_1 < t_1 < b_1 $
			&  	0.012	 \\
nuSP$_2$[N5e]
		& $ \chi^0_2 < \chi^\pm_1 < t_1 < H^0$
			&  	0.008	 \\
\hline
\hline
nuSP$_2$[$\tau$1a]
		& $ \tau_1 < \chi^0_2 < \chi^\pm_1 < \nu_\tau $
			&  	0.550	 \\
nuSP$_2$[$\tau$1b]
		& $ \tau_1 < \chi^0_2 < \chi^\pm_1 < H^0 $
			&  	0.345	 \\
nuSP$_2$[$\tau$1c]
		& $ \tau_1 < \chi^0_2 < \chi^\pm_1 < \mu_r$
			&  	0.091	 \\
nuSP$_2$[$\tau$1d]
		& $ \tau_1 < \chi^0_2 < \chi^\pm_1 < t_1$
			&  	0.021	 \\
\hline
nuSP$_2$[$\tau$2a]
		& $ \tau_1 < \chi^\pm_1 < \chi^0_2 <  \nu_\tau $
			&  	0.098	 \\
nuSP$_2$[$\tau$2b]
		& $ \tau_1 < \chi^\pm_1< \chi^0_2  < \mu_r $
			&  	0.014	 \\
nuSP$_2$[$\tau$2c]
		& $ \tau_1 < \chi^\pm_1< \chi^0_2  < H^0$
			&  	0.006	 \\
nuSP$_2$[$\tau$2d]
		& $ \tau_1 < \chi^\pm_1 < \chi^0_2 < t_1$
			&  	0.001	 \\
\hline
nuSP$_2$[$\tau$3]
		& $ \tau_1 < \chi^\pm_1 < \nu_\tau < \chi^0_2$
			&  	0.001	 \\
\hline
nuSP$_2$[$\tau$4a]
		& $ \tau_1 < \mu_r < e_r <  \chi^0_2 $
			&  	0.058	 \\
nuSP$_2$[$\tau$4b]
		& $ \tau_1 < \mu_r < e_r < \nu_\tau $
			&  	0.047	 \\
nuSP$_2$[$\tau$4c]
		& $ \tau_1 < \mu_r < e_r < t_1  $
			&  	0.001	 \\
\hline
nuSP$_2$[$\tau$5]
		& $ \tau_1 < H^0 < A^0 < H^\pm $
			&  	0.004	 \\
\hline
nuSP$_2$[$\tau$6]
		& $ \tau_1 <  \chi^0_2 < \mu_r < \chi^\pm_1$
			&  	0.001	 \\
\hline
\multicolumn{3}{c}{\vspace{-1 mm}}  \\
\end{tabular}
}
\caption{\label{nusugra_lightchargino_mass_patterns}\label{tab4}
Sparticle mass hierarchies for the nuSUGRA light chargino case (Model [2]). 
The high scale parameters lie in the range  $\mo \in [0.1,10] \TeV$, $M_1=M_3=\mhf \in [0.1,1.5] \TeV$, $M_2 = \alpha \mhf$, $\alpha \in [\frac{1}{2},1]$, $\frac{\az}{\mo} \in [-5,5]$, $\tan\beta \in [2,50]$, $\mu > 0$, with the constraints $\Omega h^2 < 0.12$, {$m_{h^0} > 120 \GeV$}.
}
\end{center}
\end{table}

\begin{table}[!]
\begin{center}
{\footnotesize
\begin{tabular}{|l|l|c|}

\hline
Pattern Label
		&  \multicolumn{1}{c|}{Mass Hierarchy}
			& \%	 \\
\hline
\hline
nuSP$_3$[C1a]
		& $ \chi^\pm_1 < \chi^0_2 < \chi^0_3 < \chi^0_4 $
			& 	63.273	 \\
nuSP$_3$[C1b]
		& $ \chi^\pm_1 < \chi^0_2 < \chi^0_3 <  g $
			& 	10.263	 \\
nuSP$_3$[C1c]
		& $ \chi^\pm_1 < \chi^0_2 < \chi^0_3 <  H^0$
			& 	4.587	 \\
nuSP$_3$[C1d]
		& $ \chi^\pm_1 < \chi^0_2 < \chi^0_3 < \tau_1 $
			& 	4.243	 \\
nuSP$_3$[C1e]
		& $ \chi^\pm_1 < \chi^0_2 < \chi^0_3 < t_1 $
			& 	4.549	 \\
nuSP$_3$[C1f]
		& $ \chi^\pm_1 < \chi^0_2 < \chi^0_3 < \chi^\pm_2 $
			& 	0.482	 \\
\hline
nuSP$_3$[C2a]
		& $ \chi^\pm_1 < \chi^0_2 < \tau_1 < \mu_r $
			& 0.854		 \\
nuSP$_3$[C2b]
		& $ \chi^\pm_1 < \chi^0_2 < \tau_1 <  \chi^0_3$
			& 	0.647	 \\
nuSP$_3$[C2c]
		& $ \chi^\pm_1 < \chi^0_2 < \tau_1 <  t_1$
			& 	0.372	 \\
nuSP$_3$[C2d]
		& $ \chi^\pm_1 < \chi^0_2 < \tau_1 <  H^0$
			& 	0.138	 \\
\hline
nuSP$_3$[C3a]
		& $ \chi^\pm_1 < \chi^0_2 < t_1 < \chi^0_3 $
			& 	0.840	 \\			
nuSP$_3$[C3b]
		& $ \chi^\pm_1 < \chi^0_2 < t_1 <  \tau_1 $
			& 	0.716	 \\	
nuSP$_3$[C3c]
		& $ \chi^\pm_1 < \chi^0_2 < t_1 <  H^0 $
			& 	0.055	 \\	
nuSP$_3$[C3d]
		& $ \chi^\pm_1 < \chi^0_2 < t_1 < g $
			& 	0.028	 \\	
\hline
nuSP$_3$[C4]
		& $ \chi^\pm_1 < \chi^0_2 < H^0 < A^0 $
			& 	0.882	 \\	
\hline
nuSP$_3$[C5]
		& $ \chi^\pm_1 < \chi^0_2 < g < \chi^0_3 $
			& 	0.523	 \\	
\hline
nuSP$_3$[C6a]
		& $ \chi^\pm_1 < \tau_1 < \chi^0_2 <  \mu_r $
			& 	0.096	 \\	
nuSP$_3$[C6b]
		& $ \chi^\pm_1 < \tau_1 < \chi^0_2 < \chi^0_3 $
			& 	0.096	 \\	
nuSP$_3$[C6c]
		& $ \chi^\pm_1 < \tau_1 < \chi^0_2 < H^0 $
			& 	0.014	 \\	
nuSP$_3$[C6d]
		& $ \chi^\pm_1 < \tau_1 < \chi^0_2 < t_1  $
			& 	0.014	 \\	
\hline
nuSP$_3$[C7]
		& $ \chi^\pm_1 < g < \chi^0_2 < \chi^0_3 $
			& 	0.055	 \\	
\hline
nuSP$_3$[C8]
		& $ \chi^\pm_1 < t_1 < \chi^0_2 < \chi^0_3 $
			& 	0.041	 \\	
\hline
\hline	
nuSP$_3$[N1a]
		& $ \chi^0_2 < \chi^\pm_1 < H^0 < A^0   $
			& 	0.978	 \\
nuSP$_3$[N1b]
		& $ \chi^0_2 < \chi^\pm_1  < A^0 < H^0   $
			& 	0.014	 \\
\hline
nuSP$_3$[N2]
		& $ \chi^0_2 < \chi^\pm_1  < g < \chi^0_3   $
			& 	0.193	 \\
\hline
nuSP$_3$[N3a]
		& $ \chi^0_2 < \chi^\pm_1  < \tau_1 < H^0   $
			& 	0.083	 \\
nuSP$_3$[N3b]
		& $ \chi^0_2 < \chi^\pm_1  < \tau_1 < \chi^0_3   $
			& 	0.028	 \\
\hline
nuSP$_3$[N4]
		& $ \chi^0_2 < \chi^\pm_1  < t_1 < H^0   $
			& 	0.041	 \\
\hline
nuSP$_3$[N5]
		& $ \chi^0_2 < \chi^\pm_1  < \chi^0_3 < H^0   $
			& 	0.014	 \\
\hline
\hline
nuSP$_3$[g1a]
		& $ g < \chi^0_2 < \chi^\pm_1 < \chi^0_3 $
			& 	3.940	 \\	
nuSP$_3$[g1b]
		& $ g < \chi^0_2 < \chi^\pm_1 < t_1 $
			& 	3.031	 \\	
nuSP$_3$[g1c]
		& $ g < \chi^0_2 < \chi^\pm_1 < H^0 $
			& 	0.138	 \\	
nuSP$_3$[g1d]
		& $ g < \chi^0_2 < \chi^\pm_1 < A^0 $
			& 	0.014	 \\	
nuSP$_3$[g1e]
		& $ g < \chi^0_2 < \chi^\pm_1 < H^\pm $
			& 0.014		 \\	
\hline
nuSP$_3$[g2a]
		& $ g < \chi^\pm_1 < \chi^0_2 < \chi^0_3  $
			& 1.529		 \\	
nuSP$_3$[g2b]
		& $ g < \chi^\pm_1 < \chi^0_2 < t_1  $
			& 0.028		 \\	
\hline
nuSP$_3$[g3]
		& $ g < t_1 < \chi^0_2 < \chi^\pm_1  $
			& 0.413	 \\	
\hline
nuSP$_3$[g4]
		& $ g < t_1 <  \chi^\pm_1 < \chi^0_2  $
			& 0.124		 \\	
\hline
nuSP$_3$[g5]
		& $ g < t_1 <  H^0 < A^0  $
			& 0.014		 \\	
\hline
nuSP$_3$[g6]
		& $ g < H^0 < A^0 < H^\pm  $
			& 0.069		 \\	
\hline
\hline
nuSP$_3$[H1]
		& $ H^0 < A^0 < \chi^0_2 < \chi^\pm_1  $
			& 	0.096	 \\
\hline
nuSP$_3$[H2]
		& $ H^0 < A^0 < \chi^0_2 < H^\pm $
			& 	0.014	 \\
\hline
nuSP$_3$[H3a]
		& $ H^0 < A^0 < H^\pm < \chi^\pm_1  $
			& 	0.069	 \\
nuSP$_3$[H3b]
		& $ H^0 < A^0 < H^\pm < \chi^0_2  $
			& 	0.055	 \\
nuSP$_3$[H3c]
		& $ H^0 < A^0 < H^\pm < t_1  $
			& 	0.014	 \\
\hline
\hline
nuSP$_3$[A1]
		& $ A^0 < H^0 < \chi^\pm_1 < \chi^0_2  $
			& 	0.014	 \\
\hline
\end{tabular}
\begin{tabular}{|l|l|c|}
\hline
Pattern Label
		&  \multicolumn{1}{c|}{Mass Hierarchy}
			& \%	 \\
\hline
\hline
nuSP$_3$[t1a]
		& $ t_1 < \chi^0_2 < \chi^\pm_1 < g  $
			& 	0.992	 \\
nuSP$_3$[t1b]
		& $ t_1 < \chi^0_2 < \chi^\pm_1 < \tau_!  $
			& 	0.427	 \\
nuSP$_3$[t1c]
		& $ t_1 < \chi^0_2 < \chi^\pm_1 < H^0  $
			& 	0.041	 \\
nuSP$_3$[t1d]
		& $ t_1 < \chi^0_2 < \chi^\pm_1 <  \chi^0_3 $
			& 	0.014	 \\
\hline
nuSP$_3$[t2a]
		& $ t_1 < \chi^\pm_1 < \chi^0_2 <  \chi^0_3 $
			& 	0.730	 \\
nuSP$_3$[t2b]
		& $ t_1 < \chi^\pm_1 < \chi^0_2 <  \tau_1$
			& 	0.069	 \\
nuSP$_3$[t2c]
		& $ t_1 < \chi^\pm_1 < \chi^0_2 <  g $
			& 	0.014	 \\
nuSP$_3$[t2d]
		& $ t_1 < \chi^\pm_1 < \chi^0_2 <  H^0$
			&   0.014		 \\
\hline
nuSP$_3$[t3]
		& $ t_1 < \tau_1 < \mu_r < e_r $
			&   0.386		 \\
\hline
nuSP$_3$[t4]
		& $ t_1 < g < \chi^0_2 < \chi^\pm_1 $
			&   0.344		 \\
\hline
nuSP$_3$[t5]
		& $ t_1 < \tau_1 < \chi^0_2 < \chi^\pm_1 $
			&   0.303		 \\
\hline
nuSP$_3$[t6a]
		& $ t_1 < \tau_1 <  \chi^\pm_1 < \chi^0_2 $
			&   0.165		 \\
nuSP$_3$[t6b]
		& $ t_1 < \tau_1 <  \chi^\pm_1 < \mu_r $
			&   0.014		 \\
\hline
nuSP$_3$[t7]
		& $ t_1 < g <  \chi^\pm_1 < \chi^0_3 $
			&   0.152		 \\
\hline
nuSP$_3$[t8]
		& $ t_1 < \tau_1 < H^0 < A^0 $
			&   0.041		 \\
\hline
nuSP$_3$[t9]
		& $ t_1 < \chi^\pm_1 < \tau_1 < \chi^0_2 $
			&   0.028		 \\
\hline
nuSP$_3$[t10]
		& $ t_1 < g < H^0 < A^0 $
			&   0.014		 \\
\hline
nuSP$_3$[t11a]
		& $ t_1 < H^0 < A^0 < H^\pm$
			&   0.014		 \\
nuSP$_3$[t11b]
		& $ t_1 <  H^0 < A^0 < \chi^0_2 $
			&   0.014		 \\
\hline
nuSP$_3$[t12]
		& $ t_1 <  \chi^\pm_1 < g < \chi^0_2 $
			&   0.014		 \\
\hline
\hline
nuSP$_3$[$\tau$1a]
		& $ \tau_1 < \mu_r < e_r < \chi^\pm_1   $
			& 	1.681	 \\
nuSP$_3$[$\tau$1b]
		& $ \tau_1 < \mu_r < e_r < t_1   $
			& 	0.234	 \\
nuSP$_3$[$\tau$1c]
		& $ \tau_1 < \mu_r < e_r <  \nu_\tau  $
			& 	0.207	 \\
nuSP$_3$[$\tau$1d]
		& $ \tau_1 < \mu_r < e_r <  \chi^0_2  $
			& 	0.207	 \\
\hline 
nuSP$_3$[$\tau$2a]
		& $ \tau_1 < \chi^\pm_1 < \chi^0_2 <  \chi^0_3  $
			& 	0.951	 \\
nuSP$_3$[$\tau$2b]
		& $ \tau_1 < \chi^\pm_1 < \chi^0_2 <  \mu_r  $
			& 	0.152	 \\
nuSP$_3$[$\tau$2c]
		& $ \tau_1 < \chi^\pm_1 < \chi^0_2 <  t_1  $
			& 	0.069	 \\
nuSP$_3$[$\tau$2d]
		& $ \tau_1 < \chi^\pm_1 < \chi^0_2 < H^0   $
			& 	0.028	 \\
\hline
nuSP$_3$[$\tau$3a]
		& $ \tau_1  < \chi^0_2< \chi^\pm_1 < H^0   $
			& 	0.661	 \\
nuSP$_3$[$\tau$3b]
		& $ \tau_1  < \chi^0_2< \chi^\pm_1 <  \chi^0_3  $
			& 	0.096	 \\
nuSP$_3$[$\tau$3c]
		& $ \tau_1  < \chi^0_2< \chi^\pm_1 < \mu_r   $
			& 	0.069	 \\
nuSP$_3$[$\tau$3d]
		& $ \tau_1  < \chi^0_2< \chi^\pm_1 <  \nu_\tau  $
			& 	0.028	 \\
\hline
nuSP$_3$[$\tau$4a]
		& $ \tau_1  < H^0 < A^0 < H^\pm  $
			& 	0.069	 \\
nuSP$_3$[$\tau$4b]
		& $ \tau_1  < H^0 < A^0 < \chi^0_2  $
			& 	0.014	 \\
\hline
nuSP$_3$[$\tau$5]
		& $ \tau_1  < t_1 < \chi^0_2 < \chi^\pm_1  $
			& 	0.124	 \\
\hline
nuSP$_3$[$\tau$6]
		& $ \tau_1  < t_1 < \mu_r< e_r  $
			& 	0.110	 \\
\hline
nuSP$_3$[$\tau$7]
		& $ \tau_1  < t_1 < \chi^\pm_1<  \chi^0_2  $
			& 	0.083	 \\
\hline
nuSP$_3$[$\tau$8]
		& $ \tau_1  < t_1 < H^0 < A^0 $
			& 	0.028	 \\
\hline
nuSP$_3$[$\tau$9]
		& $ \tau_1  <  \chi^\pm_1 < \mu_r < e_r  $
			& 	0.028	 \\
\hline
\multicolumn{3}{c}{\vspace{2.65cm}}  \\
\end{tabular}
}
\caption{\label{nusugra_lightgluino_mass_patterns}\label{tab5}
Sparticle mass hierarchies for the nuSUGRA light gluino case (Model [3]). 
The high scale parameters  lie in the range 
 $\mo \in [0.1,10] \TeV$, $M_1=M_2=\mhf \in [0.1,1.5] \TeV$, $M_3 = \alpha \mhf$, $\alpha \in [\frac{1}{6},1]$, $\frac{\az}{\mo} \in [-5,5]$, $\tan\beta \in [2,50]$, $\mu > 0$, with the constraints $\Omega h^2 < 0.12$, {$m_{h^0} > 120 \GeV$}.
}
\end{center}
\end{table}

\begin{table}[!]
\begin{center}
{\footnotesize
\begin{tabular}{|l|l|c|}
\hline
Pattern Label
		&  \multicolumn{1}{c|}{Mass Hierarchy}
			& \%	 \\
\hline
\hline
nuSP$_4$[C1a]
		& $ \chi^\pm_1  < \chi^0_2 <\chi^0_3 <\chi^0_4 $
			& 	55.065		 \\
nuSP$_4$[C1b]
		& $ \chi^\pm_1  < \chi^0_2 <\chi^0_3 < \tau_1 $
			& 		12.727	 \\
nuSP$_4$[C1c]
		& $ \chi^\pm_1  < \chi^0_2 <\chi^0_3 < A^0 $
			& 		6.205	 \\
nuSP$_4$[C1d]
		& $ \chi^\pm_1  < \chi^0_2 <\chi^0_3 < H^0 $
			& 		2.870	 \\
nuSP$_4$[C1e]
		& $ \chi^\pm_1  < \chi^0_2 <\chi^0_3 < t_1 $
			& 	0.886		 \\
nuSP$_4$[C1f]
		& $ \chi^\pm_1  < \chi^0_2 <\chi^0_3 <  \chi^\pm_2$
			& 	0.570		 \\
nuSP$_4$[C1g]
		& $ \chi^\pm_1  < \chi^0_2 <\chi^0_3 < H^\pm $
			& 	0.507		 \\
\hline
nuSP$_4$[C2a]
		& $ \chi^\pm_1  < \chi^0_2 < A^0 < H^0  $
			& 	1.224		 \\
nuSP$_4$[C2b]
		& $ \chi^\pm_1  < \chi^0_2 < H^0 < A^0  $
			& 	0.549		 \\
\hline
nuSP$_4$[C3a]
		& $ \chi^\pm_1  < \chi^0_2 < \tau_1 <  \chi^0_3 $
			& 		1.203	 \\
nuSP$_4$[C3b]
		& $ \chi^\pm_1  < \chi^0_2 < \tau_1 < H^0 $
			& 	0.042		 \\
nuSP$_4$[C3c]
		& $ \chi^\pm_1  < \chi^0_2 < \tau_1 < A^0 $
			& 	0.021		 \\
\hline
nuSP$_4$[C4]
		& $ \chi^\pm_1  < \chi^0_2 < t_1 < \chi^0_3 $
			& 	0.169		 \\
\hline
nuSP$_4$[C5]
		& $ \chi^\pm_1 < A^0 < H^0 < \chi^0_2  $
			& 	0.084		 \\
\hline
nuSP$_4$[C6]
		& $ \chi^\pm_1  < \chi^0_2 < g < \chi^0_3 $
			& 	0.063		 \\
\hline
nuSP$_4$[C7]
		& $ \chi^\pm_1  < \tau_1 < \chi^0_2 < \chi^0_3 <  $
			& 	0.021		 \\
\hline
nuSP$_4$[C8]
		& $ \chi^\pm_1  <  \chi^0_2 < H^\pm < A^0 $
			& 	0.021		 \\
\hline
\hline
nuSP$_4$[$\tau$1a]
		& $ \tau_1 < \chi^0_2 < \chi^\pm_1 < H^0 $
			&  2.617 \\
nuSP$_4$[$\tau$1b]
		& $ \tau_1 < \chi^0_2 < \chi^\pm_1 < \mu_r  $
			& 0.971  \\
nuSP$_4$[$\tau$1c]
		& $ \tau_1 < \chi^0_2 < \chi^\pm_1 < \nu_\tau $
			& 0.802  \\
nuSP$_4$[$\tau$1d]
		& $ \tau_1 < \chi^0_2 < \chi^\pm_1 < t_1 $
			&  0.106 \\
nuSP$_4$[$\tau$1e]
		& $ \tau_1 < \chi^0_2 < \chi^\pm_1 < \chi^0_3 $
			& 0.021  \\
\hline
nuSP$_4$[$\tau$2a]
		& $ \tau_1 < \mu_r < e_r < \chi^0_2  $
			& 2.849  \\
nuSP$_4$[$\tau$2b]
		& $ \tau_1 < \mu_r < e_r <  \nu_\tau $
			&  2.153 \\
nuSP$_4$[$\tau$2c]
		& $ \tau_1 < \mu_r < e_r < \chi^\pm_1  $
			& 1.224  \\
nuSP$_4$[$\tau$2d]
		& $ \tau_1 < \mu_r < e_r <  H^0 $
			&  0.021 \\
\hline
nuSP$_4$[$\tau$3a]
		& $ \tau_1 < \chi^\pm_1 < \chi^0_2 < \chi^0_3 $
			&  1.625 \\
nuSP$_4$[$\tau$3b]
		& $ \tau_1 < \chi^\pm_1 < \chi^0_2 < \mu_r $
			&  0.295 \\
nuSP$_4$[$\tau$3c]
		& $ \tau_1 < \chi^\pm_1 < \chi^0_2 < H^0 $
			& 0.084  \\
\hline
nuSP$_4$[$\tau$4a]
		& $ \tau_1 < H^0 < A^0 < H^\pm $
			& 0.148  \\
nuSP$_4$[$\tau$4b]
		& $ \tau_1 < H^0 < A^0 < \chi^0_2 $
			& 0.063  \\
\hline
nuSP$_4$[$\tau$5]
		& $ \tau_1 < t_1 < \chi^0_2 < \chi^\pm_1$
			& 0.021  \\
\hline
nuSP$_4$[$\tau$6]
		& $ \tau_1 < \mu_r < \chi^0_2 < e_r $
			& 0.021  \\
\hline
\end{tabular}
\begin{tabular}{|l|l|c|}
\hline
Pattern Label
		&  \multicolumn{1}{c|}{Mass Hierarchy}
			& \%	 \\
\hline
\hline
nuSP$_4$[N1a]
		& $ \chi^0_2 < \chi^\pm_1 < H^0 < A^0$
			& 2.216  \\
nuSP$_4$[N1b]
		& $ \chi^0_2 < \chi^\pm_1 < A^0 < H^0 $
			& 0.443  \\
\hline
nuSP$_4$[N2a]
		& $ \chi^0_2 < \chi^\pm_1 < \tau_1 < H^0$
			& 0.401  \\
nuSP$_4$[N2b]
		& $ \chi^0_2 < \chi^\pm_1 < \tau_1 < A^0$
			& 0.042  \\
nuSP$_4$[N2c]
		& $ \chi^0_2 < \chi^\pm_1 < \tau_1 < \chi^0_3 $
			& 0.021  \\
\hline
nuSP$_4$[N3]
		& $ \chi^0_2 < \chi^\pm_1 < g < \chi^0_3 $
			& 0.106  \\
\hline
nuSP$_4$[N4a]
		& $ \chi^0_2 < \chi^\pm_1 <  \chi^0_3 <  \chi^0_4 $
			& 0.0 63 \\
nuSP$_4$[N4b]
		& $ \chi^0_2 < \chi^\pm_1 <  \chi^0_3 < \tau_1 $
			& 0.042  \\
nuSP$_4$[N4c]
		& $ \chi^0_2 < \chi^\pm_1 <  \chi^0_3 < A^0 $
			& 0.042  \\
nuSP$_4$[N4d]
		& $ \chi^0_2 < \chi^\pm_1 <  \chi^0_3 < H^\pm $
			& 0.021  \\
\hline
nuSP$_4$[N5]
		& $ \chi^0_2 < \chi^\pm_1 <  t_1 < g $
			& 0.021  \\
\hline
\hline
nuSP$_4$[t1a]
		& $ t_1 < \chi^0_2 < \chi^\pm_1 <  g $
			&  0.359 \\
nuSP$_4$[t1b]
		& $ t_1 < \chi^0_2 < \chi^\pm_1 < \chi^0_3 $
			& 0.317 \\
nuSP$_4$[t1c]
		& $ t_1 < \chi^0_2 < \chi^\pm_1 < \tau_1  $
			& 0.127 \\
nuSP$_4$[t1d]
		& $ t_1 < \chi^0_2 < \chi^\pm_1 < H^0 $
			& 0.042 \\
nuSP$_4$[t1e]
		& $ t_1 < \chi^0_2 < \chi^\pm_1 < b_1  $
			& 0.021  \\
\hline
nuSP$_4$[t2a]
		& $ t_1 < \chi^\pm_1 < \chi^0_2 <  \chi^0_3  $
			& 0.106  \\
nuSP$_4$[t2b]
		& $ t_1 < \chi^\pm_1 < \chi^0_2 <  H^0  $
			& 0.021  \\
\hline
nuSP$_4$[t3a]
		& $ t_1 < H^0 < A^0 < H^\pm  $
			& 0.021  \\
nuSP$_4$[t3a]
		& $ t_1 < H^0 < A^0 < \chi^\pm_1  $
			& 0.021  \\
\hline
\hline
nuSP$_4$[H1a]
		& $ H^0 < A^0 < H^\pm < \chi^\pm_1 $
			& 	0.127	 \\
nuSP$_4$[H1b]
		& $ H^0 < A^0 < H^\pm < \chi^0_2 $
			& 	0.021	 \\
\hline
nuSP$_4$[H2a]
		& $ H^0 < A^0 < \chi^0_2 < \chi^\pm_1 $
			& 	0.084	 \\
nuSP$_4$[H2b]
		& $ A^0 < H^0 < \chi^0_2 < \chi^\pm_1 $
			& 	0.021	 \\
\hline
nuSP$_4$[H3]
		& $ A^0 < H^0 <  \chi^\pm_1  < H^\pm$
			& 	0.021	 \\
\hline
nuSP$_4$[H4]
		& $ H^0 < \chi^0_2 < \chi^\pm_1 < H^\pm $
			& 	0.021	 \\
\hline
\multicolumn{3}{c}{\vspace{2.25 cm}}  \\
\end{tabular}
}
\caption{
	\label{nusugra_nuhiggs_mass_patterns}\label{tab6}
	Sparticle mass hierarchies for the nuSUGRA nonuniversal Higgs case (Model [4]).
	The high scale parameters 
lie in the range 	
 $\mo \in [0.1,10] \TeV$, $\mhf \in [0.1,1.5] \TeV$, $\frac{\az}{\mo} \in [-5,5]$, $\tan\beta \in [2,50]$, $\mu > 0$, with the constraints $\Omega h^2 < 0.12$, {$m_{h^0} > 120 \GeV$}. However, the Higgs masses at the GUT scale are nonuniversal, $m_{H_i}(\textrm{M}_G)= \mo(1+\delta_i)$, $i=1,2$ where $\delta_i \in [-0.9,1]$.
}
\end{center}
\end{table}

\begin{table}[!]
\begin{center}
{\footnotesize
\begin{tabular}{|l|l|c|}

\hline
Pattern Label
		&  \multicolumn{1}{c|}{Mass Hierarchy}
			& \%	 \\
\hline
\hline
nuSP$_5$[C1a]
		& $\chi^{\pm}_{1} < \chi^{0}_{2} < \chi^{0}_{3}< \tau_1$
			&  39.4 \\
nuSP$_5$[C1b]
		& $\chi^{\pm}_{1} < \chi^{0}_{2} < \chi^{0}_{3}< \chi^{0}_{4}$
			&  12.5 \\
nuSP$_5$[C1c]
		& $\chi^{\pm}_{1} <  \chi^{0}_{2}  < \chi^{0}_{3} < t_1 $
			&  	1.39	 \\
nuSP$_5$[C1d]
		& $\chi^{\pm}_{1} <  \chi^{0}_{2}  < \chi^{0}_{3} < H^0 $
			&  	0.37	 \\
\hline
nuSP$_5$[C2a]
		& $\chi^{\pm}_{1} <  \chi^{0}_{2}  < \tau_1 < \chi^{0}_{3} $
			&  	17.95	 \\
nuSP$_5$[C2b]
		& $\chi^{\pm}_{1} <  \chi^{0}_{2}  < \tau_1 <  H^0 $
			&  	0.04	 \\
nuSP$_5$[C2c]
		& $\chi^{\pm}_{1} <  \chi^{0}_{2}  < \tau_1 < t_1 $
			&  	0.04	 \\
\hline
nuSP$_5$[C3]
		& $\chi^{\pm}_{1} <  \tau_1  < \chi^{0}_{2} < \chi^{0}_{3} $
			&  	0.49	 \\
\hline
nuSP$_5$[C4]
		& $\chi^{\pm}_{1} <  \chi^{0}_{2}  < H^0 < A^0 $
			&  	0.22	 \\
\hline
nuSP$_5$[C5a]
		& $\chi^{\pm}_{1} <  \chi^{0}_{2}  <  t_1  < \chi^{0}_{3} $
			&  	0.07	 \\
nuSP$_5$[C5b]
		& $\chi^{\pm}_{1} <  \chi^{0}_{2}  <  t_1  < \tau_1 $
			&  	0.04	 \\
\hline
\hline
nuSP$_5$[e1]
		& $e_r < \mu_r < \nu_e < \nu_\mu $
			&  	0.19	 \\
\hline
\hline
nuSP$_5$[H1]
		& $H^0 < A^0 < \chi^{0}_{2} < \chi^{\pm}_{1} $
			&  	0.07	 \\
\hline
\hline
nuSP$_5$[$\mu$1]
		& $ \mu_r < e_r < \nu_\mu <\nu_e $
			&  	3.37	 \\
\hline
nuSP$_5$[$\mu$2a]
		& $ \mu_r < e_r < \tau_1 < \nu_\mu $
			&  2.81		 \\
nuSP$_5$[$\mu$2b]
		& $ \mu_r < e_r < \tau_1 < \chi^{0}_{2} $
			&  	0.56	 \\
nuSP$_5$[$\mu$3]
		& $ \mu_r < e_r < \chi^{0}_{2} <  \chi^{\pm}_{1}$
			&  0.71		 \\
\hline
nuSP$_5$[$\mu$4]
		& $ \mu_r < e_r < \nu_e< \nu_\mu $
			&  0.04		 \\
\hline
nuSP$_5$[$\mu$5]
		& $ \mu_r < \tau_1 < e_r < \nu_\mu $
			&  	0.04	 \\
\hline
\hline
nuSP$_5$[N1]
		& $ \chi^{0}_{2} < \chi^{\pm}_{1} < H^0 < A^0 $
			&  0.90		 \\
\hline
nuSP$_5$[N2]
		& $ \chi^{0}_{2} < \chi^{\pm}_{1} < \tau_1  < H^0 $
			&  0.49		 \\
\hline
nuSP$_5$[N3]
		& $ \chi^{0}_{2} < \chi^{\pm}_{1} < \mu_r < e_r $
			&  0.04		 \\
\hline
\end{tabular}
\begin{tabular}{|l|l|c|}
\hline
Pattern Label
		&  \multicolumn{1}{c|}{Mass Hierarchy}
			& \%	 \\
\hline
\hline
nuSP$_5$[$\tau$1a]
		& $ \tau_1 < \chi^{\pm}_{1} < \chi^{0}_{2} < \chi^{0}_{3} $
			&  3.04		 \\
nuSP$_5$[$\tau$1b]
		& $ \tau_1 < \chi^{\pm}_{1} < \chi^{0}_{2} < \nu_\tau $
			&  	0.07	 \\
nuSP$_5$[$\tau$1c]
		& $ \tau_1 < \chi^{\pm}_{1} < \chi^{0}_{2} < H^0$
			&  	0.04	 \\
\hline
nuSP$_5$[$\tau$2a]
		& $ \tau_1 < \mu_r < e_r < \chi^{0}_{2} $
			&  4.39		 \\
nuSP$_5$[$\tau$2b]
		& $ \tau_1 < \mu_r < e_r < \nu_\mu $
			&  3.90		 \\
nuSP$_5$[$\tau$2c]
		& $ \tau_1 < \mu_r < e_r < \nu_\tau $
			&  	0.49	 \\
nuSP$_5$[$\tau$2d]
		& $ \tau_1 < \mu_r < e_r < t_1 $
			&  0.04		 \\
\hline
nuSP$_5$[$\tau$3a]
		& $ \tau_1 < \chi^{0}_{2} < \chi^{\pm}_{1} < H^0 $
			& 3.07 		 \\
nuSP$_5$[$\tau$3b]
		& $ \tau_1 < \chi^{0}_{2} < \chi^{\pm}_{1} < \mu_r $
			&  	1.12	 \\
nuSP$_5$[$\tau$3c]
		& $ \tau_1 < \chi^{0}_{2} < \chi^{\pm}_{1} < \nu_\tau $
			&  	1.05	 \\
nuSP$_5$[$\tau$3d]
		& $ \tau_1 < \chi^{0}_{2} < \chi^{\pm}_{1} <  \chi^{0}_{3}$
			&  	0.19	 \\
nuSP$_5$[$\tau$3e]
		& $ \tau_1 < \chi^{0}_{2} < \chi^{\pm}_{1} < t_1$
			&  	0.15	 \\
\hline
nuSP$_5$[$\tau$4a]
		& $ \tau_1 < \nu_\tau < \tau_2 < \chi^{0}_{2} $
			&  0.19		 \\
nuSP$_5$[$\tau$4b]
		& $ \tau_1 < \nu_\tau < \tau_2 < \chi^{\pm}_{1} $
			&  	0.04	 \\
\hline
nuSP$_5$[$\tau$5]
		& $ \tau_1 < \nu_\tau < \chi^{0}_{2} < \chi^{\pm}_{1} $
			&  	0.07	 \\
\hline
nuSP$_5$[$\tau$6]
		& $ \tau_1 < H^0 < A^0 < H^{\pm} $
			&  	0.07	 \\
\hline
nuSP$_5$[$\tau$7]
		& $ \tau_1 < t_1 < \nu_\tau < \tau_2 $
			&  	0.04	 \\
\hline
\hline
nuSP$_5$[t1]
		& $ t_1 < \chi^{\pm}_{1} < \chi^{0}_{2} < \chi^{0}_{3}  $
			&  	0.19	 \\
\hline
nuSP$_5$[t2]
		& $ t_1 < \tau_1 < \chi^{0}_{2} < \chi^{\pm}_{1} $
			&  	0.11	 \\
\hline
nuSP$_5$[t3]
		& $ t_1 < \tau_1 < < \tau_2 $
			&  	0.04	 \\
\hline
nuSP$_5$[t4]
		& $ t_1 < \tau_1 < \chi^{\pm}_{1} < \chi^{0}_{2} $
			&  	0.04	 \\
\hline

\multicolumn{3}{c}{\vspace{2.8 mm}}  \\
\end{tabular}
}
\caption{\label{nusugra_light3rdgen_mass_patterns}\label{tab7}
Sparticle mass hierarchies for the nuSUGRA light third generation case {(Model [5])}. 
The high scale parameters lie in the range $\mo^{(1)}=\mo^{(2)}=\mo \in [0.1,10] \TeV$, {$\mo^{(3)}=\frac{\mo}{1 \TeV+\mo}$}, $\mhf \in [0.1,1.5] \TeV$, $\frac{\az}{\mo} \in [-5,5]$, $\tan\beta \in [2,50]$, $\mu > 0$, with the constraints $\Omega h^2 < 0.12$, {$m_{h^0} > 120 \GeV$}.
}
\end{center}
\end{table}

\begin{table}[t!]
\begin{center}
{\tiny
\begin{tabular}{|l||c|c|c|c||c|c|c|c||c|c|}
\hline
\multicolumn{1}{|c||}{SUGRA}
		& $\mo$
			&  $\mhf$
				& $\az$
					&  $\tb$		
						& \multirow{2}{*}{$\delta_{M_2}$}
							& \multirow{2}{*}{$\delta_{M_3}$}
							 	& \multirow{2}{*}{($\delta_{H_1}$, $\delta_{H_2}$)} 
							 		& 	\multirow{2}{*}{$\delta_{M_{q_3}}$}	
							 			&  $\mu$
							 				&   $m_{h^0}$				\\
\multicolumn{1}{|c||}{Hierarchy}
		& ($\GeV$)
			& ($\GeV$) 
				& ($\GeV$)
					&  ($\frac{v_2}{v_1}$)
						&  
							&
								& 
									&
										&  ($\GeV$) 
											&	 ($\GeV$) 		\\
\hline
mSP[C1a]
		& 	6183
			&  470
				&  −4469
					&  52 
						& -
							&	-
								&  -
									& -	 
								 		& 269	
								 			& 	126.1	\\
mSP[C1a]
		& 	3715
			&  1080
				&  706
					&  52
						&  -
							&	-
								&  -
									& -	 
								 		&	569
								 			& 		123.1	\\
nuSP$_2$[C1a]
		& 	5464
			&  1049
				&  4845
					&  52
						&   -0.063
							&	-
								& -
									& - 	 
								 		& 	583
								 			&	124.0	\\							 					
nuSP$_2$[C2a]
		& 	2005
			&  1234
				&  -3105
					&  32
						&   -0.446
							&	-
								& -
									& - 	 
								 		& 1999
								 			&	125.4		\\
nuSP$_3$[C1a]
		& 	5446
			&  500
				&  -3940
					&  24
						&   -
							&	-0.524
								& -
									& -	 
								 		&	285	
								 			&	125.3	\\
nuSP$_3$[$g$1a]
		& 	5019
			&  846
				&  7759
					&  15
						&   -
							&	-0.819
								& -
									& 	-
								 		&	2066
								 			&	123.4	\\
nuSP$_4$[C1a]
		& 	1210
			&  848
				&  -1656
					&  26
						&   -
							&	-
								& (-0.830, -1.205)
									&  -
								 		&	571
								 			&	123.5	\\
nuSP$_4$[$\tau$2a]
		& 	591
			&  901
				&  -1746
					&  31
						&  -
							&	-
								& (-2.089,	-6.704)
									&  -
								 		& 1419 
								 			&	123.6	\\
nuSP$_5$[C1a]
		& 	2007
			&  1155
				&  -989
					&  48
						&   -
							&	-
								& -
									& -0.361 	 
								 		&	589
								 			&	123.3	\\
nuSP$_5$[C2a]
		& 	2301
			&  1241
				&  -2185
					&  31
						&   -
							&	-
								& -
									& -0.584	 
								 		&	541
								 			& 126.0		\\
								 								 											 		
\hline
\end{tabular}
}
\caption{\label{benchmarks_HEparams}
Benchmarks are given for SUGRA Models [1]-[5]. The sparticle mass hierarchy is specified for each benchmark. These particular model points are chosen due to having a mass pattern which has an especially large percentage of occurrence and also having passed collider, flavor and cosmological constraints. Further, they satisfy $(m_{\textrm{NLSP}}+m_{\textrm{NNLSP}})/2 < 600 \GeV$ and have a maximum NLSP-LSP mass gap. The nonuniversalities are defined as the following: $M_2=\mhf(1+\delta_{M_2})$ for Model [2], $M_3=\mhf(1+\delta_{M_3})$ for Model [3], $m_{H_i}^2=\mo^2(1+\delta_{H_i})$, where $i=1,2$ for Model [4] and $m_{q_3} =\mo(1+\delta_{m_{q_3}})$ for Model [5]. 
}
\end{center}
\end{table}

 \begin{table}[b!]
 \renewcommand{\arraystretch}{1.3}
 \begin{center}
 \hspace*{-1.35cm}
{\tiny
 \begin{tabular}{|l||c|c|c|c||c|c|c|c||c|c|c|}
 \hline
  & & & & & & & & & & & \\
 \multicolumn{1}{|c||}{\multirow{-2}{*}{SUGRA}}
 		& \multirow{-2}{*}{$\mo$}
 			&  \multirow{-2}{*}{$\mhf$}
 				& \multirow{-2}{*}{$\az$}
 					&  \multirow{-2}{*}{$\tb$}
 						& \multirow{1}{*}{$\delta_{M_2}$}
 							& \multirow{1}{*}{$\delta_{M_3}$}
 							 	& \multirow{1}{*}{($\delta_{H_1}$, $\delta_{H_2}$)} 
 							 		& \multirow{1}{*}{$\delta_{M_{q_3}}$}	
 							 			&  \multirow{-2}{*}{$R\times \sigma^\text{SI}_{p,\na}$}
 							 				&   \multirow{-2}{*}{LSP Bino}			
 							 					&   \multirow{-2}{*}{LSP Higgsino}				\\
 \multicolumn{1}{|c||}{\multirow{-2}{*}{Hierarchy}}
 		& \multirow{-2}{*}{($\GeV$)}
 			& \multirow{-2}{*}{($\GeV$)}
 				& \multirow{-2}{*}{($\GeV$)}
 					&  \multirow{-2}{*}{($\frac{v_2}{v_1}$)}
 						&  
 							&
 								& 
 									&
 										&  \multirow{-1}{*}{$\left(\textrm{cm}^2\right)$ }
 											&	 \multirow{-2}{*}{Fraction}	
 												&	\multirow{-2}{*}{Fraction}		\\
 \hline
 mSP[$H$1b]
 		& 	2221 		
 			&  424
 				&  -3862
 					&  56
 						& -
 							&	-
 								&  -
 									& -
 								 		& 3.00 $\times 10^{-45}$ 
 								 			& 	0.9997
 								 				&	0.0213	\\
 mSP[$H$2]
 		& 	3762		
 			&  416
 				&  -6727
 					&  54
 						& -
 							&	-
 								&  -
 									& -
 								 		& 7.76$\times 10^{-46}$ 
 								 			& 	0.9994
 								 				&	0.0331			\\
 mSP[$t$1a]
 		& 	3075			
 			&  1706
 				&  13772
 					&  18
 						& -
 							&	-
 								&  -
 									& - 
 								 		& 6.33$\times 10^{-49}$
 								 			& 	0.9999	
 								 				&	0.0122	\\	
 mSP[$t$1b]
 		& 	4032 		
 			&  1808
 				&  15522
 					&  12
 						& -
 							&	-
 								&  -
 									& -
 								 		& 1.17$\times 10^{-48}$
 								 			&  	0.9999	
 								 				&	0.0108	\\	
 nuSP$_2$[$N$1a]
 		& 	9874
 			& 455
 				&  -13049
 					&  25
 						&  -0.480
 							&	-
 								&  -
 									& -
 								 		& 4.70$\times 10^{-49}$
 								 			&  	0.9998
 								 				&		0.0132		\\	
 nuSP$_2$[$N$1b]
 		& 	4334
 			& 535
 				&  -7758
 					&  10
 						& -0.458
 							&	-
 								&  -
 									& -
 								 		& 5.01$\times 10^{-48}$
 								 			&  	 0.9996
 								 				&   0.0171	\\	
 nuSP$_3$[$g$1a]
 		& 	9072
 			& 1963
 				&  -11390
 					&  27
 						& -
 							&	-0.821
 								&  -
 									& -
 								 		& 5.01$\times 10^{-48}$
 								 			&   0.9998
 								 				&	0.0184			\\	
 nuSP$_3$[$g$1b]
 		& 	6695
 			& 1959
 				&  -10789
 					&  31
 						& -
 							&	-0.832
 								&  -
 									& -
 								 		& 4.98$\times 10^{-49}$
 								 			&  	0.9998
 								 				&	0.0156				\\	
 nuSP$_4$[$t$1a]
 		& 	6083
 			& 1665
 				&  17544
 					&  12
 						& -
 							&	-
 								&  (1.032, -1.692)
 									& -
 								 		& 9.93$\times 10^{-49}$
 								 			&  	0.9999	
 								 				&	0.0100			\\	
 nuSP$_4$[$\tau$2a]
 		& 	3204
 			& 1585
 				&  -7123
 					&  53
 						& -
 							&	-
 								&  (-2.182, -2.439)
 									& -
 								 		& 3.34$\times 10^{-48}$
 								 			&  	 	0.9999	
 								 				&	0.0125			\\	
 nuSP$_5$[$\tau$1b]
 		& 	465
 			& 1077
 				&  -3457
 					&  46
 						& -
 							&	-
 								&  -
 									& 2.055
 								 		& 6.56$\times 10^{-48}$
 								 			&  	0.9997	
 								 				&	0.0217	\\
 nuSP$_5$[$\tau$2b]
 		& 	504
 			& 601
 				&  -2630
 					&  35
 						& -
 							&	-
 								&  -
 									& 0.611
 								 		& 9.37$\times 10^{-48}$
 								 			&  	0.9994
 								 				&	0.0334			\\

 \hline
 \end{tabular}
 }
 \caption{\label{sugra_nmix}
 {A sample of mSUGRA and nuSUGRA model} points with especially low spin-independent neutralino-proton cross section are given, accompanied by the sparticle mass pattern to which they belong. As displayed, the gaugino-Higgsino content of the LSP is almost entirely bino for these parameter points. Since these points exhibit a nearly 100\%  bino-like LSP, this gives reason for the smallness of the observed cross sections. The Higgsino fraction given in the 
 last column is defined as $\sqrt{n_{13}^2+ n_{14}^2}$. 
 }
 \end{center}
 \end{table}

\clearpage
\bibliography{master}

\providecommand{\href}[2]{#2}\begingroup\raggedright\begin{thebibliography}{10}

\bibitem{Chatrchyan:2012ufa}
{\bf CMS Collaboration} , S.~Chatrchyan et~al., {\it {Observation of a new
  boson at a mass of 125 GeV with the CMS experiment at the LHC}},  {\em Phys.
  Lett. B} {\bf 716} (2012) 30--61, [\href{http://arxiv.org/abs/1207.7235}{{\tt
  arXiv:1207.7235}}].

\bibitem{Aad:2012tfa}
{\bf ATLAS Collaboration} , G.~Aad et~al., {\it {Observation of a new particle
  in the search for the Standard Model Higgs boson with the ATLAS detector at
  the LHC}},  {\em Phys. Lett. B} {\bf 716} (2012) 1--29,
  [\href{http://arxiv.org/abs/1207.7214}{{\tt arXiv:1207.7214}}].

\bibitem{Englert:1964et}
F.~Englert and R.~Brout, {\it {Broken Symmetry and the Mass of Gauge Vector
  Mesons}},  {\em Phys. Rev. Lett.} {\bf 13} (1964) 321--323.

\bibitem{Higgs:1964pj}
P.~W. Higgs, {\it {Broken Symmetries and the Masses of Gauge Bosons}},  {\em
  Phys. Rev. Lett.} {\bf 13} (1964) 508--509.

\bibitem{Guralnik:1964eu}
G.~Guralnik, C.~Hagen, and T.~Kibble, {\it {Global Conservation Laws and
  Massless Particles}},  {\em Phys. Rev. Lett.} {\bf 13} (1964) 585--587.

\bibitem{Akula:2011aa}
S.~Akula, B.~Altunkaynak, D.~Feldman, P.~Nath, and G.~Peim, {\it {Higgs Boson
  Mass Predictions in SUGRA Unification, Recent LHC-7 Results, and Dark
  Matter}},  {\em Phys. Rev. D} {\bf 85} (2012) 075001,
  [\href{http://arxiv.org/abs/1112.3645}{{\tt arXiv:1112.3645}}].

\bibitem{Baer:2011ab}
H.~Baer, V.~Barger, and A.~Mustafayev, {\it {Implications of a 125 GeV Higgs
  scalar for LHC SUSY and neutralino dark matter searches}},  {\em Phys. Rev.
  D} {\bf 85} (2012) 075010, [\href{http://arxiv.org/abs/1112.3017}{{\tt
  arXiv:1112.3017}}].

\bibitem{Arbey:2011ab}
A.~Arbey, M.~Battaglia, A.~Djouadi, F.~Mahmoudi, and J.~Quevillon, {\it
  {Implications of a 125 GeV Higgs for supersymmetric models}},  {\em Phys.
  Lett. B} {\bf 708} (2012) 162--169,
  [\href{http://arxiv.org/abs/1112.3028}{{\tt arXiv:1112.3028}}].

\bibitem{Draper:2011aa}
P.~Draper, P.~Meade, M.~Reece, and D.~Shih, {\it {Implications of a 125 GeV
  Higgs for the MSSM and Low-Scale SUSY Breaking}},  {\em Phys. Rev. D} {\bf
  85} (2012) 095007, [\href{http://arxiv.org/abs/1112.3068}{{\tt
  arXiv:1112.3068}}].

\bibitem{Carena:2011aa}
M.~Carena, S.~Gori, N.~R. Shah, and C.~E. Wagner, {\it {A 125 GeV SM-like Higgs
  in the MSSM and the $\gamma \gamma$ rate}},  {\em JHEP} {\bf 1203} (2012)
  014, [\href{http://arxiv.org/abs/1112.3336}{{\tt arXiv:1112.3336}}].

\bibitem{Akula:2012kk}
S.~Akula, P.~Nath, and G.~Peim, {\it {Implications of the Higgs Boson Discovery
  for mSUGRA}},  {\em Phys. Lett. B} {\bf 717} (2012) 188--192,
  [\href{http://arxiv.org/abs/1207.1839}{{\tt arXiv:1207.1839}}].

\bibitem{Arbey:2012dq}
A.~Arbey, M.~Battaglia, A.~Djouadi, and F.~Mahmoudi, {\it {The Higgs sector of
  the phenomenological MSSM in the light of the Higgs boson discovery}},  {\em
  JHEP} {\bf 1209} (2012) 107, [\href{http://arxiv.org/abs/1207.1348}{{\tt
  arXiv:1207.1348}}].

\bibitem{Strege:2012bt}
C.~Strege, G.~Bertone, F.~Feroz, M.~Fornasa, R.~Ruiz~de Austri, et~al., {\it
  {Global Fits of the cMSSM and NUHM including the LHC Higgs discovery and new
  XENON100 constraints}},  {\em JCAP} {\bf 1304} (2013) 013,
  [\href{http://arxiv.org/abs/1212.2636}{{\tt arXiv:1212.2636}}].

\bibitem{Chamseddine:1982jx}
A.~H. Chamseddine, R.~L. Arnowitt, and P.~Nath, {\it {Locally Supersymmetric
  Grand Unification}},  {\em Phys. Rev. Lett.} {\bf 49} (1982) 970.

\bibitem{Nath:1983aw}
P.~Nath, R.~L. Arnowitt, and A.~H. Chamseddine, {\it {Gauge Hierarchy in
  Supergravity Guts}},  {\em Nucl. Phys. B} {\bf 227} (1983) 121.

\bibitem{Hall:1983iz}
L.~J. Hall, J.~D. Lykken, and S.~Weinberg, {\it {Supergravity as the Messenger
  of Supersymmetry Breaking}},  {\em Phys. Rev. D} {\bf 27} (1983) 2359--2378.

\bibitem{Arnowitt:1992aq}
R.~L. Arnowitt and P.~Nath, {\it {SUSY mass spectrum in SU(5) supergravity
  grand unification}},  {\em Phys. Rev. Lett.} {\bf 69} (1992) 725--728.

\bibitem{Ibanez:2007pf}
L.~Ibanez and G.~Ross, {\it {Supersymmetric Higgs and radiative electroweak
  breaking}},  {\em Comptes Rendus Physique} {\bf 8} (2007) 1013--1028,
  [\href{http://arxiv.org/abs/hep-ph/0702046}{{\tt hep-ph/0702046}}].

\bibitem{Buchmueller:2011ab}
O.~Buchmueller, R.~Cavanaugh, A.~De~Roeck, M.~Dolan, J.~R. Ellis, H.~Flacher,
  S.~Heinemeyer, G.~Isidori, J.~Marrouche, D.~Martinez~Santos, K.~A. Olive,
  S.~Rogerson, F.~J. Ronga, K.~J.~d. Vries, and G.~Weiglein, {\it {Higgs and
  Supersymmetry}},  {\em Eur. Phys. J. C} {\bf 72} (2012) 2020,
  [\href{http://arxiv.org/abs/1112.3564}{{\tt arXiv:1112.3564}}].

\bibitem{Baer:2012mv}
H.~Baer, V.~Barger, P.~Huang, D.~Mickelson, A.~Mustafayev, et~al., {\it
  {Post-LHC7 fine-tuning in the mSUGRA/CMSSM model with a 125 GeV Higgs
  boson}},  {\em Phys. Rev. D} {\bf 87} (2013), no.~3 035017,
  [\href{http://arxiv.org/abs/1210.3019}{{\tt arXiv:1210.3019}}].

\bibitem{Akula:2013ioa}
S.~Akula and P.~Nath, {\it {Gluino-driven radiative breaking, Higgs boson mass,
  muon g-2, and the Higgs diphoton decay in supergravity unification}},  {\em
  Phys. Rev. D} {\bf 87} (2013), no.~11 115022,
  [\href{http://arxiv.org/abs/1304.5526}{{\tt arXiv:1304.5526}}].

\bibitem{SMSnowmass}
A.~Avetisyan, J.~M. Campbell, T.~Cohen, N.~Dhingra, J.~Hirschauer, et~al., {\it
  {Methods and Results for Standard Model Event Generation at $\sqrt{s}$ = 14
  TeV, 33 TeV and 100 TeV Proton Colliders (A Snowmass Whitepaper)}},
  \href{http://arxiv.org/abs/1308.1636}{{\tt arXiv:1308.1636}}.

\bibitem{Feldman:2007zn}
D.~Feldman, Z.~Liu, and P.~Nath, {\it {The Landscape of Sparticle Mass
  Hierarchies and Their Signature Space at the LHC}},  {\em Phys. Rev. Lett.}
  {\bf 99} (2007) 251802, [\href{http://arxiv.org/abs/0707.1873}{{\tt
  arXiv:0707.1873}}].

\bibitem{Feldman:2007fq}
D.~Feldman, Z.~Liu, and P.~Nath, {\it {Light Higgses at the Tevatron and at the
  LHC and Observable Dark Matter in SUGRA and D Branes}},  {\em Phys. Lett. B}
  {\bf 662} (2008) 190--198, [\href{http://arxiv.org/abs/0711.4591}{{\tt
  arXiv:0711.4591}}].

\bibitem{Feldman:2008hs}
D.~Feldman, Z.~Liu, and P.~Nath, {\it {Sparticles at the LHC}},  {\em JHEP}
  {\bf 0804} (2008) 054, [\href{http://arxiv.org/abs/0802.4085}{{\tt
  arXiv:0802.4085}}].

\bibitem{Chen:2010kq}
N.~Chen, D.~Feldman, Z.~Liu, P.~Nath, and G.~Peim, {\it {Low Mass Gluino within
  the Sparticle Landscape, Implications for Dark Matter, and Early Discovery
  Prospects at LHC-7}},  {\em Phys. Rev. D} {\bf 83} (2011) 035005,
  [\href{http://arxiv.org/abs/1011.1246}{{\tt arXiv:1011.1246}}].

\bibitem{Berger:2008cq}
C.~F. Berger, J.~S. Gainer, J.~L. Hewett, and T.~G. Rizzo, {\it {Supersymmetry
  Without Prejudice}},  {\em JHEP} {\bf 0902} (2009) 023,
  [\href{http://arxiv.org/abs/0812.0980}{{\tt arXiv:0812.0980}}].

\bibitem{Conley:2010du}
J.~A. Conley, J.~S. Gainer, J.~L. Hewett, M.~P. Le, and T.~G. Rizzo, {\it
  {Supersymmetry Without Prejudice at the LHC}},  {\em Eur. Phys. J. C} {\bf
  71} (2011) 1697, [\href{http://arxiv.org/abs/1009.2539}{{\tt
  arXiv:1009.2539}}].

\bibitem{Altunkaynak:2010tn}
B.~Altunkaynak, B.~D. Nelson, L.~L. Everett, Y.~Rao, and I.-W. Kim, {\it
  {Landscape of Supersymmetric Particle Mass Hierarchies in Deflected Mirage
  Mediation}},  {\em Eur. Phys. J. Plus} {\bf 127} (2012) 2,
  [\href{http://arxiv.org/abs/1011.1439}{{\tt arXiv:1011.1439}}].

\bibitem{Nath:2010zj}
P.~Nath, B.~D. Nelson, H.~Davoudiasl, B.~Dutta, D.~Feldman, et~al., {\it {The
  Hunt for New Physics at the Large Hadron Collider}},  {\em Nucl. Phys. Proc.
  Suppl.} {\bf 200-202} (2010) 185--417,
  [\href{http://arxiv.org/abs/1001.2693}{{\tt arXiv:1001.2693}}].

\bibitem{Feng:2013mea}
W.-Z. Feng and P.~Nath, {\it {Higgs diphoton rate and mass enhancement with
  vectorlike leptons and the scale of supersymmetry}},  {\em Phys. Rev. D} {\bf
  87} (2013), no.~7 075018, [\href{http://arxiv.org/abs/1303.0289}{{\tt
  arXiv:1303.0289}}].

\bibitem{ArkaniHamed:2007fw}
N.~Arkani-Hamed, P.~Schuster, N.~Toro, J.~Thaler, L.-T. Wang, et~al., {\it
  {MARMOSET: The Path from LHC Data to the New Standard Model via On-Shell
  Effective Theories}},  \href{http://arxiv.org/abs/hep-ph/0703088}{{\tt
  hep-ph/0703088}}.

\bibitem{Alwall:2008ag}
J.~Alwall, P.~Schuster, and N.~Toro, {\it {Simplified Models for a First
  Characterization of New Physics at the LHC}},  {\em Phys. Rev. D} {\bf 79}
  (2009) 075020, [\href{http://arxiv.org/abs/0810.3921}{{\tt
  arXiv:0810.3921}}].

\bibitem{Alwall:2008va}
J.~Alwall, M.-P. Le, M.~Lisanti, and J.~G. Wacker, {\it {Model-Independent Jets
  plus Missing Energy Searches}},  {\em Phys. Rev. D} {\bf 79} (2009) 015005,
  [\href{http://arxiv.org/abs/0809.3264}{{\tt arXiv:0809.3264}}].

\bibitem{Alves:2011sq}
D.~S. Alves, E.~Izaguirre, and J.~G. Wacker, {\it {Where the Sidewalk Ends:
  Jets and Missing Energy Search Strategies for the 7 TeV LHC}},  {\em JHEP}
  {\bf 1110} (2011) 012, [\href{http://arxiv.org/abs/1102.5338}{{\tt
  arXiv:1102.5338}}].

\bibitem{Alves:2011wf}
{\bf LHC New Physics Working Group} , D.~Alves et~al., {\it {Simplified Models
  for LHC New Physics Searches}},  {\em J. Phys. G} {\bf 39} (2012) 105005,
  [\href{http://arxiv.org/abs/1105.2838}{{\tt arXiv:1105.2838}}].

\bibitem{Papucci:2011wy}
M.~Papucci, J.~T. Ruderman, and A.~Weiler, {\it {Natural SUSY Endures}},  {\em
  JHEP} {\bf 1209} (2012) 035, [\href{http://arxiv.org/abs/1110.6926}{{\tt
  arXiv:1110.6926}}].

\bibitem{Mahbubani:2012qq}
R.~Mahbubani, M.~Papucci, G.~Perez, J.~T. Ruderman, and A.~Weiler, {\it {Light
  Nondegenerate Squarks at the LHC}},  {\em Phys. Rev. Lett.} {\bf 110} (2013),
  no.~15 151804, [\href{http://arxiv.org/abs/1212.3328}{{\tt
  arXiv:1212.3328}}].

\bibitem{Chatrchyan:2013sza}
{\bf CMS Collaboration} , S.~Chatrchyan et~al., {\it {Interpretation of
  Searches for Supersymmetry with simplified Models}},  {\em Phys. Rev. D} {\bf
  88} (2013), no.~5 052017, [\href{http://arxiv.org/abs/1301.2175}{{\tt
  arXiv:1301.2175}}].

\bibitem{Cohen:2013xda}
T.~Cohen, T.~Golling, M.~Hance, A.~Henrichs, K.~Howe, et~al., {\it {SUSY
  Simplified Models at 14, 33, and 100 TeV Proton Colliders}},
  \href{http://arxiv.org/abs/1311.6480}{{\tt arXiv:1311.6480}}.

\bibitem{Anderson:1996bg}
G.~Anderson, C.~Chen, J.~Gunion, J.~D. Lykken, T.~Moroi, et~al., {\it
  {Motivations for and implications of nonuniversal GUT scale boundary
  conditions for soft SUSY breaking parameters}},  {\em eConf} {\bf C960625}
  (1996) SUP107, [\href{http://arxiv.org/abs/hep-ph/9609457}{{\tt
  hep-ph/9609457}}].

\bibitem{Nath:1997qm}
P.~Nath and R.~L. Arnowitt, {\it {Nonuniversal soft SUSY breaking and dark
  matter}},  {\em Phys. Rev. D} {\bf 56} (1997) 2820--2832,
  [\href{http://arxiv.org/abs/hep-ph/9701301}{{\tt hep-ph/9701301}}].

\bibitem{Ellis:2002wv}
J.~R. Ellis, K.~A. Olive, and Y.~Santoso, {\it {The MSSM parameter space with
  nonuniversal Higgs masses}},  {\em Phys. Lett. B} {\bf 539} (2002) 107--118,
  [\href{http://arxiv.org/abs/hep-ph/0204192}{{\tt hep-ph/0204192}}].

\bibitem{Anderson:1999uia}
G.~Anderson, H.~Baer, C.-h. Chen, and X.~Tata, {\it {The Reach of Fermilab
  Tevatron upgrades for SU(5) supergravity models with nonuniversal gaugino
  masses}},  {\em Phys. Rev. D} {\bf 61} (2000) 095005,
  [\href{http://arxiv.org/abs/hep-ph/9903370}{{\tt hep-ph/9903370}}].

\bibitem{Huitu:1999vx}
K.~Huitu, Y.~Kawamura, T.~Kobayashi, and K.~Puolamaki, {\it {Phenomenological
  constraints on SUSY SU(5) GUTs with nonuniversal gaugino masses}},  {\em
  Phys. Rev. D} {\bf 61} (2000) 035001,
  [\href{http://arxiv.org/abs/hep-ph/9903528}{{\tt hep-ph/9903528}}].

\bibitem{Corsetti:2000yq}
A.~Corsetti and P.~Nath, {\it {Gaugino mass nonuniversality and dark matter in
  SUGRA, strings and D-brane models}},  {\em Phys. Rev. D} {\bf 64} (2001)
  125010, [\href{http://arxiv.org/abs/hep-ph/0003186}{{\tt hep-ph/0003186}}].

\bibitem{Chattopadhyay:2001mj}
U.~Chattopadhyay and P.~Nath, {\it {b - tau unification, g(mu) - 2, the b $\to$
  s + gamma constraint and nonuniversalities}},  {\em Phys. Rev. D} {\bf 65}
  (2002) 075009, [\href{http://arxiv.org/abs/hep-ph/0110341}{{\tt
  hep-ph/0110341}}].

\bibitem{Chattopadhyay:2001va}
U.~Chattopadhyay, A.~Corsetti, and P.~Nath, {\it {Supersymmetric dark matter
  and Yukawa unification}},  {\em Phys. Rev. D} {\bf 66} (2002) 035003,
  [\href{http://arxiv.org/abs/hep-ph/0201001}{{\tt hep-ph/0201001}}].

\bibitem{Martin:2009ad}
S.~P. Martin, {\it {Non-universal gaugino masses from non-singlet F-terms in
  non-minimal unified models}},  {\em Phys. Rev. D} {\bf 79} (2009) 095019,
  [\href{http://arxiv.org/abs/0903.3568}{{\tt arXiv:0903.3568}}].

\bibitem{Feldman:2009zc}
D.~Feldman, Z.~Liu, and P.~Nath, {\it {Gluino NLSP, Dark Matter via Gluino
  Coannihilation, and LHC Signatures}},  {\em Phys. Rev. D} {\bf 80} (2009)
  015007, [\href{http://arxiv.org/abs/0905.1148}{{\tt arXiv:0905.1148}}].

\bibitem{Gogoladze:2012yf}
I.~Gogoladze, F.~Nasir, and Q.~Shafi, {\it {Non-Universal Gaugino Masses and
  Natural Supersymmetry}},  {\em Int. J. Mod. Phys. A} {\bf 28} (2013) 1350046,
  [\href{http://arxiv.org/abs/1212.2593}{{\tt arXiv:1212.2593}}].

\bibitem{Ajaib:2013zha}
M.~Adeel~Ajaib, I.~Gogoladze, Q.~Shafi, and C.~S. Un, {\it {A Predictive Yukawa
  Unified SO(10) Model: Higgs and Sparticle Masses}},  {\em JHEP} {\bf 1307}
  (2013) 139, [\href{http://arxiv.org/abs/1303.6964}{{\tt arXiv:1303.6964}}].

\bibitem{Kaufman:2013oaa}
B.~Kaufman and B.~D. Nelson, {\it {Mirage Models Confront the LHC: II.
  Flux-Stabilized Type IIB String Theory}},  {\em Phys. Rev. D} {\bf 89} (2014)
  085029, [\href{http://arxiv.org/abs/1312.6621}{{\tt arXiv:1312.6621}}].

\bibitem{susykit}
S.~Akula, {\em SusyKit}, \url{http://freeboson.org/software/}.

\bibitem{Allanach:2001kg}
B.~C. Allanach, {\it {SOFTSUSY: a program for calculating supersymmetric
  spectra}},  {\em Comput. Phys. Commun.} {\bf 143} (2002) 305--331,
  [\href{http://arxiv.org/abs/hep-ph/0104145}{{\tt hep-ph/0104145}}].

\bibitem{Heinemeyer:1998yj}
S.~Heinemeyer, W.~Hollik, and G.~Weiglein, {\it {FeynHiggs: A Program for the
  calculation of the masses of the neutral CP even Higgs bosons in the MSSM}},
  {\em Comput. Phys. Commun.} {\bf 124} (2000) 76--89,
  [\href{http://arxiv.org/abs/hep-ph/9812320}{{\tt hep-ph/9812320}}].

\bibitem{Hahn:2010te}
T.~Hahn, S.~Heinemeyer, W.~Hollik, H.~Rzehak, and G.~Weiglein, {\it {FeynHiggs
  2.7}},  {\em Nucl. Phys. Proc. Suppl.} {\bf 205-206} (2010) 152--157,
  [\href{http://arxiv.org/abs/1007.0956}{{\tt arXiv:1007.0956}}].

\bibitem{Hahn:2013ria}
T.~Hahn, S.~Heinemeyer, W.~Hollik, H.~Rzehak, and G.~Weiglein, {\it
  {High-precision predictions for the light CP-even Higgs Boson Mass of the
  MSSM}},  {\em Phys. Rev. Lett.} {\bf 112} (2014) 141801,
  [\href{http://arxiv.org/abs/1312.4937}{{\tt arXiv:1312.4937}}].

\bibitem{Belanger:2010st}
G.~Belanger, N.~D. Christensen, A.~Pukhov, and A.~Semenov, {\it {SLHAplus: a
  library for implementing extensions of the standard model}},  {\em Comput.
  Phys. Commun.} {\bf 182} (2011) 763--774,
  [\href{http://arxiv.org/abs/1008.0181}{{\tt arXiv:1008.0181}}].

\bibitem{Fowlie:2014awa}
A.~Fowlie and M.~Raidal, {\it {Prospects for constrained supersymmetry at
  $\sqrt{s}={33}\,\text {TeV} $ and $\sqrt{s}={100}\,\text {TeV} $
  proton-proton super-colliders}},  {\em Eur. Phys. J. C} {\bf 74} (2014) 2948,
  [\href{http://arxiv.org/abs/1402.5419}{{\tt arXiv:1402.5419}}].

\bibitem{Kim:2013uxa}
D.~Kim, P.~Athron, C.~Bal\'azs, B.~Farmer, and E.~Hutchison, {\it {Bayesian
  naturalness of the CMSSM and CNMSSM}},  {\em Phys. Rev. D} {\bf 90} (2014)
  055008, [\href{http://arxiv.org/abs/1312.4150}{{\tt arXiv:1312.4150}}].

\bibitem{Fowlie:2014faa}
A.~Fowlie, {\it {Is the CNMSSM more natural than the CMSSM?}},
  \href{http://arxiv.org/abs/1407.7534}{{\tt arXiv:1407.7534}}.

\bibitem{Roszkowski:2014wqa}
L.~Roszkowski, E.~M. Sessolo, and A.~J. Williams, {\it {What next for the CMSSM
  and the NUHM: Improved prospects for superpartner and dark matter
  detection}},  {\em JHEP} {\bf 1408} (2014) 067,
  [\href{http://arxiv.org/abs/1405.4289}{{\tt arXiv:1405.4289}}].

\bibitem{susy2014}
P.~Nath, ``Guts and susy guts.'' SUSY2014, Manchester, UK, July 20-26, 2014.

\bibitem{Chamseddine:1983eg}
A.~H. Chamseddine, P.~Nath, and R.~L. Arnowitt, {\it {Experimental Signals for
  Supersymmetric Decays of the $W$ and $Z$ Bosons}},  {\em Phys. Lett. B} {\bf
  129} (1983) 445.

\bibitem{Dicus:1983cb}
D.~A. Dicus, S.~Nandi, and X.~Tata, {\it {$W$ Decay in Supergravity {GUTs}}},
  {\em Phys. Lett. B} {\bf 129} (1983) 451.

\bibitem{Baer:1986vf}
H.~Baer, K.~Hagiwara, and X.~Tata, {\it {Gauginos as a Signal for Supersymmetry
  at p anti-p Colliders}},  {\em Phys. Rev. D} {\bf 35} (1987) 1598.

\bibitem{Nath:1987sw}
P.~Nath and R.~L. Arnowitt, {\it {Supersymmetric Signals at the Tevatron}},
  {\em Mod. Phys. Lett. A} {\bf 2} (1987) 331--341.

\bibitem{Baer:2012vr}
H.~Baer, V.~Barger, A.~Lessa, and X.~Tata, {\it {Discovery potential for SUSY
  at a high luminosity upgrade of LHC14}},
  \href{http://arxiv.org/abs/1207.4846}{{\tt arXiv:1207.4846}}.

\bibitem{Altunkaynak:2013xya}
B.~Altunkaynak, C.~Kao, and K.~Yang, {\it {Unveiling the MSSM Neutral Higgs
  Bosons with Leptons and a Bottom Quark}},
  \href{http://arxiv.org/abs/1312.3011}{{\tt arXiv:1312.3011}}.

\bibitem{Chan:1997bi}
K.~L. Chan, U.~Chattopadhyay, and P.~Nath, {\it {Naturalness, weak scale
  supersymmetry and the prospect for the observation of supersymmetry at the
  Tevatron and at the CERN LHC}},  {\em Phys. Rev. D} {\bf 58} (1998) 096004,
  [\href{http://arxiv.org/abs/hep-ph/9710473}{{\tt hep-ph/9710473}}].

\bibitem{Chattopadhyay:2003qh}
U.~Chattopadhyay, A.~Corsetti, and P.~Nath, {\it {WMAP data and recent
  developments in supersymmetric dark matter}},  {\em Phys. Atom. Nucl.} {\bf
  67} (2004) 1188--1194, [\href{http://arxiv.org/abs/hep-ph/0310228}{{\tt
  hep-ph/0310228}}].

\bibitem{Baer:2003wx}
H.~Baer, C.~Balazs, A.~Belyaev, T.~Krupovnickas, and X.~Tata, {\it {Updated
  reach of the CERN LHC and constraints from relic density, $b \to s \gamma$
  and a($\mu$) in the mSUGRA model}},  {\em JHEP} {\bf 0306} (2003) 054,
  [\href{http://arxiv.org/abs/hep-ph/0304303}{{\tt hep-ph/0304303}}].

\bibitem{Akula:2011jx}
S.~Akula, M.~Liu, P.~Nath, and G.~Peim, {\it {Naturalness, Supersymmetry and
  Implications for LHC and Dark Matter}},  {\em Phys. Lett. B} {\bf 709} (2012)
  192--199, [\href{http://arxiv.org/abs/1111.4589}{{\tt arXiv:1111.4589}}].

\bibitem{Liu:2013ula}
M.~Liu and P.~Nath, {\it {Higgs boson mass, proton decay, naturalness, and
  constraints of the LHC and Planck data}},  {\em Phys. Rev. D} {\bf 87}
  (2013), no.~9 095012, [\href{http://arxiv.org/abs/1303.7472}{{\tt
  arXiv:1303.7472}}].

\bibitem{Aad:2014nua}
{\bf ATLAS Collaboration} , G.~Aad et~al., {\it {Search for direct production
  of charginos and neutralinos in events with three leptons and missing
  transverse momentum in $\sqrt{s} =$ 8TeV $pp$ collisions with the ATLAS
  detector}},  {\em JHEP} {\bf 1404} (2014) 169,
  [\href{http://arxiv.org/abs/1402.7029}{{\tt arXiv:1402.7029}}].

\bibitem{Altunkaynak:2010we}
B.~Altunkaynak, M.~Holmes, P.~Nath, B.~D. Nelson, and G.~Peim, {\it {SUSY
  Discovery Potential and Benchmarks for Early Runs at $\sqrt s =7$ TeV at the
  LHC}},  {\em Phys. Rev. D} {\bf 82} (2010) 115001,
  [\href{http://arxiv.org/abs/1008.3423}{{\tt arXiv:1008.3423}}].

\bibitem{Sjostrand:2006za}
T.~Sjostrand, S.~Mrenna, and P.~Z. Skands, {\it {PYTHIA 6.4 Physics and
  Manual}},  {\em JHEP} {\bf 0605} (2006) 026,
  [\href{http://arxiv.org/abs/hep-ph/0603175}{{\tt hep-ph/0603175}}].

\bibitem{deFavereau:2013fsa}
{\bf DELPHES 3} , J.~de~Favereau et~al., {\it {DELPHES 3, A modular framework
  for fast simulation of a generic collider experiment}},  {\em JHEP} {\bf
  1402} (2014) 057, [\href{http://arxiv.org/abs/1307.6346}{{\tt
  arXiv:1307.6346}}].

\bibitem{Beenakker:1996ed}
W.~Beenakker, R.~Hopker, and M.~Spira, {\it {PROSPINO: A Program for the
  production of supersymmetric particles in next-to-leading order QCD}},
  \href{http://arxiv.org/abs/hep-ph/9611232}{{\tt hep-ph/9611232}}.

\bibitem{Aad:2014lra}
{\bf ATLAS Collaboration} , G.~Aad et~al., {\it {Search for strong production
  of supersymmetric particles in final states with missing transverse momentum
  and at least three $b$-jets at $\sqrt{s} =$ 8 TeV proton-proton collisions
  with the ATLAS detector}},  {\em JHEP} {\bf 1410} (2014) 24,
  [\href{http://arxiv.org/abs/1407.0600}{{\tt arXiv:1407.0600}}].

\bibitem{Cushman:2013zza}
P.~Cushman, C.~Galbiati, D.~McKinsey, H.~Robertson, T.~Tait, et~al., {\it
  {Working Group Report: WIMP Dark Matter Direct Detection}},
  \href{http://arxiv.org/abs/1310.8327}{{\tt arXiv:1310.8327}}.

\bibitem{Chattopadhyay:1998wb}
U.~Chattopadhyay, T.~Ibrahim, and P.~Nath, {\it {Effects of CP violation on
  event rates in the direct detection of dark matter}},  {\em Phys. Rev. D}
  {\bf 60} (1999) 063505, [\href{http://arxiv.org/abs/hep-ph/9811362}{{\tt
  hep-ph/9811362}}].

\bibitem{Ellis:2000ds}
J.~R. Ellis, A.~Ferstl, and K.~A. Olive, {\it {Reevaluation of the elastic
  scattering of supersymmetric dark matter}},  {\em Phys. Lett. B} {\bf 481}
  (2000) 304--314, [\href{http://arxiv.org/abs/hep-ph/0001005}{{\tt
  hep-ph/0001005}}].

\bibitem{Belanger:2008sj}
G.~Belanger, F.~Boudjema, A.~Pukhov, and A.~Semenov, {\it {Dark matter direct
  detection rate in a generic model with micrOMEGAs 2.2}},  {\em Comput. Phys.
  Commun.} {\bf 180} (2009) 747--767,
  [\href{http://arxiv.org/abs/0803.2360}{{\tt arXiv:0803.2360}}].

\end{thebibliography}\endgroup

\end{document}